\documentclass[12pt,preprint,trackchanges]{aastex62}
\usepackage{natbib}
\usepackage{xcolor}

\newcommand{\sqdeg}{deg$^{2}$}
\newcommand{\sqam}{arcmin$^{2}$}
\newcommand{\sqas}{arcsec$^{2}$}

\graphicspath{{./}{figures/}}
\turnoffeditone

\accepted{September 29, 2021}
\submitjournal{AJ}

\shorttitle{HETDEX Instrumentation}
\shortauthors{Hill et al.}
\begin{document}

\title{The HETDEX Instrumentation: Hobby-Eberly Telescope Wide Field Upgrade and VIRUS
\footnote{Based on observations obtained with the Hobby-Eberly Telescope, which is a joint project of the University of Texas at Austin, the Pennsylvania State University, Ludwig-Maximilians-Universit\"at M\"unchen, and Georg-August-Universit\"at G\"ottingen.}
}

\correspondingauthor{Gary J. Hill}
\email{hill@astro.as.utexas.edu}

\author{Gary J. Hill}
\affiliation{McDonald Observatory, University of Texas at Austin, 2515 Speedway, Stop C1402, Austin, TX 78712, USA}
\affiliation{Department of Astronomy, University of Texas at Austin, 2515 Speedway, Stop C1400, Austin, Texas 78712, USA}

\author{Hanshin Lee}
\affiliation{McDonald Observatory, University of Texas at Austin, 2515 Speedway, Stop C1402, Austin, TX 78712, USA}

\author{Phillip J. MacQueen}
\affiliation{McDonald Observatory, University of Texas at Austin, 2515 Speedway, Stop C1402, Austin, TX 78712, USA}

\author{Andreas Kelz}
\affiliation{Leibniz Institute for Astrophysics (AIP), An der Sternwarte 16, 14482 Potsdam, Germany}
\affiliation{innoFSPEC Potsdam, An der Sternwarte 16, 14482 Potsdam, Germany}

\author{Niv Drory}
\affiliation{McDonald Observatory, University of Texas at Austin, 2515 Speedway, Stop C1402, Austin, TX 78712, USA}

\author{Brian L. Vattiat}
\affiliation{McDonald Observatory, University of Texas at Austin, 2515 Speedway, Stop C1402, Austin, TX 78712, USA}

% engineers, authors of design and construction
\author{John M. Good}
\affiliation{McDonald Observatory, University of Texas at Austin, 2515 Speedway, Stop C1402, Austin, TX 78712, USA}

\author{Jason Ramsey}
\affiliation{McDonald Observatory, University of Texas at Austin, 2515 Speedway, Stop C1402, Austin, TX 78712, USA}

\author{Herman Kriel}
\affiliation{Hobby-Eberly Telescope, 32 Mt. Locke Rd., McDonald Observatory, TX 79734, USA}

\author{Trent Peterson}
\affiliation{McDonald Observatory, University of Texas at Austin, 2515 Speedway, Stop C1402, Austin, TX 78712, USA}

\author{D.L. DePoy}
\affiliation{Department of Physics and Astronomy, Texas A\&M University, 4242 TAMU, College Station, TX  77843, USA}

\author{Karl Gebhardt}
\affiliation{Department of Astronomy, University of Texas at Austin, 2515 Speedway, Stop C1400, Austin, Texas 78712, USA}

\author{J.L. Marshall}
\affiliation{Department of Physics and Astronomy, Texas A\&M University, 4242 TAMU, College Station, TX  77843, USA}

\author{Sarah E. Tuttle}
\affiliation{McDonald Observatory, University of Texas at Austin, 2515 Speedway, Stop C1402, Austin, TX 78712, USA}
\affiliation{Department of Astronomy, University of Washington, Seattle, WA 98195, USA}

% key technical staff & others
\author{Svend M. Bauer}
\affiliation{Leibniz Institute for Astrophysics (AIP), An der Sternwarte 16, 14482 Potsdam, Germany}
%\affiliation{innoFSPEC Potsdam, An der Sternwarte 16, 14482 Potsdam, Germany}

\author{Taylor S. Chonis}
\affiliation{Department of Astronomy, University of Texas at Austin, 2515 Speedway, Stop C1400, Austin, Texas 78712, USA}
\affiliation{Ball Aerospace, Boulder, CO, USA}

\author{Maximilian H. Fabricius}
\affiliation{Universit{\" a}ts-Sternwarte München, Ludwig-Maximilians-Universit\"at M\"unchen, Scheinerstr. 1, 81679 M{\" u}nchen, Germany}
\affiliation{Max-Planck-Institut f{\" u}r Extraterrestriche-Physik, Giessenbachstrasse, D-85748 Garching b. M{\" u}nchen, Germany}

\author{Cynthia Froning}
\affiliation{McDonald Observatory, University of Texas at Austin, 2515 Speedway, Stop C1402, Austin, TX 78712, USA}

\author{Marco H{\" a}user}
\affiliation{Universit{\" a}ts-Sternwarte München, Ludwig-Maximilians-Universit\"at M\"unchen, Scheinerstr. 1, 81679 M{\" u}nchen, Germany}
\affiliation{Max-Planck-Institut f{\" u}r Extraterrestriche-Physik, Giessenbachstrasse, D-85748 Garching b. M{\" u}nchen, Germany}

\author{Briana L. Indahl}
\affiliation{Department of Astronomy, University of Texas at Austin, 2515 Speedway, Stop C1400, Austin, Texas 78712, USA}

\author{Thomas Jahn}
\affiliation{Leibniz Institute for Astrophysics (AIP), An der Sternwarte 16, 14482 Potsdam, Germany}
\affiliation{innoFSPEC Potsdam, An der Sternwarte 16, 14482 Potsdam, Germany}

\author{Martin Landriau}
\affiliation{McDonald Observatory, University of Texas at Austin, 2515 Speedway, Stop C1402, Austin, TX 78712, USA}
\affiliation{Max-Planck-Institut f{\" u}r Extraterrestriche-Physik, Giessenbachstrasse, D-85748 Garching b. M{\" u}nchen, Germany}
\affiliation{Lawrence Berkeley National Laboratory, 1 Cyclotron Road Mailstop 50R5008, Berkeley, CA
94720, USA}

\author{Ron Leck}
\affiliation{McDonald Observatory, University of Texas at Austin, 2515 Speedway, Stop C1402, Austin, TX 78712, USA}

\author{Francesco Montesano}
%\affiliation{Universit{\" a}ts-Sternwarte München, Scheinerstr. 1, 81679 M{\" u}nchen, Germany}
\affiliation{Max-Planck-Institut f{\" u}r Extraterrestriche-Physik, Giessenbachstrasse, D-85748 Garching b. M{\" u}nchen, Germany}

\author{Travis Prochaska}
\affiliation{Department of Physics and Astronomy, Texas A\&M University, 4242 TAMU, College Station, TX  77843, USA}

\author{Jan M. Snigula}
\affiliation{Universit{\" a}ts-Sternwarte München, Ludwig-Maximilians-Universit\"at M\"unchen, Scheinerstr. 1, 81679 M{\" u}nchen, Germany}
\affiliation{Max-Planck-Institut f{\" u}r Extraterrestriche-Physik, Giessenbachstrasse, D-85748 Garching b. M{\" u}nchen, Germany}

\author{Gregory R. Zeimann}
\affiliation{Hobby-Eberly Telescope, 32 Mt. Locke Rd., McDonald Observatory, TX 79734, USA}
\affiliation{McDonald Observatory, University of Texas at Austin, 2515 Speedway, Stop C1402, Austin, TX 78712, USA}

% HET staff
\author{Randy Bryant}
\affiliation{Hobby-Eberly Telescope, 32 Mt. Locke Rd., McDonald Observatory, TX 79734, USA}

\author{George Damm}
\affiliation{Hobby-Eberly Telescope, 32 Mt. Locke Rd., McDonald Observatory, TX 79734, USA}

\author{J.R. Fowler}
\affiliation{Hobby-Eberly Telescope, 32 Mt. Locke Rd., McDonald Observatory, TX 79734, USA}

\author{Steven Janowiecki}
\affiliation{Hobby-Eberly Telescope, 32 Mt. Locke Rd., McDonald Observatory, TX 79734, USA}

\author{Jerry Martin}
\affiliation{Hobby-Eberly Telescope, 32 Mt. Locke Rd., McDonald Observatory, TX 79734, USA}

\author{Emily Mrozinski}
\affiliation{Hobby-Eberly Telescope, 32 Mt. Locke Rd., McDonald Observatory, TX 79734, USA}

\author{Stephen Odewahn}
\affiliation{Hobby-Eberly Telescope, 32 Mt. Locke Rd., McDonald Observatory, TX 79734, USA}

\author{Sergey Rostopchin}
\affiliation{Hobby-Eberly Telescope, 32 Mt. Locke Rd., McDonald Observatory, TX 79734, USA}

\author{Matthew Shetrone}
\affiliation{Hobby-Eberly Telescope, 32 Mt. Locke Rd., McDonald Observatory, TX 79734, USA}
\affiliation{University of California Observatories, UC Santa Cruz, 1156 High Street, Santa Cruz, CA 95064 }

\author{Renny Spencer}
\affiliation{Hobby-Eberly Telescope, 32 Mt. Locke Rd., McDonald Observatory, TX 79734, USA}

\author{Erin Mentuch Cooper}
\affiliation{Department of Astronomy, University of Texas at Austin, 2515 Speedway, Stop C1400, Austin, Texas 78712, USA}

% Administrators, facilitators, others
\author{Taft Armandroff}
\affiliation{McDonald Observatory, University of Texas at Austin, 2515 Speedway, Stop C1402, Austin, TX 78712, USA}
\affiliation{Department of Astronomy, University of Texas at Austin, 2515 Speedway, Stop C1400, Austin, Texas 78712, USA}

\author{Ralf Bender}
\affiliation{Universit{\" a}ts-Sternwarte München, Ludwig-Maximilians-Universit\"at
M\"unchen, Scheinerstr. 1, 81679 M{\" u}nchen, Germany}
\affiliation{Max-Planck-Institut f{\" u}r Extraterrestriche-Physik, Giessenbachstrasse, D-85748 Garching b. M{\" u}nchen, Germany}

\author{Gavin Dalton}
\affiliation{Department of Physics, University of Oxford, Keble Road, Oxford, OX1 3RH, UK }
\affiliation{ STFC RALSpace, HSIC, Didcot, OX11 0QX, UK}

\author{Ulrich Hopp}
\affiliation{Universit{\" a}ts-Sternwarte München, Ludwig-Maximilians-Universit\"at M\"unchen, Scheinerstr. 1, 81679 M{\" u}nchen, Germany}
\affiliation{Max-Planck-Institut f{\" u}r Extraterrestriche-Physik, Giessenbachstrasse, D-85748 Garching b. M{\" u}nchen, Germany}

\author{Eiichiro Komatsu}
\affiliation{Max-Planck-Institut f{\" u}r Astrophysik, Karl-Schwarzschild-Strasse 1, D-85748 Garching b. M{\" u}nchen, Germany}
\affiliation{Kavli Institute for the Physics and Mathematics of the Universe, Todai Institutes for Advanced Study, the University of Tokyo, Kashiwa, Japan 277-8583 (Kavli IPMU, WPI)}

%\author{David L. Lambert}
%\affiliation{Department of Astronomy, University of Texas at Austin, 2515 Speedway, Stop C1400, Austin, Texas 78712, USA}
%\affiliation{McDonald Observatory, University of Texas at Austin, 2515 Speedway, Stop C1402, Austin, TX 78712, USA}

\author{Harald Nicklas}
\affiliation{Institut f{\" u}r Astrophysik G{\" o}ttingen, Friedrich-Hund-Platz 1, 37077 G{\" o}ttingen, Germany}

\author{Lawrence W. Ramsey}
\affiliation{Department of Astronomy, Pennsylvania State University, 516 Davey Lab, University Park, PA 16802, USA}

\author{Martin M. Roth}
\affiliation{Leibniz Institute for Astrophysics (AIP), An der Sternwarte 16, 14482 Potsdam, Germany}
\affiliation{innoFSPEC Potsdam, An der Sternwarte 16, 14482 Potsdam, Germany}

\author{Donald P. Schneider}
\affiliation{Department of Astronomy, Pennsylvania State University, 516 Davey Lab, University Park, PA 16802, USA}

\author{Chris Sneden}
\affiliation{Department of Astronomy, University of Texas at Austin, 2515 Speedway, Stop C1400, Austin, Texas 78712, USA}

\author{Matthias Steinmetz}
\affiliation{Leibniz Institute for Astrophysics (AIP), An der Sternwarte 16, 14482 Potsdam, Germany}
%\affiliation{innoFSPEC Potsdam, An der Sternwarte 16, 14482 Potsdam, Germany}



\begin{abstract}

%this counts out to 250 words - says 2 words left upon submission
The Hobby-Eberly Telescope (HET) Dark Energy Experiment (HETDEX) is undertaking a blind wide-field low-resolution spectroscopic survey of 540 \sqdeg~ of sky 
to identify and derive redshifts for a million Lyman-$\alpha$ emitting galaxies (LAEs) in the redshift range $1.9 < z < 3.5$.  The ultimate goal is to measure the 
expansion rate of the Universe at this epoch, to sharply constrain cosmological parameters and thus the nature of dark energy.
A major multi-year wide field upgrade (WFU) of the HET was completed in 2016 that substantially increased the field of view to 22 arcminutes diameter and the pupil to 10 meters, by replacing the optical corrector, tracker, and prime focus instrument package and by developing a new telescope control system. The new, wide-field HET now feeds the Visible Integral-field Replicable Unit Spectrograph (VIRUS), a new low-resolution integral field spectrograph (LRS2), and the Habitable Zone Planet Finder (HPF), a precision near-infrared radial velocity spectrograph.  VIRUS consists of 156 identical spectrographs fed by almost 35,000 fibers in 78 integral field units arrayed at the focus of the upgraded HET.  VIRUS operates in a bandpass of 3500$-$5500 \AA~ with resolving power $R~\simeq$~800. 
VIRUS is the first example of large-scale replication applied to instrumentation in optical astronomy to achieve spectroscopic surveys of very large areas of sky.  This paper presents technical details of the HET WFU and VIRUS, as flowed-down from the HETDEX science requirements, along with experience from commissioning this major telescope upgrade and the innovative instrumentation suite for HETDEX.

\end{abstract}



\section{INTRODUCTION} \label{sec:intro}

The Hobby-Eberly Telescope Dark Energy Experiment (HETDEX\footnote{\url{http://www.hetdex.org/}};
\citealt{hil08b, hil16a, geb21}) aims to tightly constrain the expansion history of the Universe and thus the evolution of dark energy by detecting and mapping the spatial distribution of about a million Lyman-$\alpha$ emitting galaxies (LAEs).  
The redshift range for LAE detection will be $1.9 < z < 3.5$ over a total of 
$\sim 540$ \sqdeg\ (11 Gpc$^3$ comoving volume)
This survey is being carried out with the Visible Integral-field Replicable Unit Spectrograph (VIRUS, \citealt{hil18a})\footnote{VIRUS is a joint project of the University of Texas at Austin, Leibniz-Institut f{\" u}r Astrophysik Potsdam (AIP), Texas A\&M University (TAMU), Max-Planck-Institut f{\" u}r Extraterrestriche-Physik (MPE), Ludwig-Maximilians-Universit{\" a}t M{\" u}nchen, Pennsylvania State University, Institut f{\" u}r Astrophysik G{\" o}ttingen, University of Oxford, and the Max-Planck-Institut f{\" u}r Astrophysik (MPA).}.
VIRUS is a highly-replicated 
integral field spectrograph \citep{hil14}, 
designed for blind spectroscopic surveys, where
``high" or ``large-scale" replication is defined as consisting of greater than 100 copies of a base instrument. 
VIRUS is composed of a set of 156 integral field spectrographs, that produce about 35,000 spectra 
with spectral range 3500$-$5500 \AA, and resolving power $R = \Delta\lambda/\lambda \simeq 800$ (at 4500 \AA, $\Delta\lambda \simeq 5.6$ \AA)
in a single observation that covers 56 \sqam\
area within an 18 arcmin diameter field (fill factor $\simeq$ 1$/$4.5).
Achieving the HETDEX goals has required both the development and implementation of VIRUS and a major rebuild and enhancement of the Hobby-Eberly Telescope\footnote{The Hobby-Eberly Telescope is operated by McDonald Observatory on behalf of the University of Texas at Austin, Pennsylvania State University, Ludwig-Maximillians-Universit{\" a}t M{\" u}nchen, and Georg-August-Universit{\" a}t, G{\" o}ttingen. The HET is named in honor of its principal benefactors, William P. Hobby and Robert E. Eberly.} with a much larger field of view of 22 arcmin diameter.  
%Description of these innovations is the task of this paper.
This paper is focused on describing these innovations and evaluating the performance of the HET Wide Field Upgrade (WFU) and VIRUS against the requirements for HETDEX. 

The HET (\citealt{lwr94,lwr98}, Figure~\ref{HETlayout}, \edit1{Table~\ref{tab-het}}) is an innovative telescope that has an 11 m hexagonal-shaped spherical primary mirror made from 91 identical 1-m hexagonal segments that points at a fixed zenith angle of 35$^\circ$.  The HET can be moved in azimuth on air-bearings to access about 70\% of the sky visible at McDonald Observatory 
(declination $-$10.3$^\circ \leq \delta \leq $ +71.6$^\circ$).
Primary mirror alignment is achieved using instruments located in the Center of Curvature Alignment System 
%(CCAS) 
tower, accessed once or twice per night at a particular azimuth.
The nature of HET requires that observations be 100\% queue-scheduled \citep{shet07}.
The pupil was originally 9.2 m in diameter, set by the design of the prime focus spherical aberration corrector, and sweeps over the primary mirror as the x-y tracker follows objects for between 50 minutes (in the south at $\delta$= $-$10.0$^\circ$) and 2.8 hours (in the north at $\delta$ = $+$67.2$^\circ$).  The original 4-mirror double-Gregorian type corrector \citep{jung99} had a 4 arcmin (50 mm) diameter science field of view. 
\edit1{Table~\ref{tab-het} presents the basic properties of the original HET and of the upgraded HET, as described in this paper. The upgrade increased telescope aperture and field of view, while accessible sky and track times remain the same between the original and upgraded HET.}

HET is located on Mt. Fowlkes at McDonald Observatory in west Texas. The site is characterized by extremely dark skies and typical continental site median seeing of $\sim$1\farcs0 ~full-width half maximum (FWHM, \citealt{barker03}). 
The original purpose of the HET was to conduct spectroscopic surveys, using its large primary mirror to enable observations of many targets in a short period of time.  However, the 4$\arcmin$ field of view limited the HET in most cases to observations of one target at a time.  This clearly was inadequate for the goals of a large program such as HETDEX.  Therefore, a consortium
was formed to re-purpose the HET with a complete redesign of all mechanical and optical components beyond the 11 m primary mirror.  In parallel the massive VIRUS instrument was designed and implemented to observe for the first time large areas of sky in a blind spectroscopic survey.

The new instrument suite for the upgraded HET emphasizes the telescope’s strengths in large surveys and
synoptic time-domain spectroscopy.
All the new instrumentation is fiber-fed so as to exploit the azimuthal scrambling
inherent to fiber transmission.
This scrambling is particularly important for a telescope with a variable pupil illumination (such as the HET).
There are two low-resolution fiber integral field  
spectrographs: VIRUS, and the second-generation Low Resolution Spectrograph
(LRS2, \citealt{chonis16, hil21}),
and two fiber-fed high-resolution spectrographs: the Habitable-zone Planet Finder (HPF, \citealt{mahadevan18}) and a forthcoming upgrade of the HET High Resolution Spectrograph (HRS, \citealt{tull98});
the latter two
instruments reside in temperature-controlled enclosures located in the basement inside the telescope pier.
LRS2 is based on two VIRUS spectrograph units with the gratings replaced by higher dispersion grisms that span 3700~-~10500~\AA~ in four spectrograph channels. The units are designated LRS2-B and LRS2-R, and each is fed by a separate lenslet-coupled integral field unit (IFU) with 6 $\times$ 12 \sqas\  field coverage. 

Section \S\ref{sec:design} presents the overall design requirements for the upgraded HET and VIRUS.  Components of the HET wide field upgrade (WFU) are discussed in \S\ref{sec:wfupgrade} and the design of VIRUS and its support infrastructure in \S\ref{sec:virus}.
Performance of VIRUS is reviewed in \S\ref{sec:Vperformance}. Observing with the HETDEX instrumentation
is described in \S\ref{sec:WFUperformance} along with the performance of the current HETDEX system. Example spectra from VIRUS are presented in \S\ref{sec:spectra}. Conclusions 
are summarized in \S\ref{sec:summary}. An appendix lists the acronyms used.

\medskip
\section{HET AND VIRUS DESIGN REQUIREMENTS FOR HETDEX}\label{sec:design}

The requirement to survey large areas of sky with VIRUS plus the need to acquire wavefront sensing stars to provide full feedback on the tracker position led us to design an ambitious new corrector employing meter-scale aspheric mirrors and covering a 22-arcmin diameter field of view. The WFU (\citealt{hil18b, lee21}) deploys the wide field corrector (WFC, \citealt{burge10,oh14,good14a,lee16a}), a new tracker \citep{good18}, a new prime focus instrument package (PFIP, \citealt{vattiat14}), new software control systems \citep{beno12,rams16,rams18}, and new metrology systems \citep{lee18a,lee18b}.  The metrology systems provide closed-loop feedback on all axes of motion and the optical configuration of the telescope. The systems include guiding, wavefront sensing, payload tilt sensing, and a distance measuring interferometer. Together these instruments control the alignment of the WFC to the primary mirror as well as providing feedback on the temperature-dependent radius of curvature of the segmented primary mirror, which is mounted on a steel truss\footnote{The WFC is named the Harold C. Simmons Dark Energy Optical System.}.
The upgrade left the primary mirror and telescope enclosure unchanged. 
The timetable for the Wide Field Upgrade is presented in Table~\ref{tab-chron}.

Table~\ref{tab-wfupgrade} presents the high-level technical requirements for the WFU, derived from the original HETDEX science requirements as established for the Preliminary Design Review in 2008. The science requirements evolved in the subsequent decade, but there are no new requirements for which there is a technical shortfall. The WFU had additional requirements driven by other science use cases, but the HETDEX criteria are summarized. VIRUS was optimized for HETDEX. 

The design of VIRUS flows directly from the requirements for HETDEX (\citealt{hil08b, hil16a, geb21}),  to maximize the number of LAEs detected in a set observing time, and to span sufficient redshift range to survey the required volume. These science requirements flow down to the following technical requirements for VIRUS:
\begin{itemize}
\item Coverage of $\Delta z \sim$ 2 and coverage into the ultraviolet to detect LAEs at the lowest feasible redshift. Analysis of the expected number of LAEs with redshift also indicates that the majority of the objects are located at $z ~<$~3.5 due to the change in distance modulus with redshift, coupled with the steepness of the LAE luminosity function. VIRUS is designed for 3500~$<$~$\lambda$~$<$~5500~\AA, or Lyman-$\alpha$ at redshift $1.9 < z < 3.5$.
\item Area coverage of at least 50 sq. arcmin. per observation.
\item Utilize fiber integral field units (IFU) to keep the weight of the spectrographs off the moving payload of the HET.
\item Fiber core diameter of 1.5 arcsec (266 $\mu$m) for optimal detection of LAEs in the typical image quality delivered by HET (1.3 to 2.0 arcsec FWHM).
\item Resolution matching the linewidth of LAEs (resolving power $R \sim$~700 or greater) to maximize detectability.
Note that $R \sim$~2000 would be required to split the [OII]$\lambda$3727 doublet associated with low-redshift interloper emission line galaxies. These objects are discriminated from LAEs through equivalent width thresholds \citep{gron07, leung17, far21, geb21}, so optimum detection of LAEs and larger survey volume (redshift range) were chosen in the trade-off against higher spectral resolution.
\item Low read-noise detectors ($\sim$3 electrons) to achieve equality between sky-background and read noise in 360 seconds at the shortest wavelengths.
\item High stability to ambient temperature variations, although not to gravity vector variations, since the VIRUS modules have a fixed altitude orientation, mounted in large enclosures.
\item Throughput sufficient to reach an emission line sensitivity of 4~$\times$~10$^{-17} {\rm erg~cm}^{-2} $s$^{-1}$ at 4500~\AA~ in 20 minutes on HET. Combined with the IFU area coverage, this sensitivity is derived from the requirement to detect $\sim$200 LAEs per observation or 2.5 LAEs per IFU, for an average of $\sim$1 LAE per \sqam.
\item Simple, inexpensive design that can be replicated in quantity.
\end{itemize}

These overarching technical requirements emphasize the ability to cover area quickly and require a highly-multiplexed spectrograph, which was achieved through large-scale replication \citep{hil14} of modular integral field spectrograph units.

Before embarking on developing the replicated
VIRUS units, a prototype of a single spectrograph was constructed in~2006 and tested on the 2.7~m Harlan J Smith Telescope at The McDonald Observatory.
The motivation for building the VIRUS prototype (the George and Cynthia  Mitchell Spectrograph,
formerly VIRUS-P, \citealt{hil08a})
was to provide an end-to-end test of the concepts behind HETDEX, both instrumental and scientific.

A number of key design choices were made and tested with the Mitchell Spectrograph, including the decision to use bare fibers rather than lenslet coupling at the IFU input and a catadioptric optical design for the spectrographs \citep{hil04, hil06}. 
The upgraded HET has a fast focal ratio in order to couple efficiently to fibers, reducing focal ratio degradation (FRD).  
The layout of the fibers in VIRUS IFUs, with 1/3 fill-factor, is most optimal for covering area, since a dither-pattern of three exposures exactly fills the field of the IFU, while maximizing the area covered per IFU. Lenslet-coupled fiber IFUs have the principal advantages of providing contiguous coverage of a small field, and allowing the slow focal-ratio beams of large telescopes to be coupled efficiently to fibers (e.g. \citealt{all-smith02}). 
However, lenslets suffer from inefficiencies in the coupling 
due to lens quality and diffraction effects. This was particularly the case with lenslet technologies available in 2006, when the properties of VIRUS were being cemented. If the fiber core is oversized to mitigate these coupling effects, then resolution is lost \citep{hil04}. For fibers fed at the same focal ratio, the bare fiber bundle will cover the same area of sky as does a lenslet system, both in three exposures, but offers higher overall throughput and lower cost. 
\edit1{Differential atmospheric refraction (DAR) has a magnitude of 0$\farcs$95 over the bandpass of VIRUS at the mean HET airmass of 1.25. The fixed elevation of HET assures that the DAR relation varies very little and is always aligned in the same direction with respect to the IFU fiber pattern.
An atmospheric dispersion corrector is not needed for VIRUS, since the dither pattern of three exposures fills in the area of the IFUs; light lost from the aperture of one fiber falls onto an adjacent fiber position in the full 3-position dither pattern. Emission line objects are positioned randomly with respect to the fiber pattern, so the DAR only affects the position of the detection, which is corrected to a common wavelength. 
%Continuum objects are extracted based on the DAR shift derived from a large number of star observations and the spectral energy distributions agree very well with objects observed in SDSS spectroscopy \citep{geb21}.
}

Another design feature of the IFUs that was established early was the number of fibers per IFU and the distribution of IFUs within the HET field of view. Each spectrograph can accommodate 224 fibers, allowing for adequate gaps to allow the spectrum of each fiber to be isolated on the detector. Packaging two spectrograph channels in a unit proved most efficient for space and other factors (see \S\ref{sec:design}), so each IFU has 448 fibers and covers 51 $\times$ 51 \sqas. With that building block size, the total field of view of the upgraded HET, and the required area and number of LAEs per observation, a grid of IFUs results with 100\arcsec\ separation and $\sim$1/4 fill-factor.
This layout has the advantage of allowing contiguous areas to be mapped in four observations, except for a central area reserved for other HET instruments (\S\ref{sec:virus}).
Non-uniformity of the window function of the observation is not important on the scale of the IFU separation, which is much smaller than scales probed by the power spectrum of LAEs to be measured by HETDEX \citep{chiang13}.

Refractive camera designs were explored initially for the VIRUS prototype spectrograph \citep{hil04}, with the aim of keeping the format of the units as small as possible. However, the requirement to work at 3500 \AA\ led to an optical design of significant complexity with calcium fluoride \edit1{and fused silica.}  
Following this design investigation, a catadioptric Schmidt camera with the charge coupled device (CCD) detector at its focus was adopted. The camera design was simpler \edit1{and less expensive}, though it necessitated a factor of two larger beam size, the image quality was better and the throughput was equivalent.

Construction and testing of the prototype verified the opto-mechanical design, the throughput, the stability, and the sensitivity, and demonstrated the utility of such an instrument for surveys of emission-line objects. It also served as a test-bed for the software development required for analyzing the data from the full VIRUS array. The Mitchell Spectrograph was used to perform the HETDEX Pilot Survey of Ly-$\alpha$ emitting galaxies (\citealt{adams11,blanc11}). The results of the Pilot Survey confirmed the sensitivity estimates on which HETDEX is based, and demonstrated the effectiveness of blind IFU spectroscopy for this scientific application.

The requirements outlined above were flowed-down to properties of the WFU and VIRUS that are described in the following sections.

\section{OVERVIEW OF THE  HET WIDE FIELD UPGRADE}\label{sec:wfupgrade} 

\begin{figure}
\epsscale{1.0}
\plotone{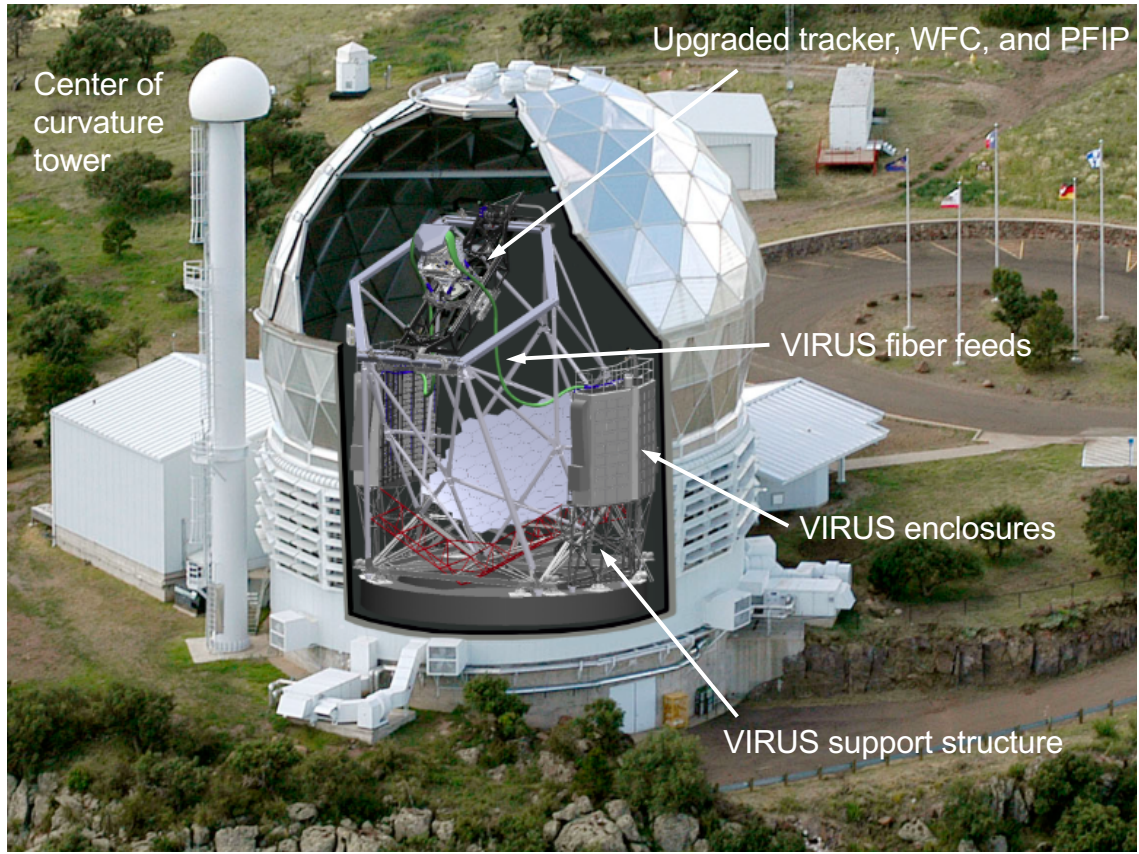}
\caption{
\label{HETlayout} \footnotesize
The layout of the HET with rendering of the WFU and VIRUS superimposed. This view is looking towards the south-east. The WFU replaces the top end of the HET with a new tracker, wide field corrector (WFC), and prime focus instrument package (PFIP). The VIRUS spectrograph units are housed in two enclosures either side of the structure, which are mounted on the VIRUS support structure, and fed by 35,000 fibers from the prime focus.
The main telescope structure, primary mirror, and alignment instruments in the Center of Curvature Alignment System tower remained unchanged from the original, except that the portions of the center of curvature tower within the HET field of view were painted black.
}
\end{figure}

The basic configuration of the HET is unchanged in the upgrade (Figure~\ref{HETlayout}), but the new tracker has a much higher payload of 3 metric tonnes to accommodate the new wide field corrector (WFC) and prime focus instrument package (PFIP), a five-fold increase. 
Of particular note is the large volume taken up by the enclosures for VIRUS and the additional structure required for their support and movement.
Detailed summaries of the development of the WFU from conception to installation are covered in \citet{hil18b} and in \citet{lee21}. The following subsections provide an overview of the project. 

\subsection{HET Principles of Operation}\label{subsec:operation}
The HET requires constant monitoring and updating of the position of the moving components relative to the optical axis of the primary mirror, in order to deliver high quality images. This axis changes constantly as the telescope tracks, so the telescope has to maintain strict tolerances on six degrees of positional freedom, plus time. Tilts of the WFC cause comatic images, and axial errors cause defocus and a change in plate scale. In addition, the global radius of curvature of the primary mirror varies slowly with temperature (as the segmented primary mirror is essentially a glass veneer on a steel truss), and must be monitored and updated during the night. 

The metrology subsystems implemented to provide the necessary feedback on these degrees of freedom are discussed in \citet{lee18b} and in more detail in \S\ref{subsec:pfip}. First, an overview is helpful before the subsystems are described in more detail. The subsystems that are involved in every track include two guideprobes, two operational wavefront sensors, a tip-tilt sensor, and a distance measuring interferometer. These are augmented by an acquisition camera, a calibration wavefront sensor, a pupil viewer camera and a bore-sight imager, that are used periodically to verify internal alignment. 
The remaining degree of freedom is the rotation angle on the sky, which is not monitored directly. The encoding of the rotation is sufficient to meet requirements.

Observing starts with primary mirror alignment, utilizing the center of curvature alignment system 
%(CCAS) 
instrumentation in the center of curvature tower. With the telescope moved in azimuth to point at the center of curvature tower, the process first involves setting the position of the 
instrument suite \citep{booth03} using a leg of the distance measuring interferometer, reflecting off the primary mirror. This process sets the radius of curvature of the primary mirror. The other center of curvature instruments are the Mirror Alignment Recovery System \citep{wolf03} that stacks the mirror segments, the 
%HET Extrafocal Imager 
Hartman Extra-focal Instrument \citep{pp04, booth04} that illuminates the primary mirror and can image the return in and out of focus, 
and a Wavescope\footnote{from Adaptive Optics Associates}, used to examine the overall figure of the primary mirror. 
The Mirror Alignment Recovery System is a Shack-Hartmann based sensor with a lenslet array matched to the 91 mirror segments.  The system utilizes an internal light source to illuminate the HET and a reference mirror to provide focused spot locations from the required spherical surface. Centroids of the HET mirror segment spots are compared to the reference spot locations to measure tip/tilt misalignments of each segment. At the start of the night the first primary mirror alignment takes about 30 minutes.
During this period, calibrations of the science instruments are obtained. Once aligned, the segment alignment maintenance system, employing inductive edge sensors, maintains the alignment between the edges of the segments and provides metrics on the quality of the mirror segment alignment over time \citep{booth03, pp04}. Additional alignment during the night is driven by large ambient temperature changes. Re-stacking of the primary mirror is at the discretion of the telescope operator, and occurs either when segment alignment maintenance system metrics indicating the quality of the stack are above acceptable levels, or when there is a segment obviously misaligned and visible on a star image. 
At most, one additional alignment is needed during the night and takes 15 minutes. 

Once set up on target, there is a hierarchy of feedback cadence for different metrology loops. The first level of metrology involves the WFC position, monitored by two guideprobes and two operational wavefront sensors on a cadence of a few to a few tens of seconds, depending on star brightness and conditions. These loops constrain sky position (and time at that position) and WFC alignment to the optical axis (focus and tilt), based on direct measurement of light from stars.
Drifts in alignment happen more slowly than changes in position and so are averaged and updated on minute timescales. 
The tip-tilt sensor and distance measuring interferometer provide secondary direct measures of the physical distance and alignment of the payload to the primary mirror. These run on a cadence of one and 10 seconds, respectively. 
Measurement of the physical separation between WFC and primary mirror when the telescope is in focus provides a constraint on the primary mirror global radius of curvature. Updates to global radius of curvature are sent to the segment alignment maintenance system between targets, as needed, driven primarily by temperature changes.  The plate scale is not monitored directly. The focal length of the system is naturally constrained by the monitoring of the radius of curvature of the primary mirror, to a level of accuracy far better than could be measured from the separation of guide stars, for example. 

\subsection{Tracker}\label{subsec:tracker}
\begin{figure}
\epsscale{1.0}
\plotone{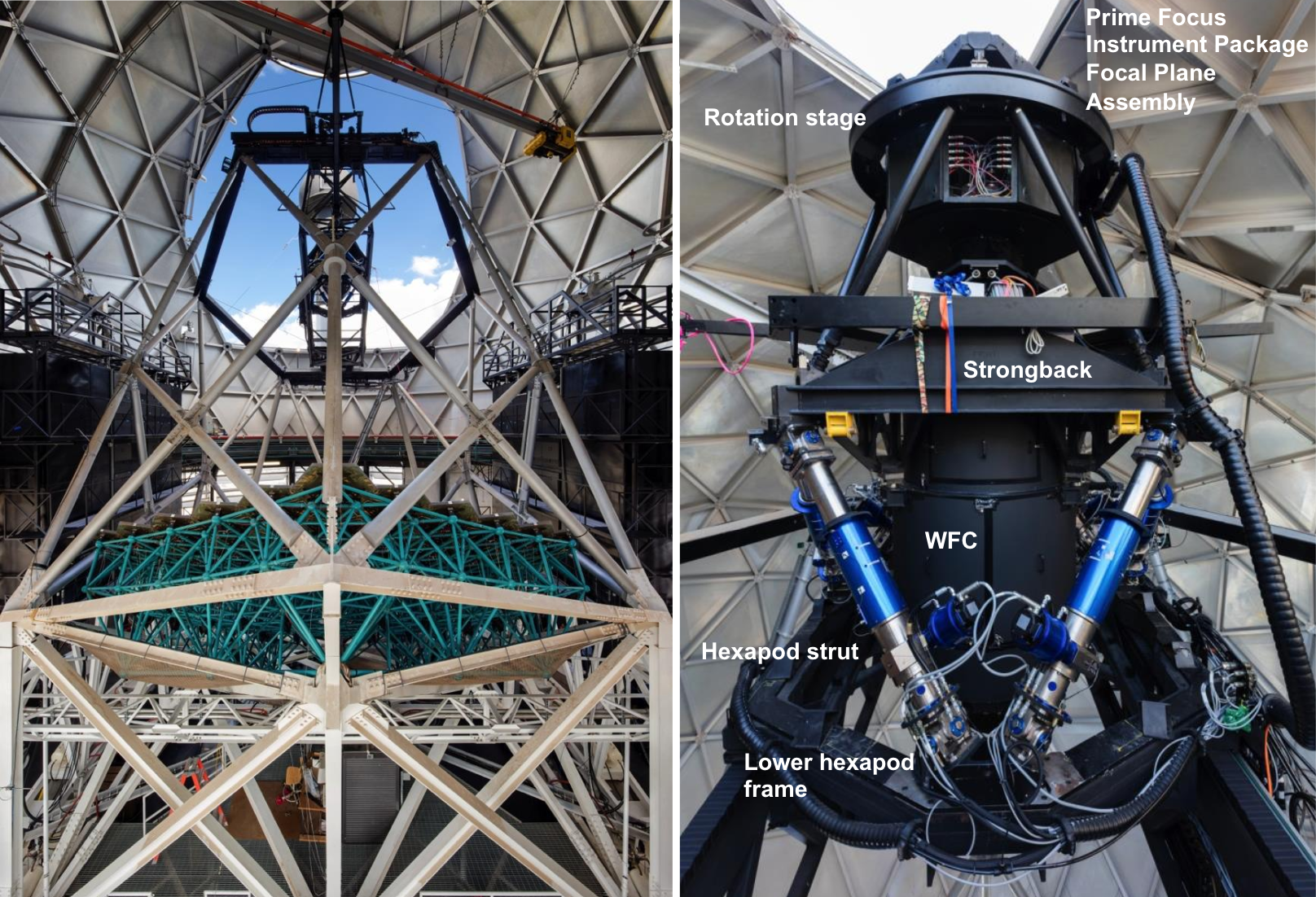}
\caption{
\label{HETupgrade} \footnotesize
Images of completed HET upgrade. Left panel: View from behind primary mirror (the mirror truss is the turquoise frame), showing the new tracker centered in the upper hexagon of the telescope structure. The VIRUS enclosures are the black-paneled structures either side of the telescope (note that the enclosures are parallel to each other in a plan view).  Right panel: The WFC and PFIP with key components indicated. The hexapod struts that orient the WFC to the primary mirror can be identified by their blue casings. The focal plane assembly (at the top of the structure) is supported by a fixed hexapod for alignment and a rotational stage that maintains the sky orientation during a track. The PFIP is shown prior to the installation of the fiber feeds.
Figure adapted from \citet{hil18b}, Fig. 2.
}
\end{figure}

The new tracker is a third-generation evolution of the trackers for HET and Southern African Large Telescope (SALT, \citealt{buckley06}), and is in essence a precision six-axis stage (Figure~\ref{HETupgrade}) with a high payload.
Its purpose is to position the WFC accurately normal to, and at the correct distance from, the primary mirror vertex as it follows the sky motion of targets.
The tracker \citep{good14b,good18} was developed jointly by The University of Texas at Austin McDonald Observatory (MDO) and Center for Electro-mechanics. Details of its deployment and commissioning are provided in \citep{hil16c}.
%The job of the tracker is to position the WFC normal to the vertex of the primary mirror and in focus as objects are observed and tracked in time. 
The tracker bridge spans the upper hexagon of the telescope structure, moving on two x-axis stages with skew sensing to maintain alignment. A carriage moves up and down the tracker bridge (the y-axis), and supports the hexapod that provides the fine adjustment in the other degrees of freedom. 
The hexapod actuators were manufactured by ADS International\footnote{located in Valmadrera, Italy} in collaboration with Center for Electro-mechanics and MDO \citep{zier12}.
The total volume of motion is about 
$7\times7\times4$~m$^3$, and the required accuracy under metrology feedback is on the order of 15 $\mu$m and 4$\arcsec$ in tilt in physical position, and 10 ms in time.

\begin{figure}
\epsscale{1.0}
\plotone{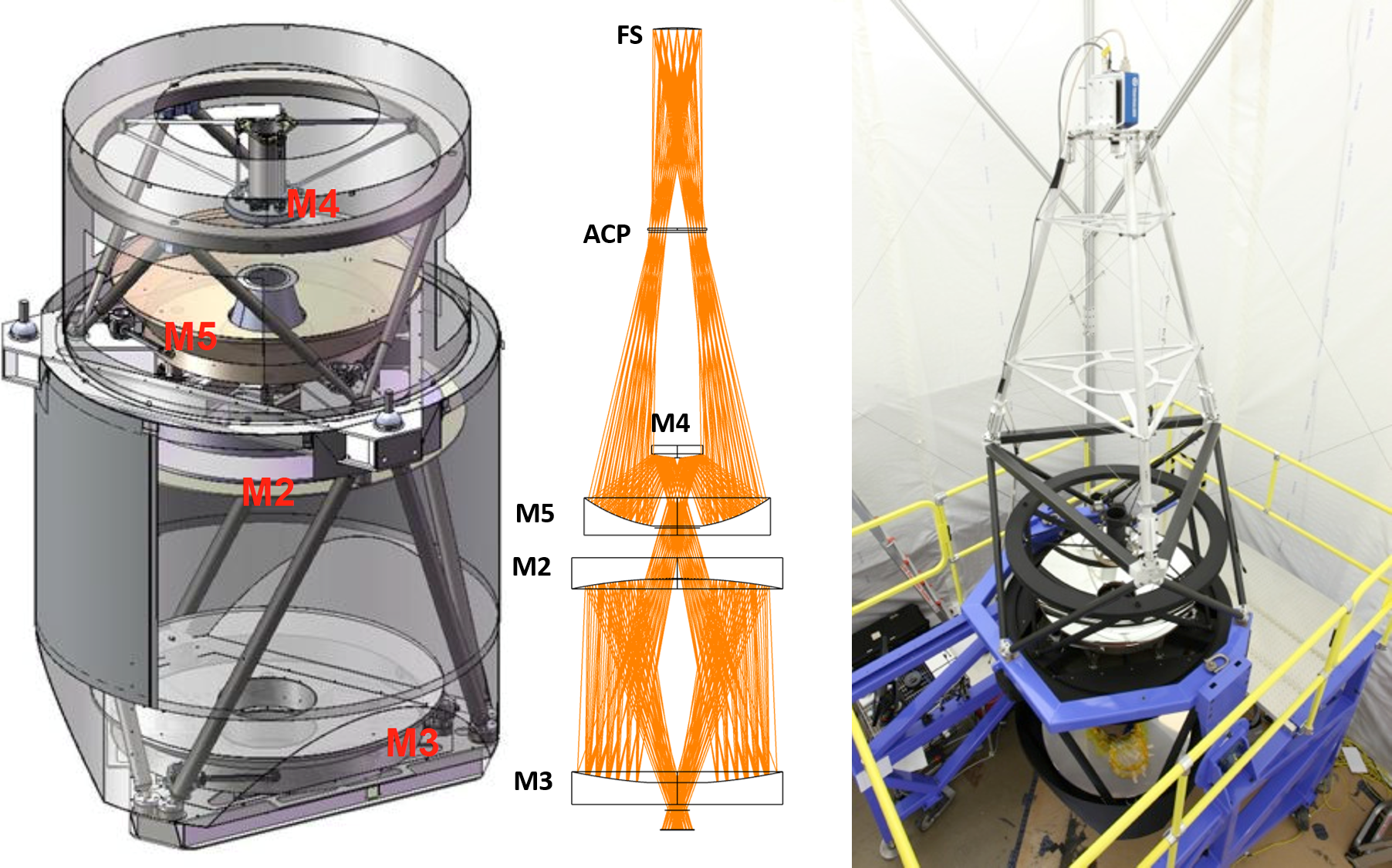}
\caption{
\label{layoutWFC} \footnotesize
Opto-mechanical layout of the  Wide Field Corrector. Left: A rendering of the four mirrors (M2-M5) and mechanical structure of the WFC. Center: The optical layout indicating the identification of the mirrors and the aspheric corrector plate (ACP) located at the exit pupil. Light enters the corrector from the primary mirror through the central hole in M3 and is focused on the concave focal surface (FS) as indicated by the ray tracing.   
Right: View of the WFC undergoing alignment and interferometric testing at the University of Arizona College of Optical Sciences in 2015. The largest mirrors (M2 and M3) are 1 meter in diameter. 
Following fabrication, the figures of all mirrors ended up being described by general aspheres.
}
\end{figure}

\subsection{Wide Field Corrector (WFC)}\label{subsec:wfc}

The new corrector (Figure~\ref{layoutWFC}) has a 22$\arcmin$ diameter field of view and a 10 m pupil diameter. The periphery of the field is used for guiding and wavefront sensing to provide the necessary feedback to maintain alignment of the payload. The WFC is a four-mirror design with two concave one-meter diameter mirrors (M2, M3), one concave 0.9~m diameter mirror (M5), and one convex 0.23~m diameter mirror (M4). \edit1{All surfaces are conics with additional general aspheric terms.
%with base conic constants of (M2,M3,M4,M5) = (0.6, -7.7, -2.1, -0.3). 
The highest general asphere orders are 8th and 10th on M5 and M3, respectively.} The corrector is designed for feeding optical fibers at $f/$3.65 to minimize focal ratio degradation, and so the chief ray from all field angles is normal to the focal surface.  This fiber alignment is achieved with a concave spherical focal surface of radius 984~mm, centered on the exit pupil. The imaging performance is 0.6$\arcsec$ FWHM or better over the entire field of view, and vignetting is minimized. The WFC has uniform plate scale over the entire field of view to ensure that the dither offsets to fill in the gaps between fibers in VIRUS IFUs are close to identical for all IFUs (variation is 0.4\%).
%The conic values are:
%M2 0.6
%M3 -7.7
%M4 -2.1
%M5 -0.3
%All mirrors have base conic sections, so the actual number of aspheric orders (from sphere) is kind of infinite, only that M3 and M5 contain additional aspheric deformation up to 10th and 8th, respectively, (starting from 6th). 
The WFC was manufactured by the University of Arizona College of Optical Sciences \citep{burge10}, with significant collaboration from MDO (\citealt{lee16a}, \citealt{good14a}).  The smaller M4 was subcontracted to Precision Asphere.

With four surfaces, reflective coatings for the WFC are required to have high reflectance (95\% or better from 3500~\AA~ to 1.8~$\mu$m), and are challenging, being based on silver and multiple dielectric layers. 
The large mirrors were coated by JDS Uniphase Corp\footnote{now Viavi Solutions}, and M4 was coated by ZeCoat. 
MDO designed and constructed the complex support fixturing needed to safely handle the large mirrors during cleaning and coating at JDS Uniphase \citep{good14a}.
Experience with coating degradation on the original HET corrector led us to adopt a sealed design for the WFC, with entrance and exit windows and careful sealing of the WFC housing. Since deployment, the WFC has been purged continuously with nitrogen gas.
Periodic visual inspection of the mirrors reveals minor changes in appearance, but direct reflectivity measurements are not possible in-situ. Monitoring of standard stars with LRS2 reveals no detectable degradation of throughput that could be traced to changes in the WFC coatings.

Details of the figuring, alignment, and testing at the College of Optical Sciences are given in \cite{burge10}, \cite{oh14}, and \cite{lee16a}.    
During initial testing, significant errors in the low-order figures of M5 and M3 were detected. Respacing the mirrors and adding an aspheric corrector plate at the exit pupil in place of the planar exit window brought the performance of the WFC back to the image quality specification \citep{lee16a}.
Final alignment utilized interferometric testing against computer generated hologram targets. 
Separate computer generated hologram tests of the M4/M5 pair, the M2/3 pair, and of the entire system were used to evaluate the alignment of the individual mirrors and the whole assembly.
A conjugate test of the M4/5 pair with a custom wavefront sensor was developed by MDO to provide an independent confirmation that the system was meeting specification, particularly in off-axis performance, which was degenerate in the computer generated hologram tests \citep{lee16a}.
Following an external review of the results, it was agreed that the WFC likely would meet specifications and the 
next step was to integrate it with the HET primary mirror and perform on-sky testing to verify the system. 
Integration and testing of the WFC on HET is described in \S\ref{subsec:wfuintegrate}.
\citet{lee16a, lee21} provide a detailed discussion of the challenges posed by the WFC and tests developed to demonstrate its performance.

\subsection{Prime Focus Instrument Package and Metrology Systems}\label{subsec:pfip}

\begin{figure}
\epsscale{1.0}
\plotone{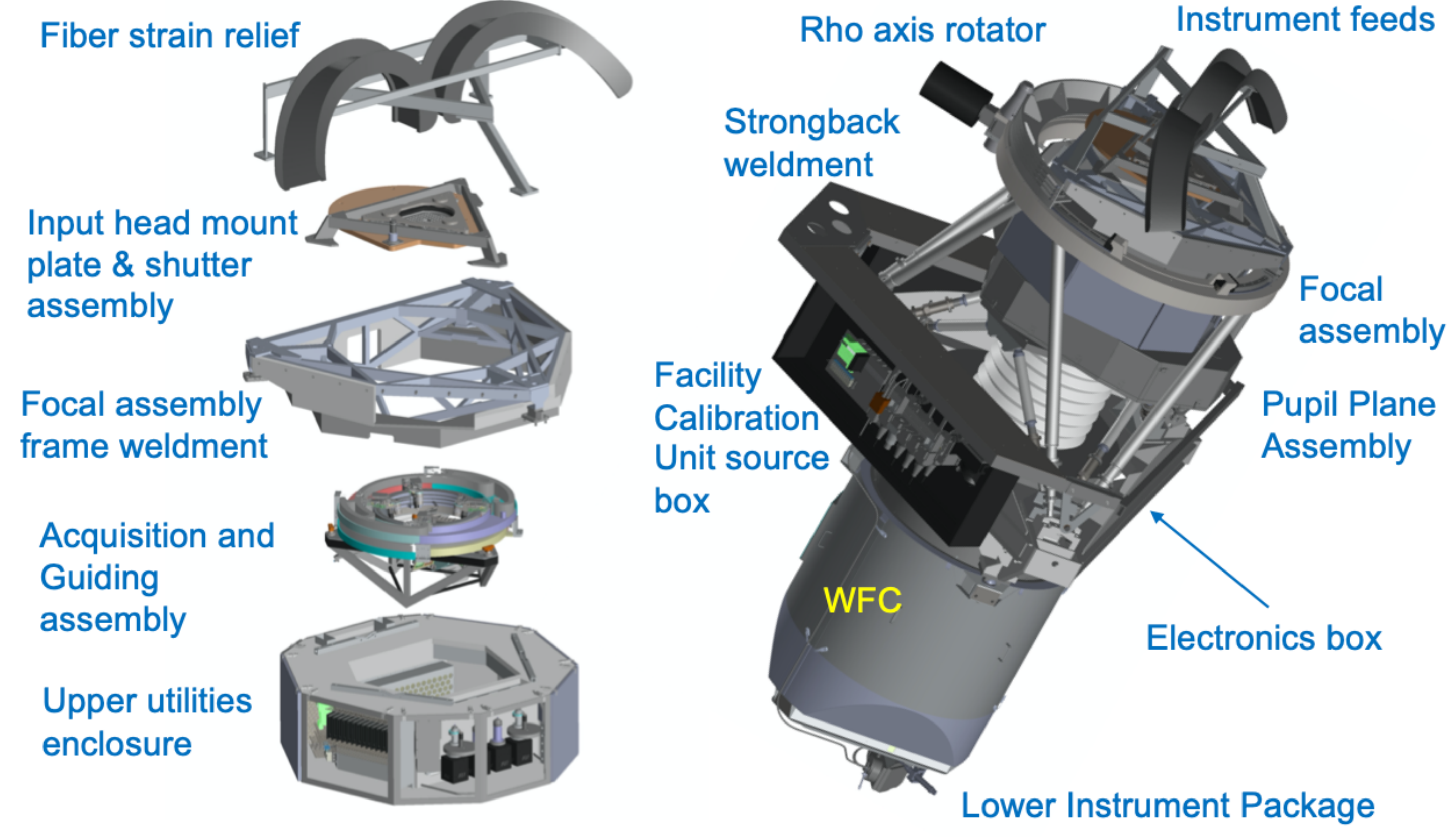}
\caption{
\label{PFIPrender} \footnotesize
Renderings of the Prime Focus Instrument Package (PFIP). Right: The full assembly with major sub-assemblies indicated. The Wide Field Corrector (WFC) is at the heart of the assembly, mounted to the strongback. The entrance to the WFC is sealed by the Lower Instrument Package (entrance) and by the Pupil Plane Assembly (exit). The Rho axis field rotator keeps the position angle of the Focal assembly fixed on sky during a track. Left: An exploded view of the Focal Assembly showing its major components. The input head mount plate mounts the VIRUS IFUs and the fiber feeds for the other instruments, which lie on the fiber strain relief trays. The Focal Assembly contains the bulk of the complexity of the instrument, including the acquisition and guiding assembly with acquisition camera, guide probes, and wavefront sensors.
Figure partly adapted from \citet{hil18b}, Fig. 4.
}
\end{figure}

The Prime Focus Instrument Package (PFIP; \citealt{vattiat12,vattiat14}) rides on the tracker and consists of several subassemblies (Figures~\ref{HETupgrade}, \ref{PFIPrender}).  The WFC mounts to a triangular steel frame (the strongback) on a three ball-in-vee kinematic mount.  
The strongback mounts to the tracker hexapod and the structure of PFIP is assembled around the corrector (Fig. \ref{PFIPrender}). The focal assembly contains all the hardware at the focus of the telescope, including the acquisition and guiding assembly, fiber instrument feeds, custom rotating-blade shutter, and electronics hardware.  The focal assembly mounts kinematically to a rotation stage (the Rho axis), supported by a fixed hexapod for position adjustment during alignment. 
A total Rho axis range of $\pm$~25$^\circ$
allows the range of sky rotation to be followed during tracking for all accessible declinations, with some margin. 
This range is much less than the $\pm$~200$^\circ$ of the original HET, due to no longer having a long-slit spectrograph and to avoid twisting the integral field fiber feeds.
Coincident with the optical focal surface there is the input head mount plate (\S\ref{sec:virus}, Fig.~\ref{ihmp}), which is a precision-machined interface for the VIRUS IFUs and the other instrument fiber feeds. The input head mount plate defines the physical focal surface of the telescope and has concave spherical shape to conform to the WFC optical focal surface\footnote{The input head mount plate  was machined and verified to high precision by the Institut f\"ur Astrophysik G\"ottingen.}. 
Trays, forming the fiber strain relief, guide the VIRUS IFUs and other fiber feeds off the payload \citep{vattiat18}.
The input head mount plate  and fiber feeds can be removed to a work position for installation and maintenance of fibers and IFUs and for access to the focal assembly. The complex systems of PFIP, and the majority of the motion control actuators, are all contained within the focal assembly, which is designed to be removable for ground service. The relationship between the fiber positions and metrology systems are maintained by kinematic mounts, ensuring rapid return to science operations following service activities.

The Lower Instrument Package is mounted to and seals the input end of the wide field corrector, and is a platform for the entrance window changer, tip-tilt camera, and facility calibration unit ( \S\ref{subsec:fcu}) output head.  A set of temperature controlled, insulated enclosures house electronics hardware and the facility calibration unit input sources, optics, and selection mechanisms.  The pupil plane assembly is located in between the wide field corrector and the focal plane assembly.  Initially, the support structure with fixed exit pupil baffle and the corrector plate have been deployed. 
Future upgrades to the pupil plane assembly have been considered and space and interfaces to accommodate a moving baffle at the exit pupil of the telescope, a platform for selectable exit windows (aspheric corrector plates), and a future atmospheric dispersion compensator have been built into the system.

As discussed in \S\ref{subsec:operation}, all degrees of freedom of the motion of the WFC, with respect to the optical axis of the primary mirror, must be monitored and maintained during a track in order to deliver good images. The feedback to maintain these alignments requires excellent metrology, which is provided by the following subsystems \citep{lee12c,lee18b}:
\begin{itemize}
\item The acquisition camera\footnote{Finger Lakes International Microline ML090000 with Kodak KAF-09000 CCD} has a 3.0$~\times$~3.0 \sqam~ field of view, offset 1.0 arcmin from the optical axis so as to also cover the positions of the LRS2 IFUs and the fiber feeds for the basement high resolution spectrographs. The acquisition camera is fed by a deployable pick-off mirror and has $B, g', r'$, and $i'$ filters\footnote{from Astrodon}.
\item Two Guide Probes to monitor the position on the sky, the plate scale of the optical system, and to monitor the image quality and atmospheric transparency. The guide probes each have five filters (clear, $B, g', r'$, and $i'$) and 23~$\times$~23 \sqas\ field of view.
\item Two operational wavefront sensors of Shack-Hartmann design with 11 sub-apertures across the pupil diameter, to monitor the focus and tilt of the WFC during tracking. Capture range of 5\arcsec\ diameter. The guide probes and operational wavefront sensors use the same cameras\footnote{Finger Lakes International Microline MLx285 with Sony ICX285AL CCD}.
\item Calibration Wavefront Sensor\footnote{Allied Vision Technology (AVT) Prosilica GC2450 camera with Sony ICX625 CCD} with 21 sub-apertures across the pupil diameter, to provide a higher resolution on-axis reference for focus and wavefront monitoring. The calibration wavefront sensor is aligned to be cofocal with the input head mount plate and provides the overall focus reference for the HET.
\item A distance measuring interferometer\footnote{The distance measuring interferometer is a custom system based on interferometry, provided by FOGALE nanotech}) operating at 1.6 $\mu$m to measure the physical distance between the WFC and primary mirror \citep{pp06}. The distance measuring interferometer projection head is fed by a fiber link and mounted on the lower instrument package. The interferometer has an additional fiber link to a head in the center of curvature tower for setting the distance at which to stack the primary mirror segments, which defines the primary mirror radius of curvature. 
\item A tip-tilt sensor operating at 1.6 $\mu$m mounts to the lower instrument package to monitor the tip/tilt of the WFC with respect to the optical axis of the primary mirror, via a reflected beam from the primary mirror. The 1.6 $\mu$m operating wavelength for the tip-tilt sensor and distance measuring interferometer was chosen so as to avoid the wavelength range of any instruments conceived for the telescope, since the HET design is poor for thermal infrared instrumentation. 
\item A pupil viewer based on a color camera\footnote{AVT Manta G-046 with SONY ICX415 CCD} to view the pupil and monitor mirror segment reflectivity.
\item The Bore-sight imager is a coherent fiber bundle from Schott mounted in the input head mount plate that feeds a camera\footnote{AVT Manta G-201B} for monitoring the relationship between the input head and the acquisition camera, periodically with bright stars.
\end{itemize}

\begin{figure}
\epsscale{1.0}
\plotone{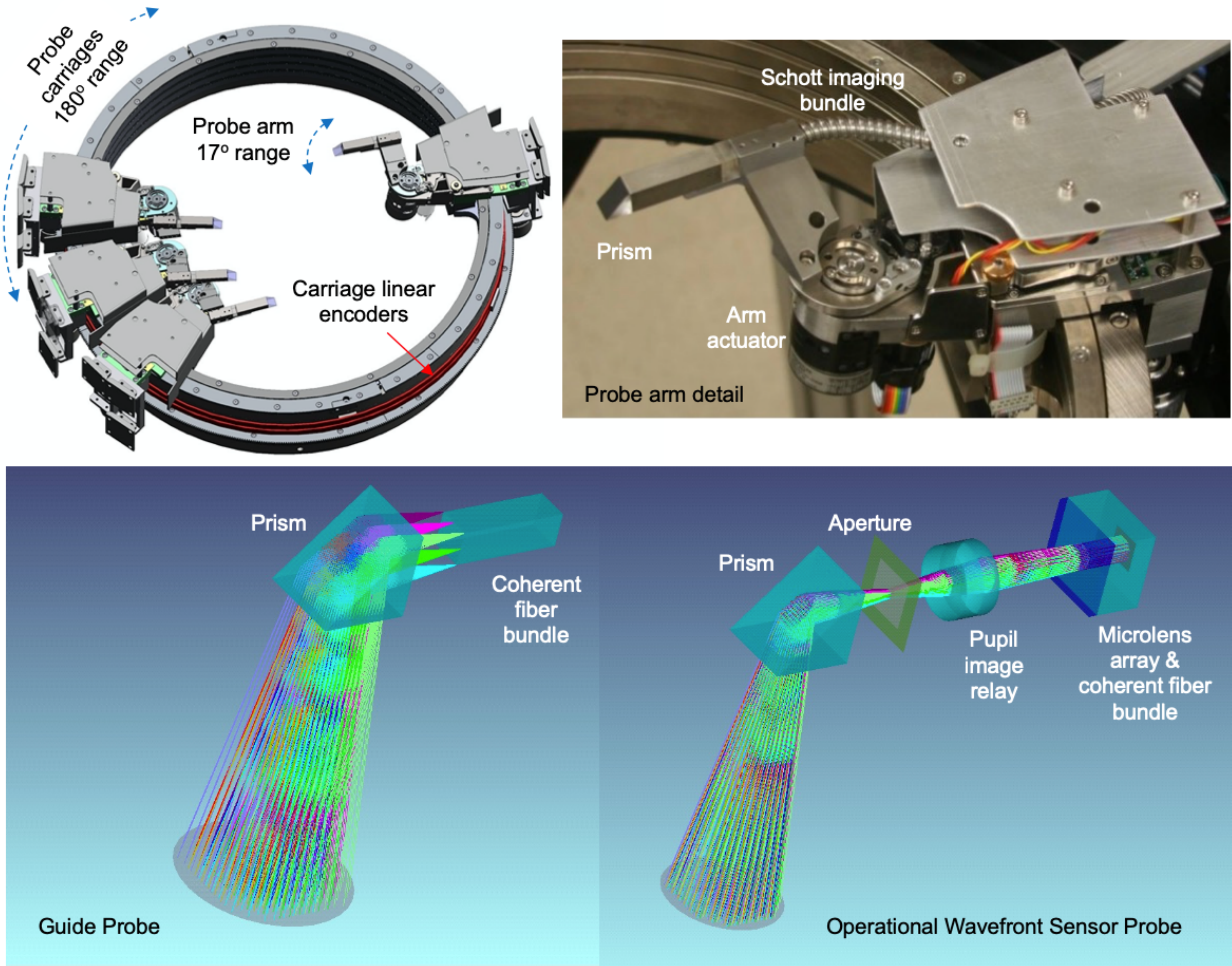}
\caption{
\label{guideprobe} \footnotesize
\edit1{Details of the PFIP guide-probe assembly optomechanical design.
Top left is a rendering of the four probe carriages (two each, guide probes and operational wavefront sensors) on the large ring bearings. The rendering indicates how the carriages can nest together without colliding. Each carriage can range over 180$^\circ$ and the probe arms swing 17$^\circ$ to access the guide field annulus between 18 and 22 arcmin. field diameter. Carriages utilize linear magnetic tape encoders and probe arms utilize single turn rotary encoders. 
Both mechanisms are driven by brushless DC servo motors through capstan drives. 
For scale reference, the inner diameter of the carriage bearing stack is 445 mm.
Top right shows detail of an individual carriage. The axis of the arm actuator passes through the center of the exit pupil of the WFC so that the prism moves concentric with the spherical image focal surface to remain in focus.
The bottom panels show the optical layout of the guide probes (left) and operational wavefront sensors (right). For the guideprobes, the prisms reflect the converging light directly onto coherent imaging fiber bundles by Schott. For the wavefront sensors the light is focused onto an aperture of 5$\arcsec$ diameter, collimated by a relay lens and then imaged by a microlens array onto the imaging bundle, in a Shack-Hartmann configuration. 
Adapted from \citet{vattiat12} and \citet{lee12c}.
}
}
\end{figure}

The heart of the metrology system for the WFU is the Acquisition and Guiding assembly, which mounts the guide probe assembly, the acquisition camera, calibration wavefront sensor and pupil viewer. The light is directed to acquisition camera, calibration wavefront sensor, and pupil viewer by deployable pickoff mirrors with pneumatic actuators. During tracking, the guide probe assembly is used for guiding of the telescope and wavefront sensing feedback to the telescope focus and WFC tilt.  Careful design of the pickoff mirrors allows the acquisition camera, calibration wavefront sensor, or pupil viewer to be used simultaneously with the guide probes and operational wavefront sensors in the guide probe assembly \citep{vattiat12,lee12c,vattiat14}. There are four probes: two imaging probes and two wave-front sensing probes, providing redundancy \edit1{(Fig.~\ref{guideprobe}}.  Each probe consists of a probe optical head, containing the necessary optics coupled to a coherent fiber bundle purchased from Schott.  
In the operational wavefront sensors, a microlens array is bonded to the input faces of the imaging fiber bundles. 
Images incident to the fiber bundle input are captured by a remote camera system at the bundle output, fed by reimaging optics and including a filter wheel for each of the guide probes.  Each probe optical head is mounted to a carriage with an arm for moving the probe radially in the field with a range of 9-11 arcminutes from the center of the telescope’s field.  The four carriages each move through 180$^\circ$ sectors on large circular bearings to access stars. 
The axis of the arm motion passes through the center of the exit pupil, so the arm naturally traces the spherical focal surface and remains perpendicular to it, without need for any focus adjustment.
\edit1{See Fig.~\ref{guideprobe} for details. }
The positioning accuracy requirement of the guide probes is 20 microns on the spherical focal surface.  To achieve this performance, both mechanical position actuation and encoding require a high level of precision.  The guide probe assembly makes $\sim$10$^4$ moves per year. Maintenance is undertaken on a regular basis during focal assembly ground service.

\begin{figure}
\epsscale{1.0}
\plotone{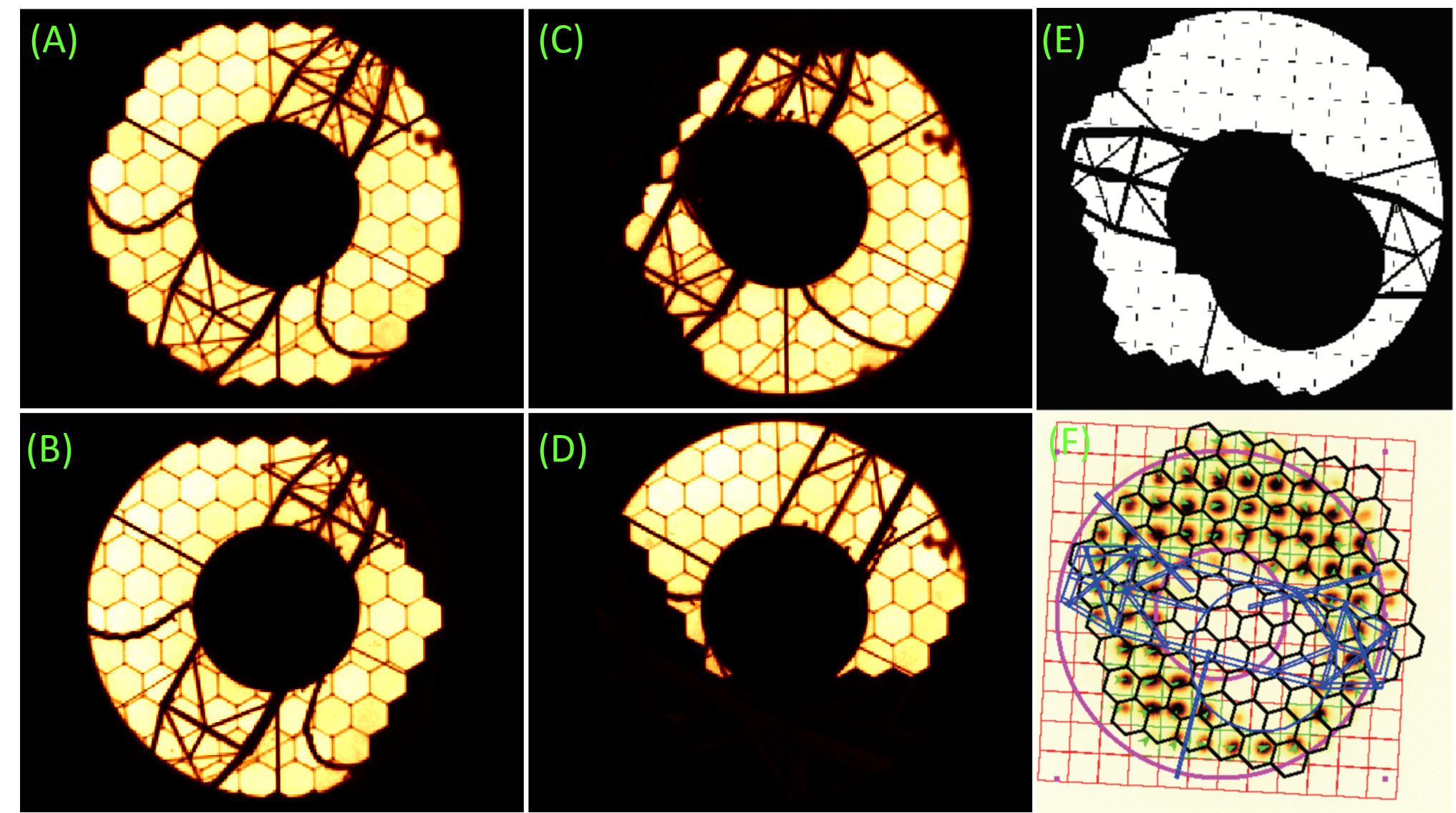}
\caption{
\label{pupil} \footnotesize
Examples of the variations in pupil illumination of the HET. The left four images show views from the Pupil Viewer camera at image field center, with the telescope positioned at different (X,Y) tracker positions expressed in meters: (A) the center of a track (0, 0), (B) (0.4, 0.4), (C) (1.5, 0), and (D) (0.4, -1.6), \edit1{all in meters}. The change in illumination of the pupil as it moves off the edge of the mirror during a track is evident. The top right image (E) illustrates the pupil illumination model that incorporates the obstructions of the WFC and the tracker bridge, as well as the pupil position on the primary mirror. Note that this view is for a field position away from the center, so the holes in the WFC mirrors are not aligned with the central obstruction of the pupil.
Bottom right (F) is an operational wavefront sensor image with the telescope pupil position (magenta), primary mirror segments (black) and illumination obstruction features (blue) overlaid. The square grid of the wavefront sensor elements is shown in red while elements with a successful measurement are highlighted in green with green arrows indicating the image centroids. The relationship between the lenslets and the mirror segments changes throughout a track.
}
\end{figure}

The upgrade adds wavefront sensing \citep{lee18a,lee18b} to the HET in order to close the control loop on all axes of the system, in conjunction with the distance measuring interferometer adapted from the original tracker metrology system and a new tip-tilt sensor \citep{vattiat14}.  The design of the wavefront sensors is straightforward, but their application to the HET, with the varying illumination of the telescope pupil during a track (Fig.~\ref{pupil}), requires development of a robust software system for analysis of the sensor data to produce reliable wavefront information \citep{lee18a,lee18b}.
There is redundancy built into the new metrology system to obtain the highest reliability. The two guide probes distributed around the periphery of the field of view provide feedback on position, rotation, and plate scale, as well as providing a record of image quality and transparency as a function of wavelength. The alignment of the corrector is monitored by the two operational wavefront sensors as well as by the distance measuring interferometer and tip-tilt sensor. The radius of curvature of the primary mirror is monitored by the combination of focus position from the operational wavefront sensors with the physical measurement from the distance measuring interferometer. 

The PFIP was deployed on HET in July 2015. It is a complex instrument in its own right, and was commissioned in phases from July 2015 to May 2016. 
%This complex system now is mature and is operating on a nightly basis. 
Pupil illumination variations during tracking are illustrated in Figure~\ref{pupil}, along with an example of the illumination model and an image from one of the wavefront sensors. The illumination model is utilized to predict variations in light reaching all points on the field of view and is used to correct the brightness observed for guide stars so that they can be used to monitor atmospheric transparency, as an input into target choices.
The model is also employed to correct the relative throughput of VIRUS IFUs when needed (e.g. \S\ref{sec:Vperformance}).

\subsection{Facility Calibration Unit}\label{subsec:fcu}
The facility calibration unit supports VIRUS and the facility instruments and consists of an enclosure (``source box") containing various calibration sources and an input head. The facility calibration unit head, connected to the source box through two liquid light guides\footnote{Lumatek Series 300 and 2000, to support the broad wavelength coverage}, is attached to the bottom of the WFC as part of the lower instrument package,
and can be deployed into the beam to inject calibration light through the WFC whenever calibration is needed. A set of Fresnel lenses and engineered diffusers are used in the facility calibration unit head to mimic the caustics of M1 as closely as possible, to re-produce both the telescope’s pupil and focal surface illumination patterns \citep{lee12}. 

The source box has a switch-yard to select the lightsource \citep{lee12}. Continuum light for flat field calibration is provided by a combination of a broad-band laser pumped Xenon laser driven lightsource\footnote{Energetiq Technology model EQ-99-FC-S} for UV to red and a quartz-tungsten halogen lamp for red to near-infrared. Wavelength calibration sources of Mercury, Cadmium, Neon, Iron-Argon, Krypton, and Thorium-Argon can be selected.
Various imaging and non-imaging optical components (e.g. Compound Parabolic Concentrators, cone reflectors, condenser lenses) are used for efficient coupling between different types of calibration lamps and the light guides, covering wavelengths from 3500 \AA~ to 1.8~$\mu$m \citep{lee12}. One of the switch-yard positions selects a calibration fiber feed for the HPF instrument.
This fiber injects light from a laser frequency comb \citep{ost12} located in the basement spectrograph room for precision wavelength calibration as well as other calibration sources specific to the needs of HPF \citep{halver14}.
Instrument-specific calibration sets are obtained at the start and end of every night and supplemented by twilight flats for VIRUS and LRS2. Note that HETDEX data processing relies exclusively on twilight flats \citep{geb21}, while data processing for other VIRUS projects utilizes the facility calibration unit laser driven lightsource, Mercury, and Cadmium light sources (\S\ref{sec:spectra}).

One of the liquid lightguides is not rated for the lowest temperatures encountered during operations (-10~$^\circ$C) and has been replaced once after five years operation. The brightness and throughput of the calibration system is monitored with the LRS2 instrument. The laser driven lightsource varies in brightness over months timescales and has been refurbished once over the same period. 
\\

\subsection{Telescope and Instrument Control System}\label{subsec:tcs}
The performance of the HET is critically dependent on metrology and the control software system, as the moving payload needs constant and precise adjustment to maintain optical alignment. The WFU has tight specifications on pointing, tracking and guiding performance (in all axes) that have been met by a combination of careful systems design and detailed analysis of performance referring to physical models of the hardware.

The new integrated software control system for the HET WFU 
uses a component architecture providing a high degree of monitoring, automation, scriptability and scalability \citep{rams16,rams18}.
It consists of a network of control systems, each of which models a sub-set of closely coupled hardware. The control systems communicate with each other using a simple but flexible messaging scheme encoding commands to subsystems and events informing of state changes. Each system is responsible for specific functions based on type or proximity to hardware, and is designed to be run autonomously. For engineering purposes, each of the subsystems can be scripted independently.

The primary software control systems for the WFU and VIRUS are the telescope control system, the prime focus instrument package control system, the payload alignment system, the VIRUS data acquisition system (\S\ref{sec: camera}), and the tracker motion control system (\citealt{beno12}), along with a centralized logging system. In addition to these control systems, graphical user interfaces for the Telescope Operator and Resident Astronomer have been developed. The telescope control system is responsible for coordinating the operation of all other control systems and can be scripted for automatic observing (\S\ref{subsec:target}); knowledge of the high-level astronomy-related state is restricted to telescope control system. 

The PFIP control system controls the hardware on PFIP, including the large rotating-blade shutter. The payload alignment system is responsible for gathering and processing metrology from the various alignment systems, such as cameras, needed to close all tracker-motion related loops. The tracker motion control system is based in the Matlab-Simulink environment in a dSPACE Inc. controller. Constraints within the tracker motion control system environment limit the ability to perform complex calculations at the 2.5 ms update rate of the motion control system, so telescope control system interprets all the higher-level functions for the tracker motion control system, such as generating the track-position data stream. Logging from tracker motion control system is at 5 Hz and from other subsystems at their native update rates. The logger accesses local databases if the central log-server is down. These local databases are synchronized automatically with the central log server when it is available. In addition to log messages, logging can be configured to record any subset of events generated by the system and thus obtain detailed execution traces. This adjustment can be performed at any time for engineering purposes without interfering with the operation of the control system, generating no additional overhead or changes in timing. 

\subsection{WFU Integration, Alignment, and Mount Model}\label{subsec:wfuintegrate}
Table~\ref{tab-chron} presents the chronology of the WFU and deployment of instrumentation. 
Following delivery of the WFC to HET, its alignment was rechecked and then the WFC and PFIP were integrated on the tracker. Initial alignment of the WFC axis to the telescope was made with an alignment telescope looking down on the primary mirror to establish the tracker position offsets for the WFC to be centered on and normal to the center mirror segment. The alignment telescope was then reversed to view along the WFC axis to the center of curvature tower instrument, which verified that the WFC was aligned with the primary mirror central axis. Details are given in \citet{hil16c} and \citet{lee16a}.

Following this alignment process, first light was achieved on 29 July 2015, with pointing within an arcminute, and excellent image quality on the acquisition camera (1.3$\arcsec$ FWHM, consistent with the expected median image quality of the telescope system). Achieving good on-axis image quality did not vindicate the internal alignment of the WFC, however (\S\ref{subsec:wfc}), so over the Fall of 2015 a series of on-sky tests were performed with miniature deployable wavefront sensors over the field of view (mounted in the VIRUS IFU seats in the input head mount plate)
as a final confirmation of the performance of the WFC \citep{lee16a}. By this point, both sidereal objects and geostationary satellites could be tracked with high reliability. Testing of the internal WFC alignment required acquisition of a geostationary satellite on each of the deployable wavefront sensors in turn, measuring the wavefront as a function of field position. These tests verified that the WFC was aligned within specifications with only a very small tilt of the focal surface with respect to the optical axis. Details of the tests and analysis are provided in \citet{lee16a}.

The original HET system utilized a heuristic mount model based on on-sky measurements, which convolved the many physical effects that contribute to the pointing and tracking accuracy of the integrated system. Adjustments to improve pointing would often result in poorer tracking accuracy, and vice-versa. For the WFU, the mount model was based on direct physical measurement of subsystems that could be combined to create a deterministic correction to the tracker position with well-understood physical underpinnings. The primary tool in this effort was a laser tracker\textcolor{red}{\footnote{Automated Precision Inc. model LTS-3000}} and spherically mounted retro-reflectors, which were utilized to understand the deflections of the tracker subsystem, using a dummy WFC to mimic the load \citep{good18}. 
At this point, the distance between the payload and primary mirror was established so the telescope would be in focus when the WFC was deployed.
These measurements created a transform with low-order terms that accounted for the deflections of the tracker relative to the ideal tracking sphere. This transform was further refined using the TTCAM and distance measuring interferometer to provide direct measurements of the payload relative to the surface of primary mirror, and thereby tie the tracker frame to the optical frame provided by the surface of the primary mirror, aligned by the center of curvature tower instrumentation\footnote{see \S\ref{subsec:imagequal} for details of the center of curvature tower instrumentation}.
Deflections of the telescope structure during the track, due to the unbalanced loading as the payload moves in X,Y, and tilts to remain normal to the primary mirror, were also measured relative to the telescope pier with the laser tracker and incorporated as a further layer of mount model terms that transform the tracker frame of reference to the projection on-sky. 
After setting in azimuth, the telescope structure sits on four feet on an exceedingly flat concrete pier.
As the telescope is moved in azimuth to access different declinations there is also a term in the pointing mount model that accounts for the small irregularities of the pier.  This approach to the mount model proved highly successful with initial pointing residuals meeting requirements ($<$~30\arcsec) in the central half of the tracker range. One final layer of low-order corrections, based on on-sky residuals, resulted in requirements and goals  being met over the entire tracker range ($\leq$~12\arcsec). 
Just as important as pointing, open-loop drift rates are low and correlate closely with pointing residuals, indicating that improvements in one will be reflected in the other.  

\section{VIRUS} \label{sec:virus}

The requirements in \S\ref{sec:design} emphasize ability to cover area quickly, require large-scale replication \citep{hil14}, and lead to the following properties for VIRUS: each of the 78 VIRUS units is fed by 448 fibers that each cover 1.8 arcsec$^2$ on the sky, split between two spectrograph channels per unit. The fibers feeding a two-channel unit are arrayed in a 51$\times$51 arcsec$^2$ IFU.
The fibers in the VIRUS IFUs have a fill-factor of 1$/$3, such that offsets of the telescope in an equilateral triangle pattern of side $1 \farcs 46$ will fill in the area. These offsets are referred to as the dither pattern. An observation for HETDEX consists of three exposures, each 360 seconds duration, with dithered offsets between. 

\begin{figure}[!ht]
\epsscale{1.18}
\plotone{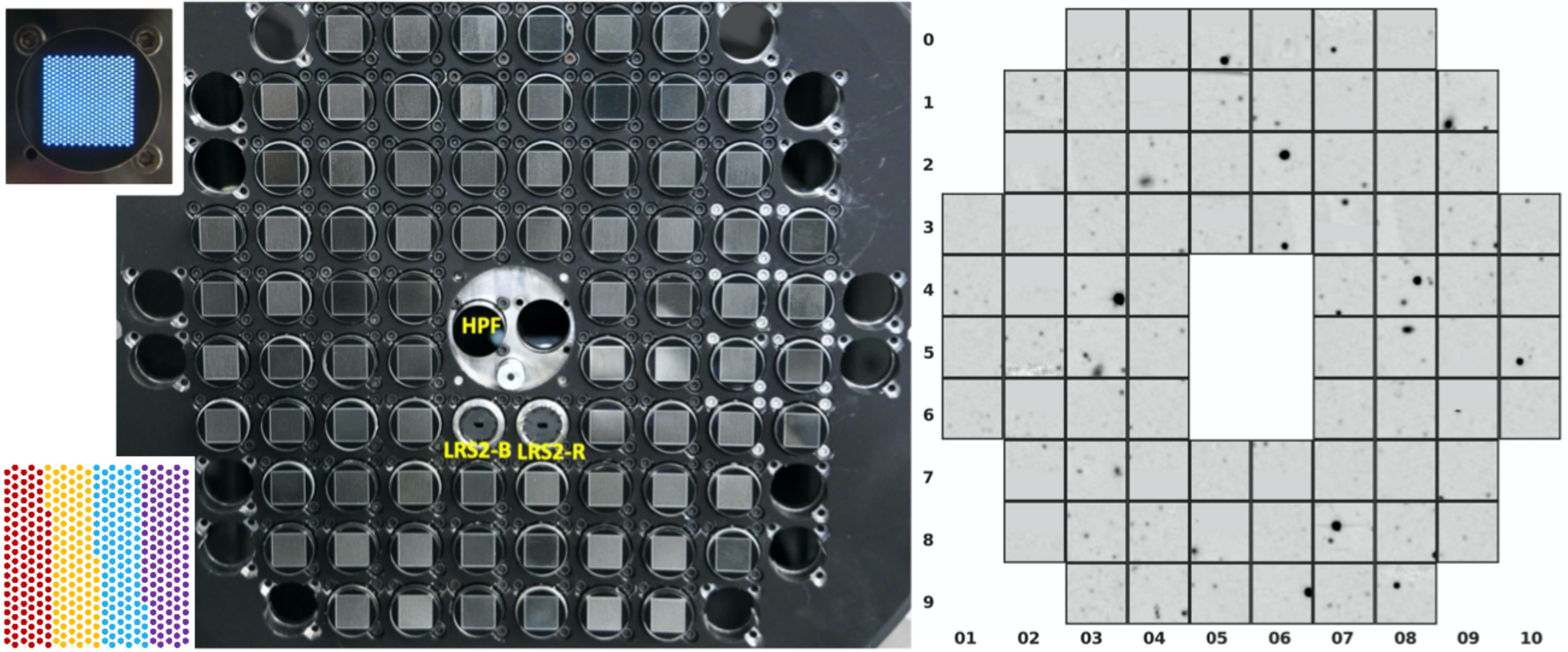}
\caption{
\label{ihmp} \footnotesize
Left – Photon’s-eye view of the deployed fiber integral field units (IFUs) for VIRUS and the fiber feeds for other HET instruments mounted in the precision input head mount plate. The two IFUs of the LRS2 instrument (LRS2-B and LRS2-R)
and the fiber feed for the Habitable-zone Planet Finder (HPF) occupy three of the four IFU locations
in the central region in the input head mount plate. The fiber feed for the future upgrade of the
High Resolution Spectrograph (HRS)
will be deployed in the remaining central position. VIRUS IFUs are arrayed in a
100\arcsec\ center-to-center grid. The full complement of 78 VIRUS IFUs are shown installed. Each IFU covers \hbox{51$''$ $\times$ 51$''$} and has 448 optical fibers.  The 16 empty seats on the periphery will not be populated. For scale, the corners of the outer VIRUS IFUs in this image are~$\approx 18'$ apart, which is~$\approx$~190~mm at the HET focus.
A magnified image of a single VIRUS IFU is shown inset at top left to illustrate the fiber layout.
The inset image at the lower left shows how the fiber layout maps to CCD amplifier readouts (with different colors for the 2 CCDs, each with 2 amplifiers). 
Right – Example of sky data from VIRUS on a typical HETDEX field with all 78 spectrograph units operating.
The spectral dimension is collapsed to create synthetic g-band images for each IFU;
%It is collapsed over the full wavelength weighted by the PanSTARRS g filter.  So it is a synthetic g band image from VIRUS spectroscopy.
this representation removes the $\approx$50\arcsec\ gaps between the IFUs and increases the size of the IFU fields by a factor of two, for better representation.
The central 6 positions are reserved for the other instruments. 
%IFUs are deployed independently of the spectrographs, so four of the IFUs shown on the left were not yet attached to spectrograph units and hence were not active on-sky in the presented data.
The direction of the parallactic angle on sky is up in this figure. Numbers represent the IFU coordinate system in the input head mount plate.
}
\end{figure}

The spectral resolution of VIRUS is 5.6~\AA~ (resolving power R~$\sim$~800 at 4500~\AA), with coverage of 3500$-$5500~\AA. The optical design is simple, using three reflective and two refractive elements. High throughput is obtained with dielectric reflective coatings optimized for the wavelength range. VIRUS units are located in large enclosures off the moving payload of HET (Figure \ref{HETlayout} and \S\ref{sec: vss}), 
while maintaining fiber length of $\sim$20 m on average, to preserve as much UV response as possible. The full VIRUS array can obtain 35,000 spectra simultaneously, with 14 million (spectral $\times$ spatial) resolution elements. In total, the VIRUS CCDs have 0.7 Gpixels (unbinned), comparable to the largest imagers yet deployed. The IFUs are arrayed in a square grid pattern within 
18 arcminutes field diameter and fill this area with $\sim$1$/$4.5 fill factor.
Figure~\ref{ihmp} shows the 78 IFUs arrayed in the input head mount plate  at the prime focus of HET, along with an image reconstructed from the spectra of 74 spectrograph units ($\sim$33,000 fibers). Note the 100\arcsec\ grid pattern of the IFU layout.

The VIRUS spectrographs have been designed and constructed by the McDonald Observatory of
the University of Texas at Austin (MDO) and Texas~A$\&$M University (TAMU),
the IFU development was led by the Leibniz Institute for Astrophysics (AIP),
and many VIRUS mechanical components were supplied by
The University of Oxford and the Institut f\"ur Astrophysik G\"ottingen. The VIRUS data processing software was led by the Max-Planck-Institut f\"ur
Extraterrestriche-Physik (MPE) and The University of Texas at Austin. 
%**
The cameras were produced at MDO, the collimators at TAMU \citep{marsh14}, and the IFUs at AIP \citep{kelz14}.
Spectrograph integration, alignment, and characterization is led by MDO \citep{tutt16,hil16a}.

\medskip
%\subsection{\hi VIRUS Spectrographs} \label{sec: spectrographs}
\subsection{VIRUS Spectrographs} \label{sec: spectrographs}
\smallskip

The details of the design, prototyping, and production of VIRUS and associated subsystems are described in
\cite{hil08a,hil18a,murphy08,chonis10,lee10,vattiat10,murphy12,chonis12,marsh14,kelz14,tutt16,vattiat18,spencer18,lee18c}.
\edit1{Table~\ref{tab-virus} summarizes the basic properties of the spectrographs, which are discussed in more detail in the following sections.
}

\begin{figure}
\epsscale{1.0}
\plotone{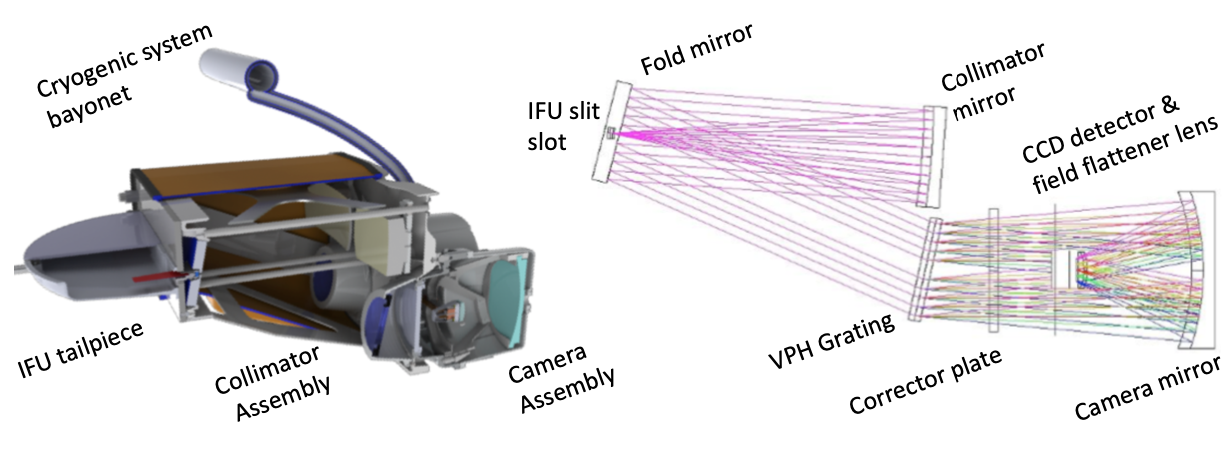}
\caption{
\label{virus} \footnotesize
Layout of the production VIRUS spectrograph unit. Each VIRUS unit has two identical spectrograph channels. A cutaway of the production mechanical design is shown on the left and the optical design of one channel is shown on the right. The volume phase holographic grating has 930 lines per mm and the wavelength coverage is fixed at 3500~-~5500 \AA.  The mechanical design cutaway shows the IFU slit assembly on the left, mounted to the collimator. The slits of the IFU assembly protrude into slots in the fold mirrors, to minimize obstruction. The distance between the fold and collimator mirrors is maintained with three invar struts for each channel to maintain focus as the ambient temperature changes. The $f/$1.25 Schmidt camera has an internal CCD in the vacuum mounted to the main bulkhead of the collimator. The two CCDs are cooled by a flexible line from above, filled with liquid nitrogen, with a breakable cryogenic bayonet connection, as shown on the left.
The spectrograph unit without IFU mounted has dimensions 970 mm long, 740 mm wide, and 460 mm deep, and weighs 73 kg.
Figure partly adapted from \citet{hil18a}, Fig. 2.
}
\end{figure}

The VIRUS instrument 
was evolved for mass-production from the Mitchell Spectrograph and reconfigured with 
three basic sub-units: the fiber IFU, the collimator assembly, and the camera assembly. 
While the Mitchell Spectrograph can be reconfigured through collimator angle and disperser changes, the production VIRUS units have no moving parts. 
Each VIRUS unit houses two spectrograph channels (Figure~\ref{virus}). The beam size of each channel is 125~mm, allowing the collimator to accept an $f$/3.32 beam from the fibers, accommodating a small amount of focal ratio degradation (FRD, e.g. \citealt{schmoll03,murphy08,murphy12}) of the $f$/3.65 input from the telescope.
Each camera is an $f$/1.25 vacuum Schmidt design with 166 mm focal length and a custom
\hbox{2064 $\times$ 2064} format CCD with 15~$\mu$m pixels at its internal focus. 
A single corrector plate suffices for both the Schmidt collimator and camera, and acts as the vacuum window of the camera. The field flattener lens in front of the CCD is also an aspheric element\footnote{aspheric elements manufactured by Asphericon Inc., Jena}. 
The three mirrors in the system have dielectric high reflectivity coatings optimized for 3500~-~7200 \AA.
The IFU, collimator, and camera subsystems are connected by kinematic interfaces at the two main plates of the collimator assembly (Fig. ~\ref{virus}). These plates incorporate precision-machined location features. The  evolution of the spectrograph design considered using castings for these plates, but computer numerical controlled machining from aluminum plate stock proved to be more cost-effective, given the capacity of machines available at the University of Oxford.   
The mechanical design and optical tolerancing of the spectrographs for mass production are described in \citet{vattiat10} and \citet{lee10}, respectively.

While the VIRUS units are mounted in fixed housings and are gravity invariant, their enclosures track ambient temperature in the telescope enclosure and they are required to operate with high stability over a temperature range of $-5$~$^\circ$C to~+25~$^\circ$C.  The instrument is specified to not require recalibrating for shifts in the positions of the fiber spectra over the temperature range encountered in an hour, with the goal of applying a single set of calibrations during an entire night. Stability is crucial since the data processing and analysis are sensitive to shifts of the spectra on the level of $\sim$0.1~ pixel. 
The requirement corresponds to shifts smaller than
0.5~unbinned pixels (one-tenth of a resolution element) at the detector for a~5~$^\circ$C temperature change.
This stability was achieved by using an all-aluminum structure with Invar-36 metering structures for the collimator mirror focus and the internal structure of the cameras. Angles within the optical path are maintained by the homologous expansion of the aluminum structure, 
while flexures accommodate the difference in coefficient of thermal expansion between the aluminum and invar.
The observed stability is a factor of five better than the requirement.

VIRUS employs volume phase holographic gratings, which offer high efficiency and low cost. Details of the grating design, development, and testing are reported in \cite{chonis12,chonis14}. The grating has a fringe frequency of 930~lines~mm$^{-1}$ and operates at order $m$~=~1 in transmission from~3500 to~5500~\AA, for unpolarized light. Efficiency is optimized for the ultraviolet with the angle of incidence of $\sim$9$^{\circ}$ and the angle of diffraction of $\sim$15.3$^{\circ}$ at 450 nm.
In addition, a~1$^{\circ}$ tilt of the fringes in the grating to ensures that the “Littrow”
recombination ghost \citep{burgh07} falls off the detector for the VIRUS configuration.  
The contract for 170 gratings was awarded to SyZyGy. MDO provided SyZyGy with a custom test instrument with which to evaluate grating performance in diffraction efficiency and scattering at three wavelengths in nine sub-apertures over the 138 mm diameter clear aperture of the gratings. 
Details of the design and results of these tests for the 170 gratings are reported in \citep{chonis14}. 
\\

\subsection{VIRUS Fiber Feed} \label{sec: IFU}
%\subsection{\hi VIRUS Fiber Feed} \label{sec: IFU}

\begin{figure}[!ht]
\epsscale{1.0}
\plotone{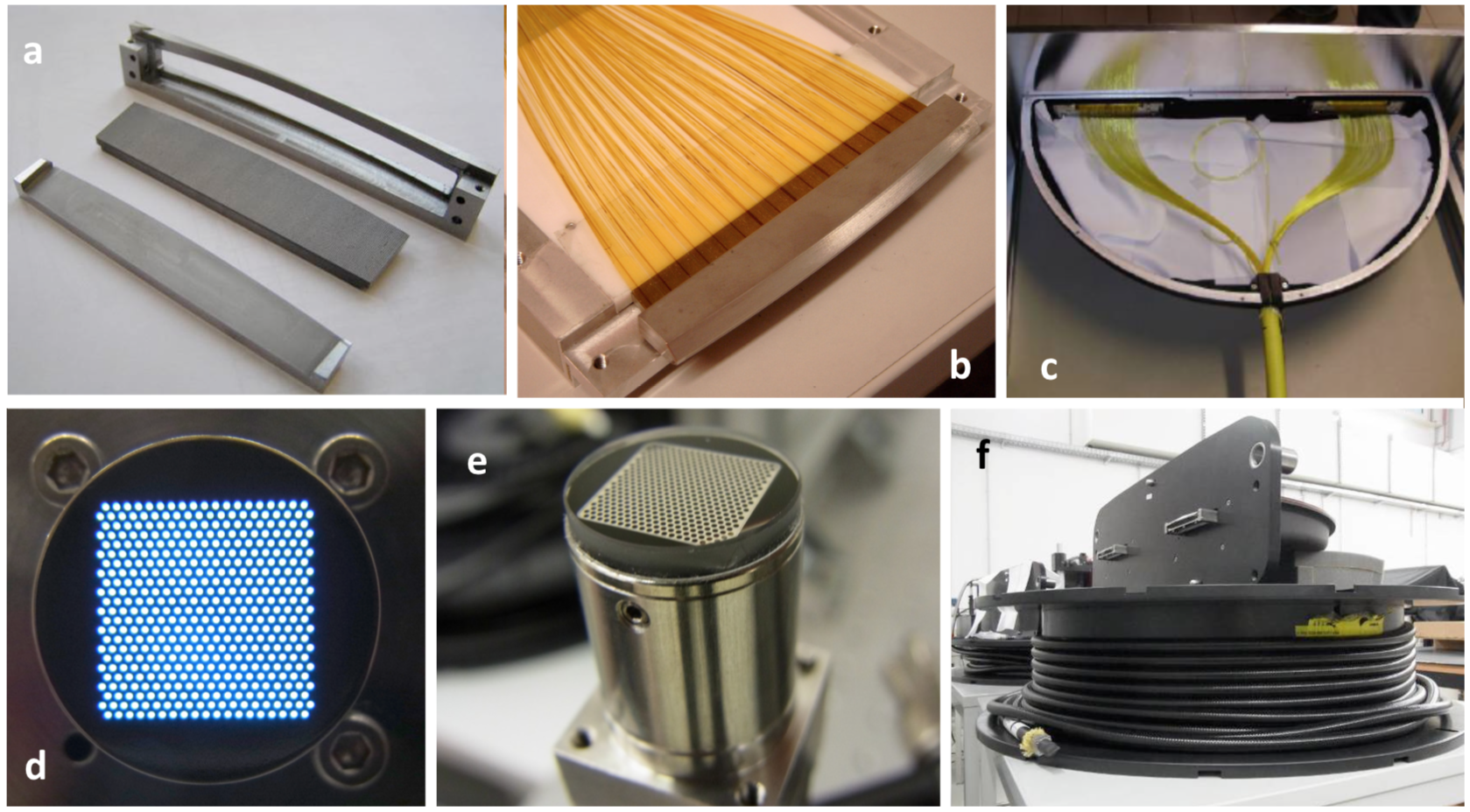}
\caption{
\label{ifuprod} \footnotesize
Production of fiber IFU cables at AIP. (a)~precision slit assembly components comprising (from lower left to upper right) the groove block that locates the fibers, the cap and the cylinder lens holder that acts as the reference for focus;
(b)~an assembled and polished slit block containing 224 fibers;
(c)~an integrated output slit assembly with two slit blocks mounted and the fan-out of fibers from the conduit;
the two spectrograph channels are separated by 320~mm; (d)~an input head, back illuminated, with
448~fibers; the fiber cores have a 266~$\mu$m diameter and one-third fill factor,
and the array is approximately 9~mm on a side. 266~$\mu$m projects to 1\farcs5 on sky.
(e)~input head with cover lens installed. The head is located within a cylindrical seat in the input head mount plate (Fig.~\ref{ihmp}) with the clocking constrained by a pin (not shown, pressed into the hole at the lower left), and secured with three machine screws, as shown; 
(f)~a completed IFU of 20~m length on transport spool. Note the protruding fiber slits and two bushings that engage on pins to guide the output slit assembly safely onto the collimator unit.
Figure adapted from \citet{hil18a}, Fig. 3.
}
\end{figure}

IFU development at AIP (\citealt{kelz14,kelz21}) and MDO \citep{murphy08,murphy12}
focused on establishing a design that minimizes FRD, maximizes throughput, and can be manufactured in quantity \citep{kelz14}.
An overview of IFU development and performance is presented in \citet{kelz21}.
Careful and rigorous apportioning of tolerances between the components aimed to
retain 95\% of the transmitted light within the spectrograph pupil
(125~mm, $f$/3.32) for the input focal ratio from the telescope of $f$/3.65. Figure~\ref{ifuprod} presents images of the slit and input ends of production fiber cables, along with other views of the IFU assembly.  
In total, the VIRUS IFUs contain $\sim$700 km of fiber, which was custom manufactured by FiberTech GmbH and CeramOptec GmbH. 
The high-OH silica fiber has core size 266 $\mu$m, cladding 290 $\mu$m, and buffer 320 $\mu$m diameter\footnote{fiber types are designated: CeramOptec UV265/292P/320 and Fibertech AS266/292UVPI/318}.
266 $\mu$m projects to 1\farcs5 on sky. Each IFU contains 448 fibers, divided equally between two slits.
The design was kept as simple and lightweight as possible. The input head consists of a precision micro-drilled block\footnote{of ARCAP AP1D from Euro Micron}, into which the fibers are fed,
which is in turn clamped within a stainless steel shell that provides the mounting features. 
At the exit, the cable bifurcates within a slit housing into two slits with the fibers glued to grooved blocks. The grooves aim the fibers such that their axes pass through the center of curvature of the spherical collimator mirror and are normal to the surface of the mirror. At input and output, the fibers are glued in position with epoxy\footnote{with Epotek 301-2} and then cut and polished.
The input and output are bonded to a thin lens and a cylindrical lens, respectively,
both of fused silica and anti-reflection coated\footnote{input bonded with Norland 63 UV curing adhesive and output with Cargille code 0607 gel}. The input lens ensures that
the chief ray of the curved focal surface is normal to the fiber input for all the fibers,
despite a flat input face.
This feature is necessary due to the curvature of the HET focal surface (984 mm radius of curvature) and the size of the IFU fiber array (12.5 mm on the diagonal).

The conduit housing the fiber cables underwent extensive design evaluation and prototyping.
It is important that the fibers not piston significantly within the conduit where they exit
the cable into the output slit assembly (Fig.~\ref{ifuprod}c). Such pistoning could occur due to changes in axial load or ambient temperature swings;
particular concerns are shipping and handling during installation, i.e., avoiding twists and torsional stress \citep{murphy08}.
To minimize the weight of the conduit, which could dominate the total weight,
a custom fully-interlocked aluminum conduit with polyvinyl chloride sheathing from Hagitec was adopted.
An inner sock of Kevlar protects the
fiber from the internal structure of the conduit. The Kevlar is tensioned during assembly, which stabilizes the length of the conduit to prevent fibers pistoning or developing axial  stress. In total, the design evolved through six versions, including those for the Mitchell Spectrograph. Before production, the final design was exercised through the full range of motions expected at the HET in a lifetime test, which simulated 10.2~years of wear (188.7~km of linear travel) on a single
fiber bundle. Results of the lifetime test are described in \cite{murphy12},
which qualified the cable design for final manufacture. During the test, there were
no signs of the fiber pistoning within the conduit, after initial settling; this behavior
has been borne out by experience with the IFUs deployed at the HET. 

During production, manufacturers were provided with kits of parts (fiber, mechanical parts, conduit, etc.) and they performed the assembly and polishing of the input and output surfaces to a strict prescription \citep{kelz14}. Final integration of the slit blocks into the output slit assembly was done at AIP. 
Three production lines were established, based on qualification work at several vendors:
AIP\footnote{Assembled by Christian Haubitz-Reinke, Berlin Fibre; www.berlin-fibre.de}, CeramOptec, and FiberWare. Acceptance testing and evaluation at AIP included confirmation of physical properties, microscope examination of polish of the input and output ends against fiducial standards, FRD testing, throughput testing, fiber mapping, and position measurement. Fiber positions within the input head were measured with a precision reimager against a fiducial head that had been measured externally on a coordinate measuring machine in order to relate the fiber core positions to the input head mount features. 
The fibers deviate from a uniform grid with a dispersion of 10 $\mu$m rms (an on-sky projected dispersion of $0 \farcs 05$).  
A final system test was performed on each IFU cable with a fiducial spectrograph unit, supplied to AIP by MDO, generating an IFU cable report and fiber mapping files that are used for data processing and record keeping.  
A detailed description of the design and production of the IFUs and on-sky performance is given in \citet{kelz21}. Design of the fiber handling and deployment of IFUs is described in \citet{vattiat18}.

\subsection{Camera and Detector System} \label{sec: camera}
%\subsection{\hi Camera and Detector System} \label{sec: camera}

The camera cryostat vacuum is shared between a pair of spectrographs;
this approach significantly reduces the component count and increases
the evacuated volume. Similarly, the cryogenic cooling system is shared within a unit,
which also reduces the component count of the VIRUS cryogenic system
and reduces losses associated with fittings and valves. The VIRUS cryogenic system and its
testing are described in Sec.~\ref{sec: vss} and in detail in \citet{chonis10} and \citet{spencer18}. The cryostat is composed of two aluminum castings, post-machined only on critical mount surfaces and flanges. The cryostats were manufactured by MKS Inc., following extensive evaluation of prototypes. An impregnating step with Locktite Resinol, following machining, is intended to seal the porosity of the cast aluminum.
However, careful leak-checking of the cryostats using a residual
gas analyzer was required in order to locate and repair
leaks that compromise the hold time. Applying two epoxy types of high and low viscosity, for larger and smaller leaks, results in consistent vacuum performance.
While the tooling cost for casting is quite significant, the price for even a single cryostat of this size is competitive with machining from bulk stock, and is much cheaper for the large VIRUS production run. 

The CCDs for VIRUS have  \hbox{2064 $\times$ 2064} format with 15~$\mu$m pixels.  The required readout time is relatively slow at 20~seconds, binned \hbox{2 $\times$ 1,} but low read noise ($\approx$~3~electrons) is required and the parallel readout of 156~CCDs distributed through the volume of the HET structure is challenging.  Each CCD is read through two amplifiers, with the serial registers parallel to the spectrograph dispersion direction to avoid splitting spectra across amplifiers. 
The 156 VIRUS CCDs
total 664~megapixels, when fully deployed, which is comparable to the VLT Multi Unit Spectroscopic Explorer (MUSE, \citealt{bacon10}) and to the largest operational
imaging mosaics. 
The single-exposure raw dataset \hbox{(binned 2 $\times$ 1 at 16 bits digitization),} from the full VIRUS array is 664 MB. 
The integrated detector system was supplied by Astronomical Research Cameras, Inc. (ARC), with the University of Arizona Imaging Technology Laboratory (ITL) providing thinned backside illuminated CCDs with anti-reflection coatings optimized for the VIRUS bandpass, as a subcontract, from wafers manufactured by Semiconductor Technology Associates, Inc (STA).  Since the CCDs are produced in custom wafer runs, an imaging area of \hbox{2064 $\times$ 2064 pixels was selected}, allowing more latitude for alignment. The device is designated STA3600.

\begin{figure}[!ht]
\epsscale{1.15}
%\plotone{camera.png}
\plotone{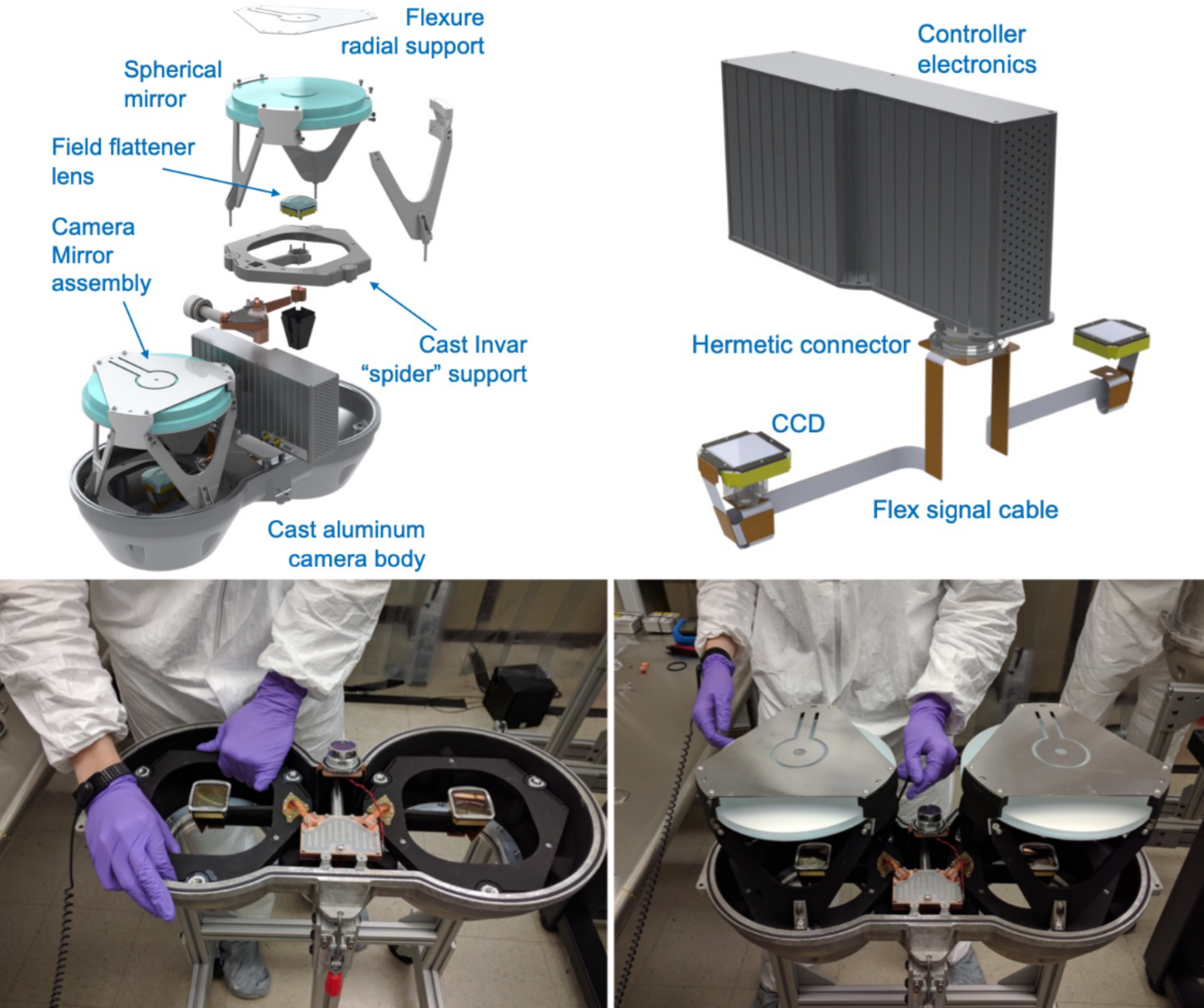}
\caption{
\label{camera} \footnotesize
The upper renderings show the anatomy of the VIRUS camera assembly. 
The upper left hand view displays how the Invar-36 camera-mirror assembly integrates with the "spider" support for the detector and field flattener lens in the cryostat housing.
All parts of the camera-mirror assembly are invar-36. The mirrors have a spherical back for lightweighting. 
The Invar blade flexures at the top provide radial support and 
are bonded to flats at the centers of the mirror backs,
with mirror adjustments at the three corners that are accessed via vacuum feed-throughs on a temporary adjuster back, during alignment.
The upper right rendering shows the electronics signal path via the double-sided flex circuit connecting each detector to the common 4-channel readout electronics (controller). The controller box mounts to a hermetic vacuum feed-through connector that mounts in the cryostat cover (not shown).
The lower two panels show two stages of the assembly of a VIRUS camera. The left image shows
two CCDs with integrated field flattener lenses mounted on cast Invar “spiders” with cold links and flex circuit, integrated in the cast aluminum camera body. The right panel displays the two camera-mirror assemblies integrated,
prior to installation of the camera cover, which is also cast from aluminum.
Figure partly adapted from \citet{hil18a}, Fig. 4.
}
\end{figure}

The design of the detector package, flex circuit, and controller were customized to the VIRUS application, since the engineering effort was spread over a large production run. 
Figure~\ref{camera} presents renderings of the camera and presents the assembly of a pair of detectors in a cryostat. The structure of the camera assembly is all Invar-36.
The CCD package, machined from Invar-36, was designed for minimum obstruction
and lies in the shadow of the field flattener lens. 
The package has a custom header board that brings the traces to a single connector on one side of the package. A custom flex circuit with complex geometry connects the two detectors to a single 55-pin hermetic bulkhead connector. The controller mounts directly to this connector, without external cables, and its form-factor was customized to fit between the cylinders of the cryostat cover.
Detector readout is performed through two diagonally opposite amplifiers out of the 4 available in the CCD design. The amplifier pair is selected via jumpers on the readout system flex circuit, within the cryostat. Detectors are vetted by ITL and then at MDO with 3650 \AA~flat field and Fe55 illumination, to verify and optimize performance. 

The CCD and field flattener lens alignment tolerances are the tightest in the system. To avoid deposition of epoxy in the corners of the CCDs during cure due to outgassing, separate alignment references are used
for the field flattener lens and CCD that allow them to be bonded separately and then assembled. Two alignment stations incorporating alignment telescopes allow a CCD and field flattener lens to be bonded simultaneously \citep{lee18c}.
The CCD is mounted in the cast invar “spider” (Fig. \ref{camera}) which has a minimum-obstruction arm to suspend the detector in the beam. 
The CCD is aligned in a second fixture by adjusting the three spider mount points which are glued to set the alignment.
The CCDs are flat to within $\pm$~10~$\mu$m
over their entire surface, which falls within the tolerance requirements. 
ITL supplies detailed metrology of the shape and location of the CCD surface in relation to the package mount points, and this information is used during alignment of the CCD and field flattener lens assembly \citep{lee18c}. After 72~hours cure time, the field flattener lens is installed onto the detector package. 

Two mirrored spider assemblies are integrated with cold links and flex circuits and then mounted in the camera body
(Figure~\ref{camera}, lower left). The three mount points of each spider interface with precise features post-machined into the aluminum casting.
The alignment of the CCD head assembly to the axis of the camera is achieved within the required tolerances of 50~$\mu$m in centration and separation,
and 0.05$^{\circ}$ in tilt \citep{lee18c}.
Since the camera mirror assembly mounts to the same points, the entire camera becomes a single unit with an Invar-36 structure, with only the aspheric corrector plate window (with much lower alignment tolerances) not as tightly integrated (Figure~\ref{camera}).

The CCD controllers have input 12 VDC power and control and data links on fiber-optic connections. In the ARC readout system, the data system requires several levels of multiplexing, utilizing Peripheral Component Interconnect (PCI) and PCI-Express (PCIe) interface cards.
Each four-channel CCD controller commands two detectors, each with two readout amplifiers.  Ten custom-built multiplexers each combine the output from a set of eight CCD controllers.  
To minimize crosstalk, the timing of the CCD clocks is synchronized to master clock signals on each multiplexer, distributed over a low-voltage differential signal system.
The output of each multiplexer is fed into a separate PCIe interface card mounted 
in the VIRUS data acquisition system computer.
The data are transferred via direct memory access from the PCIe interface cards
into the VIRUS data acquisition system computer memory. 
The software for VIRUS data acquisition controls monitoring and readout of VIRUS units and the new low resolution integral field spectrograph, LRS2. 
It is written in C$^{++}$ and integrated within the
HET Telescope Control System (\S\ref{subsec:tcs}, \citealt{rams16,rams18}).
Controllers have proven unreliable due to a combination of design, component choice, and build quality. Failures of analog-to-digital converter integrated circuits and power supply components have had the greatest impact. 
The clock driver channels needed substantial modification to drive the CCD clock capacitance without generating spurious charge that raises the effective readout noise.
These issues have been addressed through careful analysis and component changes and will be discussed in a future paper. 

As reported in \cite{hil16a}, the original VIRUS detectors suffered from failures in the back side surface treatment, triggered by triggered by chemical contamination from shipping and lab storage containers, that caused significant quantum efficiency depressions as well as clusters of low quantum efficiency pixels, particularly at the corners. This issue led to the majority of the original delivered generation 1 detectors being unusable, but 
enough were identified
with sufficiently good cosmetics to deploy 16 units in 2016, in order to advance understanding of the system and start commissioning the instrument.  
New wafer runs were procured from STA in 2017 and 2019, and ITL has been processing these runs.
These generation 2 detectors have a thicker epitaxial layer of higher resistivity silicon that 
has yielded better results\footnote{CCDs are thinned and backside illuminated. The generation 1 VIRUS CCDs are nominally 17 um thick;  the generation 2 are nominally 27 um thick. The epitaxial silicon has resistivity of 150 ohm-cm for generation 1 and 1000 ohm-cm for generation 2. The CCDs are operated with multi pinned phase readout.}. 

As of May 2021, 7 of the generation 1 units remain in the VIRUS array.
Some of their cosmetic features are detrimental to data quality, but the devices have remained stable.
The small clusters of bad pixels present problems, in particular, since they are of similar size to the instrument resolution elements. Some of the generation 1 CCDs also show many charge traps. In a blind detection experiment of single emission lines, such features must be corrected or eliminated to prevent detection of spurious objects \citep{geb21}.
Units with generation 2 CCDs have been steadily delivered to bring the total to 74 units on sky, as of May 2021 (Fig. ~\ref{ihmp}).
As a gauge of the impact of the CCD cosmetic and readout system issues, 4.4\% of amplifiers have readout issues and 2\% of the resolution elements are masked and eliminated from the data used to date in the HETDEX survey.
This percentage is dropping as controllers are repaired and generation 1 CCDs are replaced.
CCD delivery is the pacing item in completing the delivery of the final spectrograph units to the HET.
A comprehensive description of the detector system for VIRUS will be given in a future paper, once deployment is complete. 

\subsection{VIRUS Alignment and Characterization} \label{sec: alignment}
%\subsection{\hi VIRUS Alignment and Characterization} \label{sec: alignment}

Following assembly of the collimators and cameras, the spectrographs are integrated and aligned. The alignment procedure involves attaching an adjustment back cover to the camera cryostat, in place of the regular cryostat back. The adjustment cover incorporates six ferrofluidic vacuum feed-throughs for manipulation and locking of the camera mirrors. Small adjustments of the collimator mirror tip, tilt, and piston are also allowed.
A test IFU with an input face mask and no cover plate is used to provide a set of 
sparse spectral line images of Hg and Cd over the full field of the CCDs for the alignment procedure.
Experience with aligning the VIRUS prototype revealed that this step was
likely to become a bottle-neck in the large-scale production,
which led to the development of a deterministic alignment procedure that utilizes moment-based wavefront sensing analysis \citep{lee18c}.
This technique relies on the geometric relation between the image shape moments and the geometric wavefront modal coefficients. Moment-based wavefront sensing allows a non-iterative determination of the modal coefficients from focus-modulated images at arbitrary spatial resolutions. The determination of image moments is a direct extension of routine centroid and image size calculation, making
its implementation straightforward in the alignment of systems such as VIRUS. The alignment procedure can be accomplished in three hours per channel, once the detectors are cold.
The resultant image quality exceeds the specifications in most cases, due to the achieved
accuracy of the alignment of the field flattener lens to the CCD.

After alignment, VIRUS units undergo a characterization \citep{indahl18}
before being packed for transport to the HET. The characterization station is located in a separate lab that can be darkened sufficiently to ensure no stray light for the tests. A lab calibration unit \citep{indahl16}
houses a broad-band laser driven lightsource (\S\ref{subsec:fcu})
for flat-fielding and Mercury and Cadmium line lamps for wavelength determination. A standard production IFU is designated for these tests so there is a uniform reference. In addition, a “pixelflat” head that mounts in place of the IFU head and provides a continuous illumination of two slits (rather than the highly spatially-modulated fiber IFU output) is utilized to provide the source for pixel-flat-fields of the detector.
This approach produces a flat field free of spatial-dimension fiber modulation,
and therefore allows characterization of the pixel-to-pixel quantum efficiency
variations and identification of bad pixels.
Bias levels are set and photon transfer curves are generated to determine the read noise and gain of each
channel. Sets of bias and dark frames are recorded to act as a reference once the units are installed at HET.

\begin{figure}[!ht]
\epsscale{1.0}
\plotone{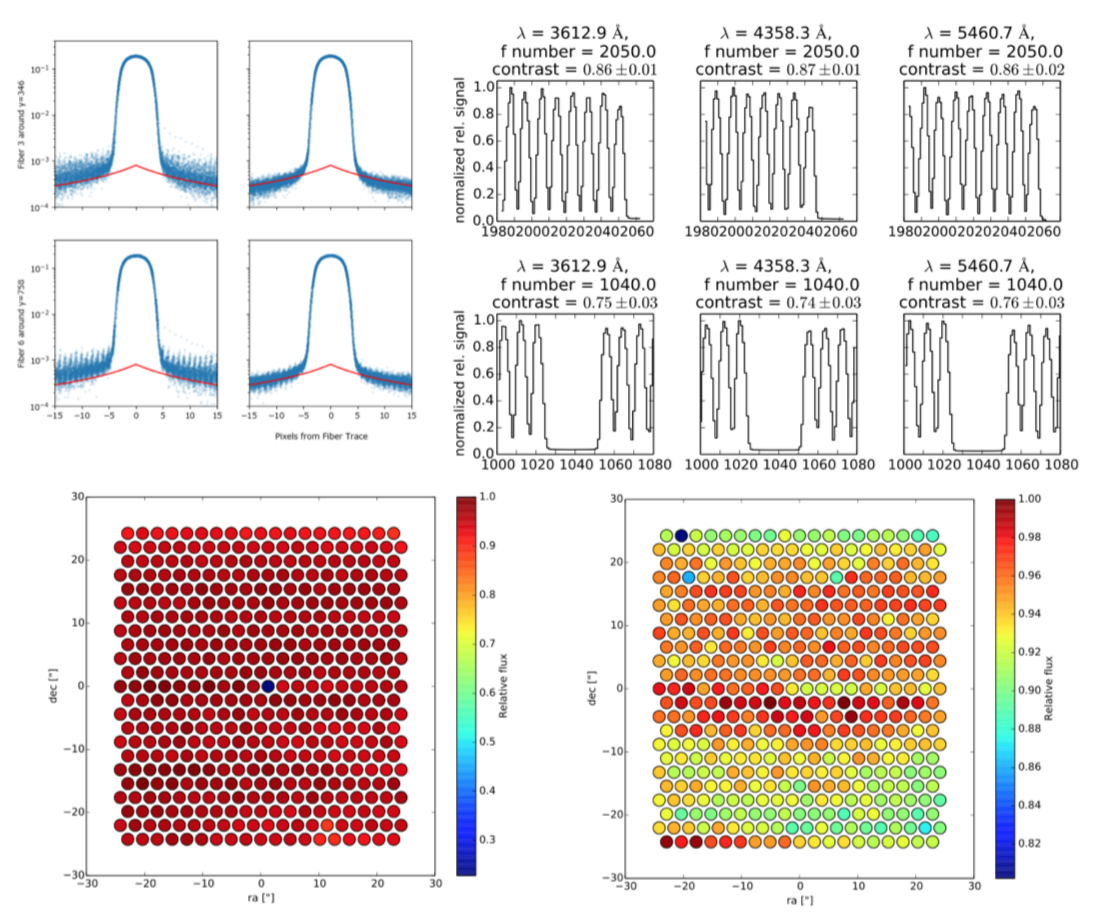}
\caption{
\label{VIRUSchar} \footnotesize
Examples of characterization data obtained on VIRUS spectrograph units and IFUs. Top left: Fiber profiles of isolated fibers in the spatial dimension measured as part of the spectrograph characterization, plotted on a logarithmic scale against offset from profile center, showing Gaussian-hermite core and in red the exponential wings. The wings sum to ~3\% of the integrated flux and account for all the light seen between fibers. Top right: Fiber profiles against pixel number in the spatial dimension measured as part of the IFU characterization at six locations on the detector. This example exceeds requirements. The profiles indicate excellent contrast between the peak and trough of the profiles. The three profiles on the lower row straddle the break between amplifiers where there is a 3-fiber gap. The lower two panels show typical examples of fiber throughput relative to the peak fiber for two IFUs. The axes are nominal fiber position from the center expressed in arcseconds projected on sky. The fibers are shown oversize compared to reality.
The top half of the fibers connect to one of the spectrograph channels and the bottom half feeds the other channel. The split between spectrograph channels occurs at the center of the IFU. Typical fiber-to-fiber non-uniformity is at the 10\% level as seen in these examples. The left hand IFU has a broken fiber near the center.
Figure adapted from \citet{hil18a}, Fig.5.
}
\end{figure}

Figure~\ref{VIRUSchar} 
provides some examples of outputs from characterization of VIRUS spectrographs and IFUs. The spectrographs produce excellent image quality and the fiber profiles are characterized using the sparsely illuminated IFU input. The profiles are well fitted by a Gaussian-hermite function with exponential wings. The wings contain about 3\% of the total light 
%in the core 
and are consistent with the scattering expected from the surface roughness specifications of the spectrograph optics. Fig.~\ref{VIRUSchar} also shows the excellent separation (contrast) between fibers that is achieved in the alignment process \citep{lee12b}. The lower panels in Fig.~\ref{VIRUSchar} present two examples of the IFU fiber-to-fiber throughput measured relative to the highest throughput fiber, which is part of the $\it{LabCure}$ report generated at AIP \citep{kelz14, kelz21}. 
Typical variations are around 10\%, peak-to-peak, with some systematics depending on the fiber location in the slits.

\subsection{VIRUS and LRS2 Support Infrastructure and Deployment} \label{sec: vss}
%\subsection{\hi VIRUS and LRS2 Support Infrastructure and Deployment} \label{sec: vss}

\begin{figure}[!ht]
\epsscale{1.1}
\plotone{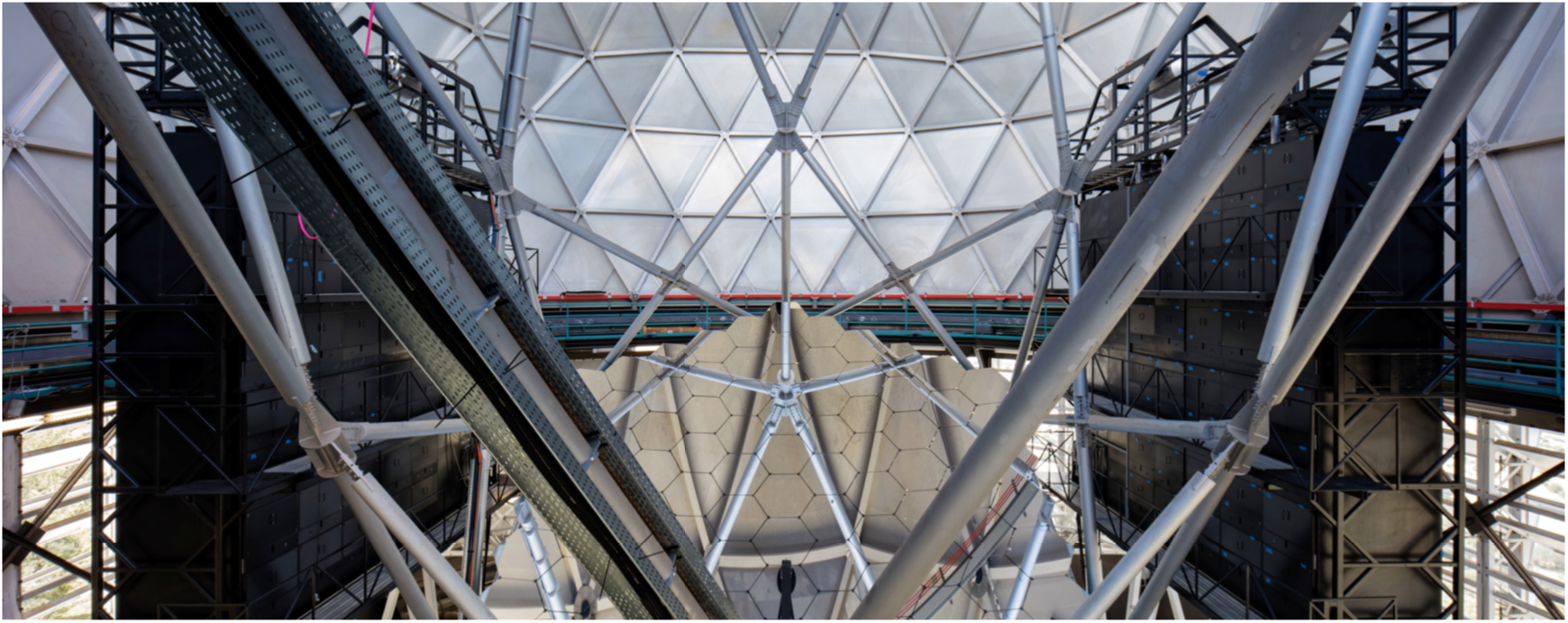}
\caption{
\label{vss} \footnotesize
View of the HET from the front showing the primary mirror and the large VIRUS enclosures either side of the telescope structure. The enclosures sit on the VIRUS support structure, which moves on air bearings to follow the azimuth setting of the telescope. The liquid nitrogen phase separator tanks of the VIRUS cryogenic system can be seen mounted to the top of each enclosure.
In total, the enclosures and VIRUS support structure weigh 42,500 kg (measured with pressure sensing film and via air-bearing pressures) 
when all VIRUS units are deployed, and each enclosure
has dimension \hbox{6.5 m wide $\times$ 1.23 m deep $\times$ 6.15 m tall.}
Figure adapted from \citet{hil18a}, Fig. 6.
}
\end{figure}

VIRUS and LRS2 are fiber-fed, which allows the mass of the spectrographs to be carried in
two enclosures \citep{prochaska14}, one on each side of the telescope structure (Figures~\ref{HETlayout} and~\ref{vss}).
Each enclosure can support 40~pairs of spectrographs, providing capacity for the
78 VIRUS units and the two units of LRS2. 
The enclosures are carried by the VIRUS Support Structure, which is a complex weldment that interleaves with the main telescope structure without applying loads to it that could couple wind induced vibration from the enclosures to the telescope. It rides on separate air-bearings that lift it during changes in telescope azimuth, and is linked to the main structure allowing it to be moved by the main azimuth drive. 
The enclosures exclude light and dust as much as possible.
%are large clean rooms with 
They have filtered air circulation and heat extraction to remove heat from the VIRUS controllers and ensure that the skin temperature of the enclosures remains close to ambient to ensure they do not impact the dome seeing. The weldments for the enclosures were procured by MDO and were outfitted with hatches, seals, cables and the heat removal system by TAMU \citep{prochaska14}.
The VIRUS support infrastructure,
installation, and maintenance procedures are described in \cite{spencer18}.

The distributed and large-scale layout of the VIRUS array presented a significant challenge
for the cryogenic design \citep{smith08, chonis10}. Allowing a 5~W heat load for each detector,
accounting for all losses and a 50\% margin, the cooling source is required to
deliver 3,600~W of cooling power. Following a trade-off between cryocoolers, small pulse-tubes, and liquid nitrogen based systems, it was clear, from a reliability and cost point of view, that liquid nitrogen was the optimum choice. 
The challenge of supplying the coolant to the distributed suite of spectrographs was overcome
by adopting a gravity siphon system fed by an 11,000 gallon external tank.

An important aspect of the cryogenic design is the ability to remove a camera cryostat or spectrograph unit from the system for service without impacting the other units. This capability is particularly difficult in a liquid distribution system. A design was developed that combines a standard flexible stainless steel vacuum jacketed line (SuperFlex) to a cryogenic bayonet incorporating copper thermal connector contacts into each side of the bayonet. When the bayonet halves are brought together they close the thermal contact. The resulting system is completely closed, i.e., it is externally dry with no liquid nitrogen exposure. The camera end is connected by a copper cold finger to the detector. This design has another desirable feature: in normal operation the SuperFlex tube slopes downwards and the bayonet is oriented vertically. Liquid evaporation flows monotonically upwards in order to avoid a vapor lock. If the bayonet is unscrewed and raised upwards, a vapor lock will occur and the bayonet will be cut off from the cooling capacity of the liquid nitrogen. This configuration effectively acts as a “gravity switch”, which passively halts cooling to that camera position, for maintenance or removal. 
This feature has been key in enabling the staged deployment of VIRUS units, while allowing the VIRUS cryogenic system to remain in continuous operation.

The VIRUS cryogenic system was constructed by Midwest Cryogenics and installed in 2012, and has been in continuous use since then. The draw on the external tank is approximately 1200 liters per day or about 2,200 W. That is approximately 60\% of the design cooling power. The external tank has capacity for one month supply and is replenished by regular deliveries of liquid nitrogen, roughly twice a month.

An essential part of VIRUS cryogenic system is its safety system \citep{spencer18}. This system continuously monitors critical variables (e.g., dome atmosphere oxygen levels and liquid nitrogen pressure, flow rates, and storage tank level). When predefined limits are exceeded the system automatically activates strategically-located audio and visual alarms, and if required closes the main liquid nitrogen supply line valve. Each afternoon the system performs an auto-test of the alert system, and sends test telephone alerts to the recipient list, ensuring that the system cannot cause an unsafe condition or go off-line for an extended period without being noticed.

The VIRUS array underwent a staged deployment of IFUs and spectrograph units, starting in late 2015 \citep{tutt16,hil16a, vattiat18,spencer18}.
The left panel of Figure~\ref{ihmp} shows the full compliment of 78 VIRUS IFUs deployed at the focus of HET,
along with the two IFUs of LRS2 and the HPF
fiber feed. As of 2021~May, 74 of the VIRUS IFUs are attached to spectrograph units.
This number is considered complete for the purposes of the HETDEX survey, but the final four units will be brought on line as they become available.

Installation and maintenance of the VIRUS spectrograph units is described in \citet{spencer18}. Primary ongoing activities are vacuum maintenance and special calibrations.
VIRUS cryostats are not outfitted with vacuum gauges, but the combination of cold block heater current and detector temperature setting allow the state of the vacuums to be monitored. CCD temperatures are held at set points between -110 and -100 $^\circ$C. When the heater power needed to maintain a set point drops below a threshold, the set point is adjusted warmer and before it reaches -100 $^\circ$C, the unit is scheduled for cold vacuum pumping. Cold pumping during the day lasts typically 4-6 hours and results in the vacuum being improved from $\sim$2$\times$10$^{-3}$ to $\sim$10$^{-5}$ mbar. Two cryostats can be pumped at once and most last in excess of 4 months between pumpings. The effort required to maintain the vacuums on 78 cryostats is significant, and an upgrade to add ion pumps to all the units is underway.
The principal special calibrations are defocused flat field images that are processed to create pixelflats that isolate the small-scale pixel-to-pixel variations in the flat field. Normal flat fields have the strong spatial modulation of the fiber traces that is on a similar scale to features in the CCD flat field (see Fig. \ref{VIRUSchar}). Spacers are introduced at the kinematic mounts between the IFU slit unit and the spectrograph collimator to sufficiently defocus the spectra that the spatial brightness modulation can be fit and removed to leave the small scale structure in the flat field \citep{geb21}. Introducing the spacers and taking the calibration data is time consuming and the HET staff work to cycle through the full array of VIRUS units once per year. That cadence is sufficient to reveal any small changes in the flat fields with time. 

\section{VIRUS Performance} \label{sec:Vperformance}

With 95\% of the VIRUS array deployed, the performance can be compared with the requirements and technical specifications. VIRUS is the first time a single spectrograph design has been replicated on such a large scale in astronomy, and it is of interest to examine the uniformity of properties across a large cohort of ostensibly identical instruments. In this section, properties of the spectral coverage and resolution, experience with the deployed IFUs, and spectrograph stability and performance are presented.

\subsection{Spectrograph coverage and resolution} \label{sec:speccomp}
%\subsection{\hi Spectrograph coverage and resolution} \label{sec:speccomp}

\begin{figure}[!ht]
\epsscale{1.0}
\plotone{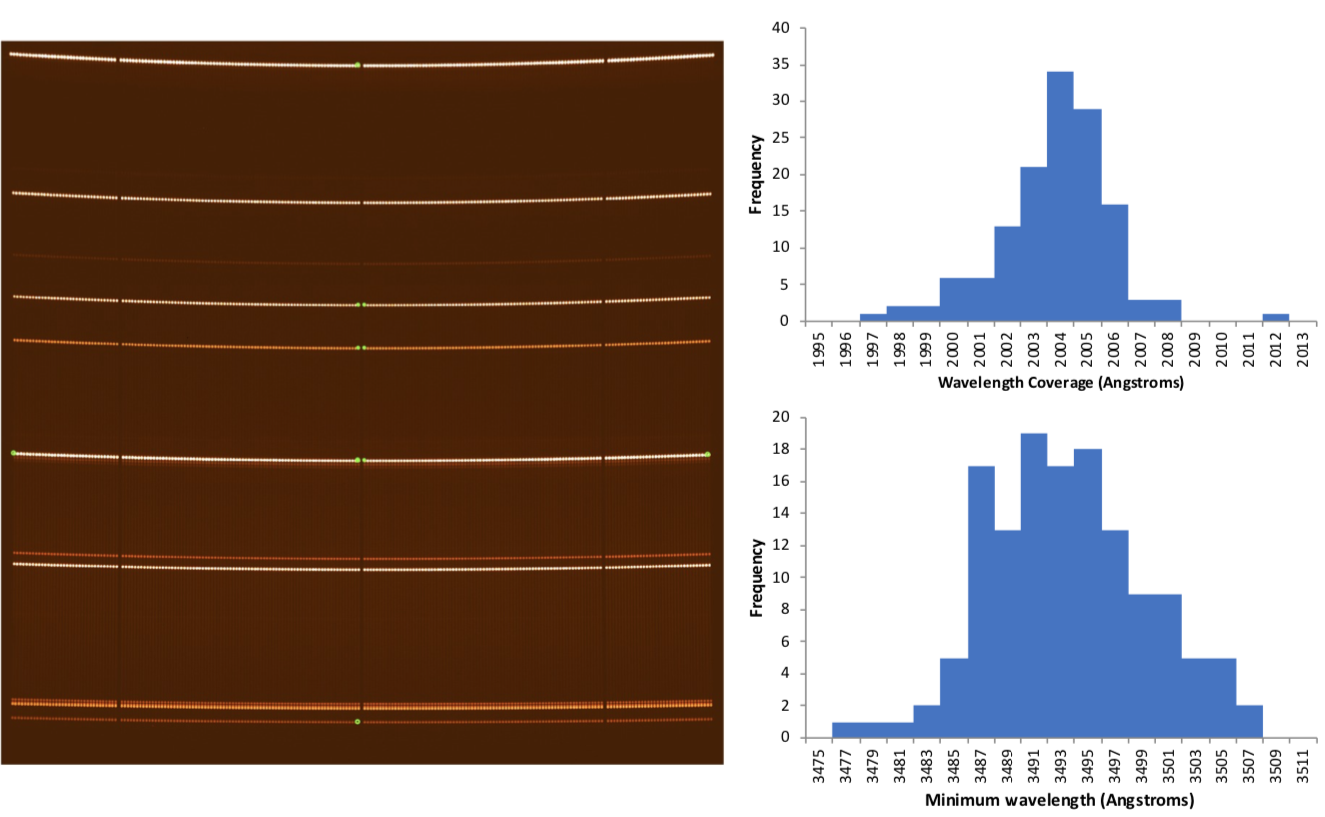}
\caption{
\label{Vcoverage} \footnotesize
VIRUS spectral coverage. Left panel: A spectrograph channel illuminated by Hg+Cd emission line lamps in the lab, following alignment. Wavelength increases going up. Green circles are measurement marks for position on the CCD. Note the intentional gaps with fibers missed in the slit layout. The central gap of 3 fibers provides margin for alignment so spectra do not bridge the (vertical) readout split between the two amplifiers per CCD. The gaps at about 2$/$3 of the slit extent and CCD pixels beyond the ends of the slit provide the ability to better separate the wings of the fiber image profiles. 
The curvature is factored into the wavelength coverage of the channels. 
Upper right: The minimum wavelength coverage accessible for all fibers for 138 spectrograph channels. 
Lower right: The minimum wavelength of the same channels. The dispersion in total wavelength coverage is 2.1~\AA~and in minimum wavelength it is 5.9~\AA. For reference, the spectral resolution is 5.6~\AA.
See text for discussion.
}
\end{figure}

Figure~\ref{Vcoverage} presents statistics on the wavelength coverage of VIRUS spectrograph channels. The image of calibration lines reveals the curvature of lines of constant wavelength with slit position, which was taken into account in the design so as to preserve a minimum of 2000 \AA~ of wavelength coverage from 3500~-~5500~\AA. The fringe frequency of the volume phase holographic gratings is the primary driver of the coverage, while the fringe frequency and angles of incidence and diffraction set by the spectrograph structure control the minimum wavelength. Uniformity of wavelength coverage of one spectral resolution element was targeted, leading to a specification on fringe frequency of $\pm$ 2 fringes and on structure accuracy after alignment of 0.025 degrees. 
The histograms in Fig.~\ref{Vcoverage} demonstrate a mean coverage of 2003.3~$\pm$~2.1~\AA~ and a mean minimum wavelength of 3492.4~$\pm$~5.9~\AA.
%for 138 spectrograph channels. 
The dispersion in minimum wavelength is one resolution element, which was the adopted specification.
This wavelength range corresponds to redshift coverage of 1.87 to 3.52 for Ly-$\alpha$, bracketing the high-level requirement of 1.90 to 3.50.

\begin{figure}[!ht]
\epsscale{1.2}
\plotone{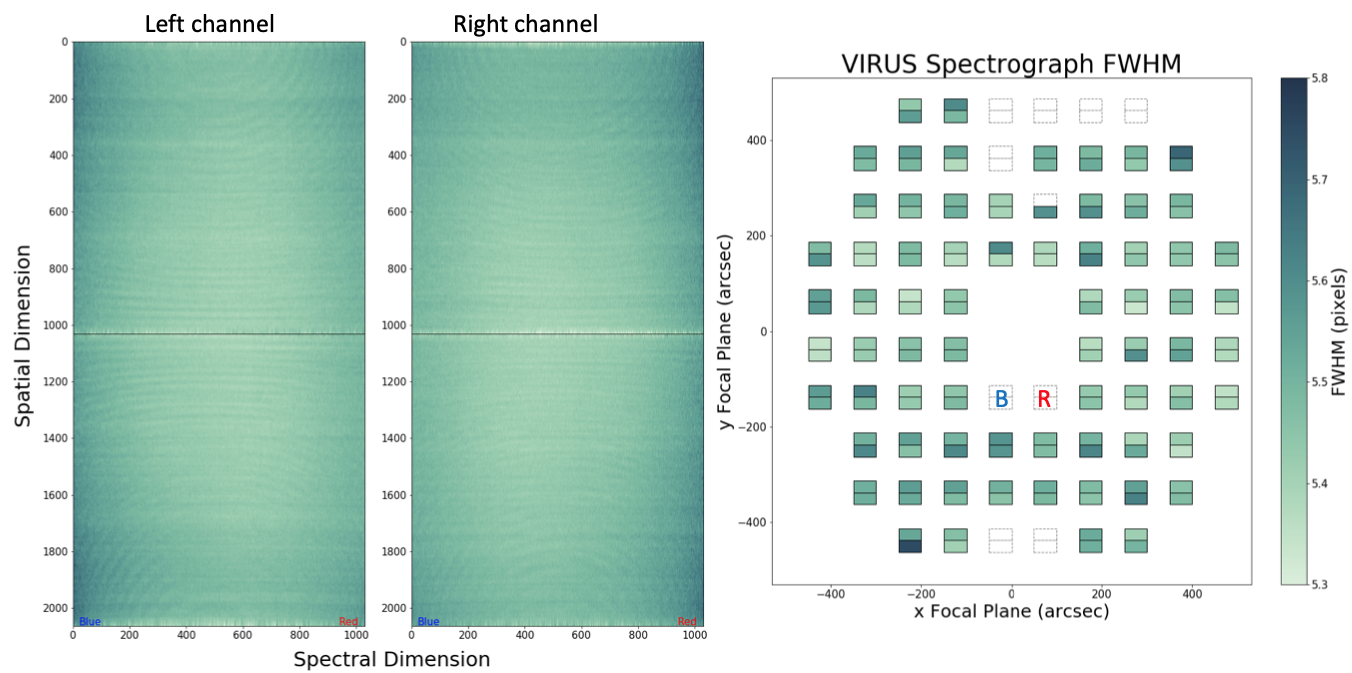}
\caption{
\label{resolution} \footnotesize
VIRUS resolution. Left panel: Median variation of instrumental resolution expressed in unbinned pixels FWHM over the areas of the Left and Right channel CCD detectors. 3500~\AA\ wavelength is to the left and 5500~\AA\ to the right in each channel. Note that the (2064$\times$2064 pixel) CCDs are binned by a factor of 2 in the spectral dimension. The horizontal line shows the position of the break between the two amplifiers for each channel detector readout.
Right panel: Median resolution (FWHM) in unbinned pixels for 141 spectrograph channels active 
on 2019 October 19.
The scale is the same as in the left panel and the color bar on the right shows the full range of values and applies to both panels. 
This projection shows the layout of the IFUs within the focal surface of the input head mount plate. The LRS2-B and LRS2-R IFU positions are marked ``B" and ``R". Positions are expressed in arcseconds projected on sky. The X axis is perpendicular to the tracker bridge and the Y axis is in the direction of the tracker bridge with positive Y being the parallactic angle. 
Each IFU feeds two spectrograph channels with the Left channel mapping to the lower half of the IFU in this projection.
IFU seat positions with no spectrograph attached are shown as dashed outlines. The IFU with only one active channel was exhibiting a readout problem at the time the data were taken. 
See text for discussion.
}
\end{figure}

\begin{figure}[!ht]
\epsscale{1.1}
\plotone{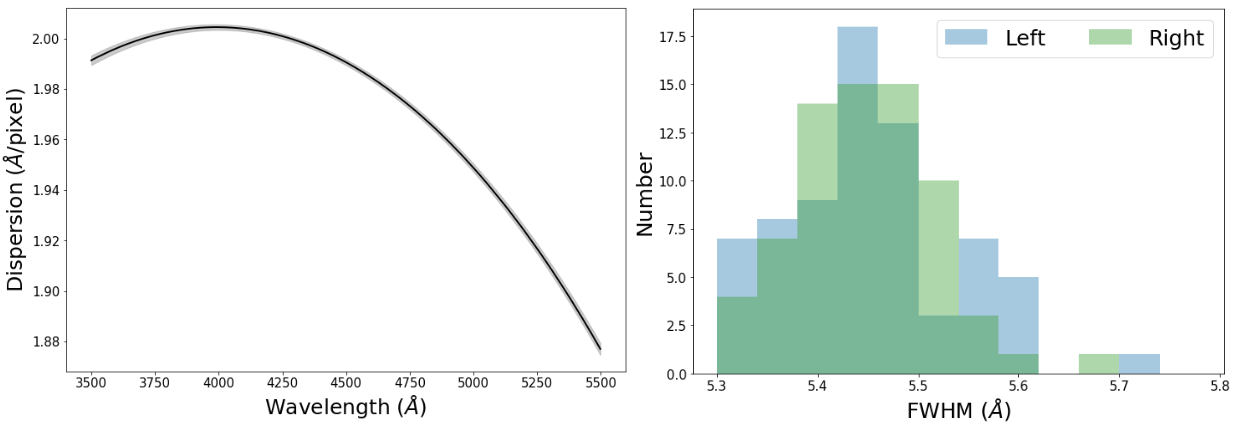}
\caption{
\label{dispersion} \footnotesize
VIRUS dispersion. Left panel: Dispersion relation for 141 spectrograph channels. Dispersion is expressed as \AA\ per binned pixel. The median relation and 1-$\sigma$ variation are shown as a black line and grey shading, respectively. The median dispersion is 1.99 \AA\ per binned pixel.
Right panel: The variation in median resolution for the channels shown in Figure~\ref{resolution} converted from pixels to \AA. 
}
\end{figure}

Instrumental resolution is very uniform from channel to channel and stable over time. 
Figure~\ref{resolution} shows the resolution over the CCDs of the Left and Right spectrograph channels. The values are color-coded as unbinned pixel FWHM  and calculated from the median over each channel. The right panel in Fig.~\ref{resolution} shows the distribution over the HET focal surface of median resolution by channel, for deployed spectrograph units. The resolution is expressed in unbinned pixels FWHM, though in practice the spectral dimension is usually binned by a factor of 2. In the right panel, the mapping of one IFU to two spectral channels is evident with small variations in spectral resolution on the two halves of the IFU. 
Figure~\ref{dispersion} shows the uniformity of the dispersion relation over the same channels (left panel) and shows histograms of the range of spectral resolution exhibited by each channel in \AA\ FWHM (right panel). These analyses show very consistent performance from channel to channel as expected from the flow-down of requirements to component specifications. 

Spectrograph resolution and long-term stability of the spectrographs can be monitored with internal calibration sources and through observations of extended emission-line regions. 
The high stability required for the position of the spectra on the CCDs of $<$0.5 (unbinned) pixels over the 5~$^\circ$C temperature swing typical of an observing night (\S\ref{sec: spectrographs}) has been met.
Analysis of 61 channels shows average (maximum) shifts of the spectra of 0.03 (0.18) pixels in the spatial dimension and 0.09 (0.23) pixels in the spectral dimension, for a 5~$^\circ$C temperature change, where the factor of 2 binning in the spectral dimension has been accounted for. 
The spectral dimension shows slightly greater variation,
corresponding to $\delta \lambda \sim$0.1~\AA\ on average,
but the total shifts even in the worst case are well below the requirement. 
As a result, calibrations can be applied over a whole night. 

\subsection{IFU Fiber Cable Performance} \label{subsec:ifuperf}
%\subsection{\hi IFU Fiber Cable Performance} \label{subsec:ifuperf}

Deployment of IFUs is discussed in \citet{vattiat18} and \citet{spencer18}. 
Performance of the IFUs is discussed in \citet{kelz21}.
During installation, care is taken to ensure that axial twists are not introduced, since experience with the Mitchell Spectrograph prototype revealed a failure mode where repeated twists produced stress in the corner fibers of the IFU input, resulting in severe FRD for those fibers. Once installed, the IFUs hang in simple catenaries between strain relief points on the PFIP and on the VIRUS enclosures. Motion of the tracker simply changes the distance between the strain relief points.
Extensive prototype testing \citep{murphy12} involving motions equivalent to 10 years operation revealed only an initial relaxing of the conduit, and lengthening relative to the fibers, as a concern after installation. 
The tailpieces of the IFUs, where the fiber branches into the two slits, have a clamp to allow adjustment of the axial position of the conduit relative to the fibers to remove any tension that might appear (Fig.~\ref{ifuprod}c). After an initial examination and adjustment of any that exhibit tight fibers, subsequent checks have not revealed significant signs of the fibers translating, axially. Checks are made on approximately a one year cadence at the same time as pixelflat calibrations are obtained. The IFUs are extremely stable, physically and in their optical properties such as transmission.
The IFU specification allows for an average of 1 broken fiber (2.2\%) and a maximum of 4, per IFU. In practice, broken fibers are rare, with 37 (0.1\%) in total in the deployed array.

\subsection{VIRUS Throughput and Sensitivity} \label{subsec:Vthuput}

The throughput requirement of the system of VIRUS units and the HET was calculated
during the Preliminary Design Review phase of HETDEX \citep{hil08b} and the baseline was established in 2010.
Sensitivities and predicted number of detected emission-line objects were derived
and compared to the project's science requirements. Typical observing conditions were included in the simulations. This analysis established the required number of fibers and hence the number of VIRUS units needed (\S\ref{sec:design}).
The overall throughput was apportioned at the individual component level, in order to establish requirements for the minimum performance of each component along with mean performance requirements for production sets of components. This approach allowed manufacturers some leeway for the inevitable range of component performance, without having to reject many components with the resulting increase in cost and project schedule. 

\begin{figure}[!ht]
\epsscale{1.18}
%\plotone{throughput_new.pdf}
\plotone{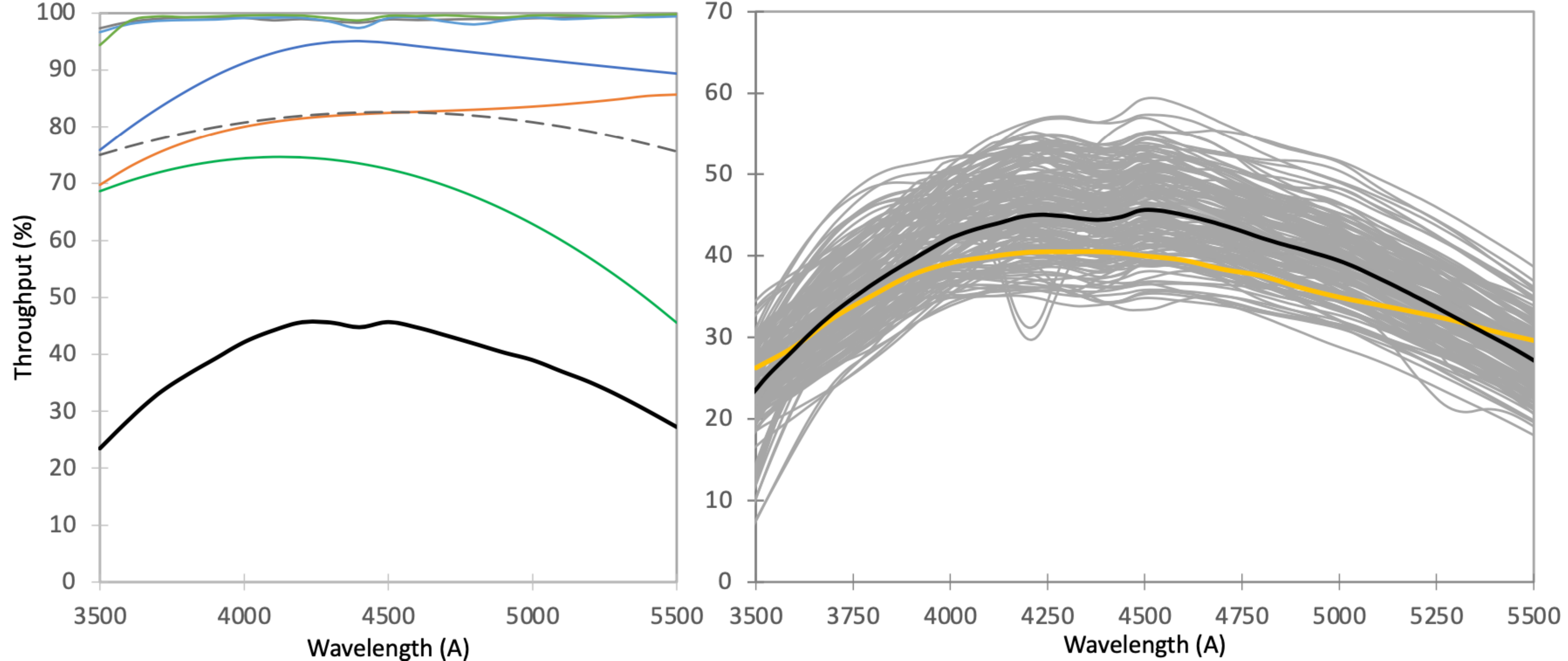}
\caption{
\label{throughput} \footnotesize
VIRUS component efficiencies and spectrograph channel throughputs. 
\edit1{Left panel: Median performance of each of the components of a VIRUS spectrograph channel as a function of wavelength, calculated from data delivered by vendors. The three curves with the highest throughput are the reflective coatings on the mirrors, the blue solid curve is the CCDs, the orange solid curve is the IFU fibers, and the green solid curve is the gratings. The grey dashed curve is the spectrograph obstruction of the detector package in the camera in relation to the central obstruction of the pupil from the telescope as transmitted through the fibers. 
This curve assumes good focal ratio degradation (FRD) performance so the central obstruction remains dark within the spectrograph and best throughput is achieved at camera field center. 
The lowest black line shows the combination of these component medians and the obstruction.}
Right panel: Predicted range of VIRUS unit throughputs without the telescope and atmosphere.
The grey curves present 156 realizations of the spectrograph channel throughput made by
multiplying randomly-selected components, along with the same model for the spectrograph obstruction presented in the left panel.
The black line is the mean of those 156 channels and can be compared to the black line in the left panel.
This ground-up average is compared to the prediction from 2010 on which the VIRUS specifications were based, shown in orange. That prediction incorporates the same spectrograph obstruction model, so the higher throughput in the middle of the wavelength range reflects the component performance being better than specifications, except at the extreme ends of the wavelength coverage.
The broad dip around 4300 - 4500 \AA\ is a feature in the reflectivity curves of the dielectric mirror coatings that was not taken into account in the prediction.
}
\end{figure}

\edit1{The left panel of Figure~\ref{throughput} presents the median efficiency (throughput) of each of the components of VIRUS. It also shows the obstruction model that accounts for the CCD package at the prime focus in the camera \citep{hil18a}. If light is transmitted through the fibers with little focal ratio degradation (FRD), as intended, then the dark central obstruction of the telescope pupil will be preserved in the far-field image at the output of the fibers. There is azimuthal scrambling of the light as it is transmitted through the fiber but ideally very little radial scrambling, which is the definition of good FRD performance. In that case, on axis at field center in the spectrograph camera, the obstruction of the detector package largely coincides with the pupil central obstruction, which is dark, and there is less light loss than off-axis in the camera where the detector package and central obstruction are not aligned. Hence, the shape of the obstruction curve shown in Figure~\ref{throughput} peaks at field (or wavelength) center. Poor FRD would depress the obstruction curve at the center due to light being scattered radially into the central obstruction of the far-field light distribution during transmission through the fibers \citep{murphy08,murphy12}.
}

The components that most affect the overall performance are the IFU fiber transmission, grating, and CCD (Fig.~\ref{throughput}). \edit1{\citet{indahl18} present the mean performance and variations of each component.}
The gratings show some blaze variation that trades efficiency at 5500~\AA~ with efficiency at 3500~\AA~\citep{chonis14,indahl18,indahl21}. This was quite well controlled and improved through production, but amounts to a variation of $\sim$10\% of the mean efficiency at 3500 \AA~ and $\sim$26\% at 5500 \AA~(2-$\sigma$). The effect on sensitivity of this variation is more pronounced at the shorter wavelengths where the overall throughput is dropping off. 
The reflectivity specifications for the multi-layer dielectric reflectors \edit1{(collimator, fold-flat, and camera mirrors)} were sufficiently stringent that they do not significantly contribute to the dispersion in properties, except for isolated dips in efficiency as seen in Figure~\ref{throughput}. A small number of coating batches had to be replaced as they were out of specification.
 
The VIRUS units incorporate 156 spectral channels that are individual realizations of the same spectrograph with the varying performance of the individual components. Mirrors, gratings, IFUs and CCDs have measured efficiencies that can be combined to compare to the on-sky throughputs of deployed spectrograph channels. \cite{indahl18, indahl21}  discuss the range in properties of the VIRUS components and examine multiple realizations of the spectrograph to explore the expected range of throughputs for the spectral channels. 
%Individual components (mirrors, gratings, IFUs and CCDs) have measured efficiencies that can be combined to compare to the on-sky throughputs of spectrograph channels. 

The right panel of Figure~\ref{throughput} presents the results of combining the efficiencies of randomly-selected manufactured components to create
156 random realizations of the spectrograph channel throughput.
The 2-$\sigma$ spread in throughput is approximately~$\pm$25\% except at the ends of the wavelength range, where it rises to $\pm$30\%. This is in line with expectations for the specifications of the components.
Figure~\ref{throughput} also compares the average of these simulated spectrographs
to the model of VIRUS throughput adopted in 2010. The model and the
derived throughput from simulated components both include the field- and wavelength-dependent correction to account for the obstruction of the detector package in the Schmidt camera of VIRUS, as discussed above \citep{hil18a}. 

The ends of the wavelength range
presented the greatest challenge for meeting throughput requirements, due to the fiber
transmission, CCD quantum efficiency, and grating blaze. 
This fact is reflected in the higher
throughput performance than required in the middle of the wavelength range
and the slight shortfall at the extreme ends. 
Overall, this comparison indicates that the predictions of component performance were realistic and the specifications supplied to the manufacturers on a batch basis could, on average, be met or exceeded over most of the bandpass.

\begin{figure}[!ht]
\epsscale{1.18}
%\plotone{onsky.pdf}
\plotone{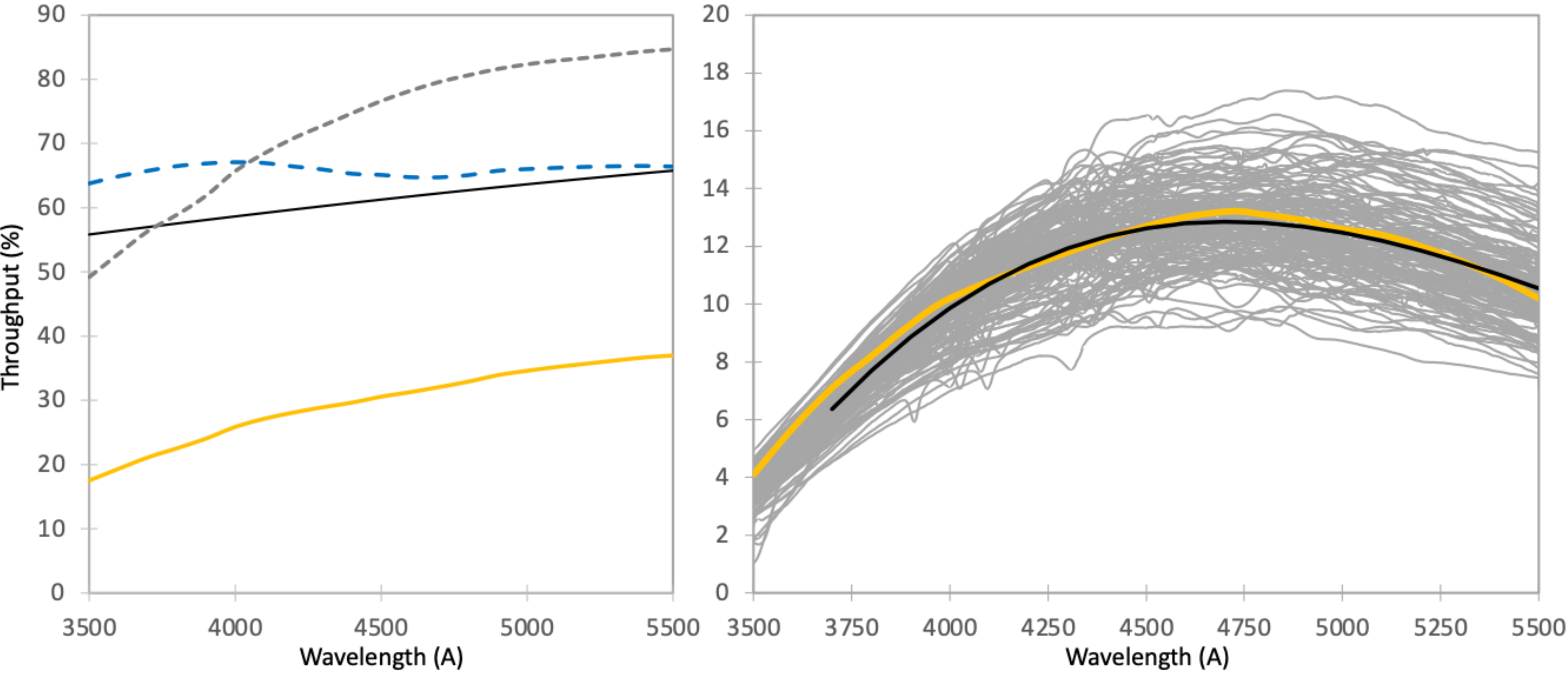}
\caption{
\label{onsky} \footnotesize
On-sky throughput of VIRUS plus HET and atmosphere, compared to expectations.
\edit1{Left panel: Components of the HET throughput model that transforms the spectrograph throughputs in Figure~\ref{throughput} to the on-sky prediction. 
The upper grey dashed curve is the atmospheric transmission for McDonald Observatory at 1.25 airmasses (the HET is a fixed elevation telescope).
The black solid curve is the prediction for the primary mirror segments that have bare aluminum coatings along with a model for the coating reflectivity degradation as a function of wavelength. The degradation model is based on limited measurements made against freshly coated segments using on-sky star observations. The degradation is shown for 20 months.
The dashed blue curve is the expected reflectivity plus on-axis obstruction for the mirrors of the wide field corrector, based on witness samples from the mirror coating deposition. 
The orange line is the combination of these components and represents the best estimate of the HET on-axis throughput at center track.}
Right panel: The orange curve shows the prediction for the average VIRUS throughput based on the 2010 model shown in Figure~\ref{throughput} combined with the model of the HET and atmosphere \edit1{presented in the left panel}.
The black curve is the average measured on-sky throughput for 135 VIRUS channels corrected
for the field vignetting of the WFC at track center by comparing to the LRS2-B throughput under the best observing conditions. LRS2-B coverage is above 3700 \AA.
Grey curves show the individual throughputs of 135 channels, measured on sky and normalized to LRS2-B in the same way, as described in the text. 
Throughput is defined as the fraction of photons incident upon an unobstructed 10 m diameter aperture above the atmosphere that result in detected electrons on the CCD. 
}
\end{figure}

A comparison can be made between these predictions and the measured on-sky performance by combining them with a model of the HET. The HET model shown in the left panel of Figure~\ref{onsky} includes the WFC obstructions and primary mirror illumination at field center with the tracker on axis, and accounts for
those losses relative to a 10~m unobstructed aperture above the atmosphere with an atmospheric extinction model at an airmass of~1.25, which is the mean value for a HET track. 
Mirror reflectivities measured for the WFC are included and the primary mirror segment reflectivity includes a wavelength-dependent degradation to account for a mean segment coating age of 20 months. 
\edit1{The primary mirror degradation model is based on limited on-sky measurements made with the original HET of stars with the primary mirror segments destacked so as to produce discrete star images for each segment. Comparison of brightness between freshly coated segments and those of varying ages results in mean degradation coefficients of between 1 and 2\% loss per month over the bandpass of VIRUS, increasing towards shorter wavelengths. These coefficients are quite uncertain and the primary mirror reflectivity is the least well understood component in this prediction of HET throughput.
}

The measured on-sky VIRUS throughput shown in the right panel of Figure~\ref{onsky} was bootstrapped from detailed measurements of standard stars made with the LRS2 instrument, which is well calibrated and stable. Spectrophotometric standards are observed every night with LRS2, and the most pristine conditions were selected to measure the throughput of the system for 3700~-~10500~\AA.
The~98\% fill-factor lenslet-coupled IFUs of LRS2, with $0 \farcs 6$ spatial elements, allow essentially
all the photons from a star point-spread-function (PSF) to be recorded. 
It is considerably more challenging to account for a star’s PSF with VIRUS data
due to the $1 \farcs 5$ fiber size and the need to acquire three dithered exposures to fill in the fiber pattern
(\S\ref{sec:virus}).
A low-order polynomial fit to the VIRUS throughput curve was obtained by comparing simultaneous observations of the sky between LRS2-B and VIRUS units, accounting for the different spatial element areas and the field illumination pattern of the HET. Wavelength regions with structure in the throughput of both instruments were avoided in the normalization.  The VIRUS throughput is the average of deployed units,
corrected to the center field, center track. This measure removes the field illumination pattern caused by vignetting in the WFC that has a maximum loss of 10\% for the edge IFUs at 9\arcmin~ field radius, with the tracker centered. Figure~\ref{onsky} demonstrates good agreement in the shape and amplitude of the system response between prediction and observation.
There is the potential for some inaccuracy in the model of the HET mirror reflectivities, particularly the primary mirror segments that are recoated on about a year cycle time\footnote{During the COVID-19 pandemic telescope access restrictions prevented mirror segment coating, and the desired 12 month cycle time has extended to more than 20 months.}, or due to fiber FRD scattering some light into the central obstruction at the pupil of the spectrograph channels. However, the solid agreement in the shape of the system response suggests that the VIRUS optics obstruction model, assuming good fiber FRD performance, is on average valid.

Figure~\ref{onsky} also presents the measured variation in on-sky throughput for deployed spectrograph channels,
corrected to field center for the HET field illumination pattern. The curves are generated from twilight sky observations to produce the normalized relative throughput for each channel compared to the average;
these normalized curves are multiplied by the average throughput derived from comparison with LRS2-B to generate the curves presented for each spectrograph channel. 
The LRS2-B comparison is for wavelengths above 3700 \AA. Below this wavelength the curve is extrapolated and hence may be subject to some error.
There is some shortfall in measured throughput below 4000~\AA, compared to the model, but overall the on-sky performance of the spectrograph channels is in good agreement with expectations with a shortfall of a factor of about 0.9 at the shortest wavelengths. 
Considerable focus in design and production was applied to maximize the throughput at 3500~\AA, in spite of the combined challenges posed by fiber length, CCD efficiency, and atmospheric transmission, since the majority of LAEs will be detected between 3500 and 4500~\AA, due to their distance modulus and luminosity function \citep{hil08b, geb21}.

HETDEX observations are calibrated directly against field stars in the field of view, and provide independent confirmation of the analysis presented here \citep{geb21}. 
The definition of throughput adopted here is different from that in the HETDEX data reduction pipeline where the number of detected electrons is compared to photons incident on a 50 m$^2$ area, rather than a 10 m diameter unobstructed aperture, and the field illumination of each specific IFU is included. The best average throughputs at 4940 \AA, measured in HETDEX data, are 18\% with that definition. Correcting the average throughput at the same wavelength in Figure~\ref{onsky} for the difference in effective aperture and the average illumination for the IFUs, which is 0.94, yields 18.6\% under the same definition.  More typical observing conditions yield observations with throughputs of 15-16\% under the HETDEX definition, but the best values are in good agreement with the calibration presented here, considering that the two methods of measuring the throughput are quite different and independent. The shape of the average system response also agrees well with that derived from the calibration of HETDEX observations over the full wavelength range, which further indicates that the cross-calibration between VIRUS and LRS2-B is well-defined and stable.

\begin{figure}[!ht]
\epsscale{1.15}
\plotone{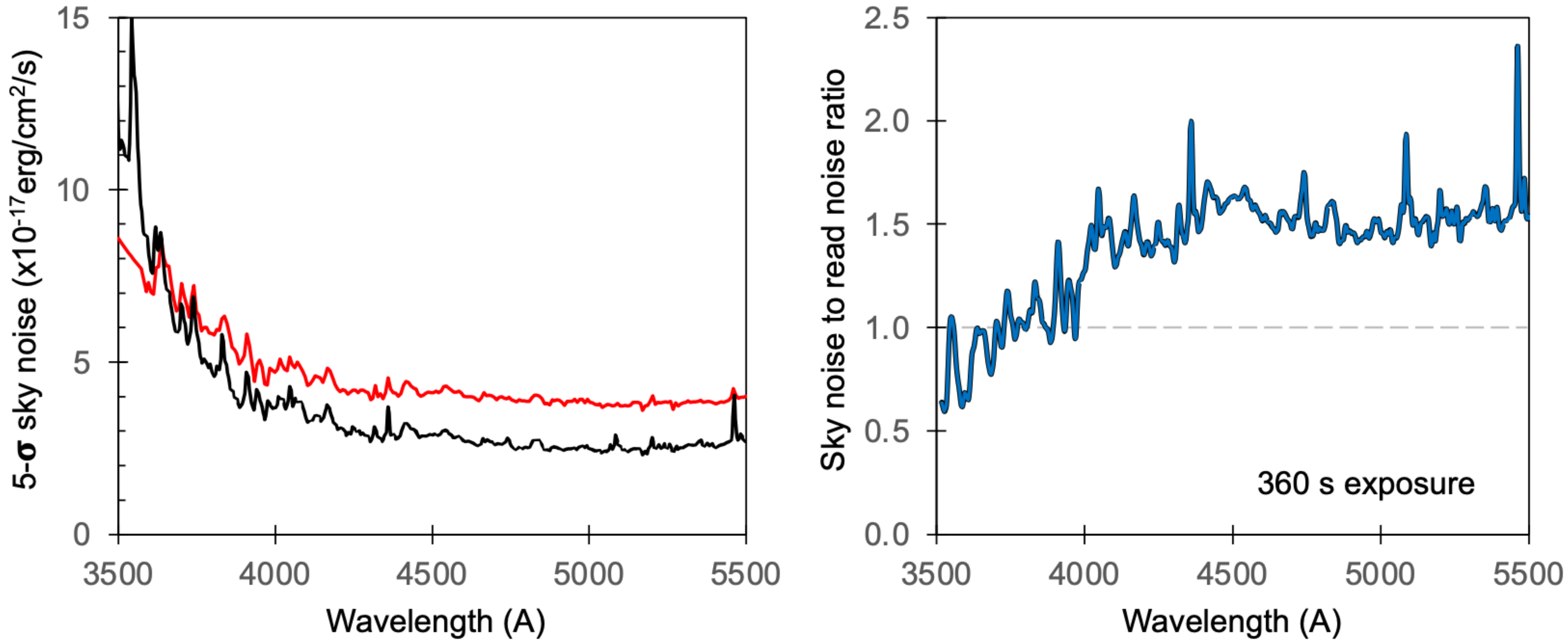}
\caption{
\label{sensitivity} \footnotesize
Left panel: Comparison between predicted 5-$\sigma$ \edit1{sky noise} flux limit (red),
which is five times the noise per spectral resolution element, to that measured from the noise in blank sky resolution elements (black).
The limit is per fiber, for three 360 second dithered exposures that comprise a HETDEX observation.
Right panel: Ratio of sky noise to read noise per resolution element in a 360~second exposure as a function of wavelength. The horizontal dashed line indicates a ratio of unity.
The ratio is measured in dark time for the mean read noise of 2.95 electrons. The requirement was to have read noise and sky noise equivalent in this exposure time, but below about 3700 \AA~ there is a shortfall. 
}
\end{figure}

Another approach to evaluate the performance is to compare detection sensitivities with predictions.
The noise in blank sky spectra is well characterized, since the vast majority of fibers in a given observation
only contain sky signal. 
\edit1{The most direct measure of sensitivity to compare with predictions is the noise level per resolution element in observations of the dark sky. For objects near the detection limit, the sky noise level determines the sensitivity.}
Figure~\ref{sensitivity} (left panel) presents the sensitivity of VIRUS, expressed as the 5-$\sigma$ noise in the sky per fiber, per resolution element, compared to an estimate derived from the throughput, sky, and atmosphere models developed in 2010. 
The sky model used in the prediction was derived from observations with the Mitchell Spectrograph for the HETDEX Pilot Survey \citep{adams11}. Some terrestrial sky features such as HgI$\lambda$5461 are noticeably stronger over the decade between the Pilot Survey and HETDEX observations, 
\edit1{but the overall sky brightness level is the same.}
%The predicted flux limit is shown for an unresolved object (both spatially and spectrally) centered on a fiber. Models of the flux limit with the object located at the vertex of three fibers in the dithered observing pattern, or centered on 7 fibers differ by only a few percent in the models. The assumed image quality was 1$\farcs$5 FWHM. 
The single fiber flux limit based on the sky is best for direct comparison between observations and the model, without having to account for variable image quality and transparency 
\edit1{that will determine the flux limit for observed sources.} 
%\added{This comparison also avoids the complexity of accounting for differential atmospheric refraction as discussed in section XX.} 
Figure~\ref{sensitivity} indicates that by this measure the system is delivering similar or better sensitivity than predicted at wavelengths longer than $\sim$3700~\AA, and poorer sensitivity for shorter wavelengths.
%The detection algorithm and limits for HETDEX are discussed in detail in \citet{geb21}.
%\edit1{The object line flux sensitivity that corresponds to the noise level in Figure~\ref{sensitivity} %will depend on how many fibers are included in the detection. Models of the flux limit with the object %located at the vertex of three fibers in the dithered observing pattern, or centered on 7 fibers differ %by only a few percent in the models. The faintest emission lines among the detections have total line %flux of \hbox{of $\sim 5 \times 10^{-17}$ erg cm$^{-2}$ s$^{-1}$,}}

Read noise measured from the overscan data of bias frames of deployed CCDs, with four amplifiers per unit, has an average value of 2.95~$\pm$~0.34 electrons, in line with specifications derived from science requirements.
The right panel of Figure~\ref{sensitivity} presents the ratio of sky noise to read noise per resolution element for a typical spectrograph channel for the short 360 second HETDEX exposure time. It is the average from all fibers in 71 VIRUS units, active in mid 2020. 
The requirement is that these two noise sources be equivalent at the shortest wavelength for a 360 second exposure and higher for longer wavelengths; this is typically borne out on sky, but there is a shortfall at wavelengths below about 3700~\AA, and this accounts for the shortfall in sensitivity versus expectations at the shortest wavelengths. Longer exposures of 1000 seconds or more, for other observing programs with VIRUS, will not suffer from this limitation, since sky noise then equals or dominates read noise at all wavelengths.
Nonetheless, while the read noise does contribute more then intended at the shortest wavelengths and this coincides with the poorer sensitivity at these wavelengths, overall the sensitivity of VIRUS appears in line with expectations except for the bluest wavelengths.
%The 5-$\sigma$ noise per resolution element measured from the sky data 

%The noise level shown in Figure~\ref{sensitivity} indicates a line flux detection threshold
%\hbox{of $\sim 4 \times 10^{-17}$ erg cm$^{-2}$ s$^{-1}$,} in the fiber aperture around 4500 \AA, which is in line with predictions. 
A detailed discussion of throughputs, detection, sensitivity and completeness limits delivered for the HETDEX survey is presented in \citet{geb21}. 
\edit1{The object line-flux sensitivity that corresponds to the noise level in Figure~\ref{sensitivity} will depend on how many fibers are included in the detection.} Models of the flux limit with the object located at the vertex of three fibers in the dithered observing pattern, or centered on 7 fibers differ by only a few percent, \edit1{so the sensitivity will not depend significantly on object position in the IFU. 
%Typical detections include data from 3 fibers.
In HETDEX, objects are extracted from fibers within a 3\arcsec~ radius with weighting by the point spread function \citep{geb21}.
}
The noise measured in sky-subtracted, extracted spectra of emission line detections is consistent with the sensitivity levels indicated in Figure~\ref{sensitivity}. 
\edit1{The faintest emission lines among the detections under good conditions have total line flux of 
\hbox{$\sim 5 \times 10^{-17}$ erg cm$^{-2}$ s$^{-1}$}.}

\section{Observing with the HETDEX Hardware}\label{sec:WFUperformance}

The HET re-entered full queue-scheduled science operations in 2016 December \citep{hil18b}.
All the metrology subsystems described in \S\ref{subsec:pfip} are working as intended (Table~\ref{tab-wfupgrade}) and the control loops of the tracker are extremely robust due to the tracker hardware design \citep{good14b,good18,lee18b}.
In this section, details of the observing framework and performance
of the HETDEX instrument system are discussed.

\subsection{Target Preparation and Acquisition}\label{subsec:target}

The new telescope control system \S\ref{subsec:tcs}  is scriptable, so almost fully automated observation is enabled.
The HET often observes a diverse set of science programs on any night, utilizing all the available instrument modes. Setups can be blind or on targets visible on the acquisition camera. Night operation maintains 
manual setup capability to accommodate this diversity.

The target submission language specifies instrument and configuration, acceptable observing conditions, and target properties. Constraints on whether the observation is obtained with the telescope azimuth set for an east or a west track, along with observation groupings and sequences, can be specified as needed. 
Once accepted, target coordinates and azimuth are processed through a target setup utility called {\it Shuffle}, which selects guide and wavefront sensor stars within the outer-field annulus.

\begin{figure}
\epsscale{1.0}
\plotone{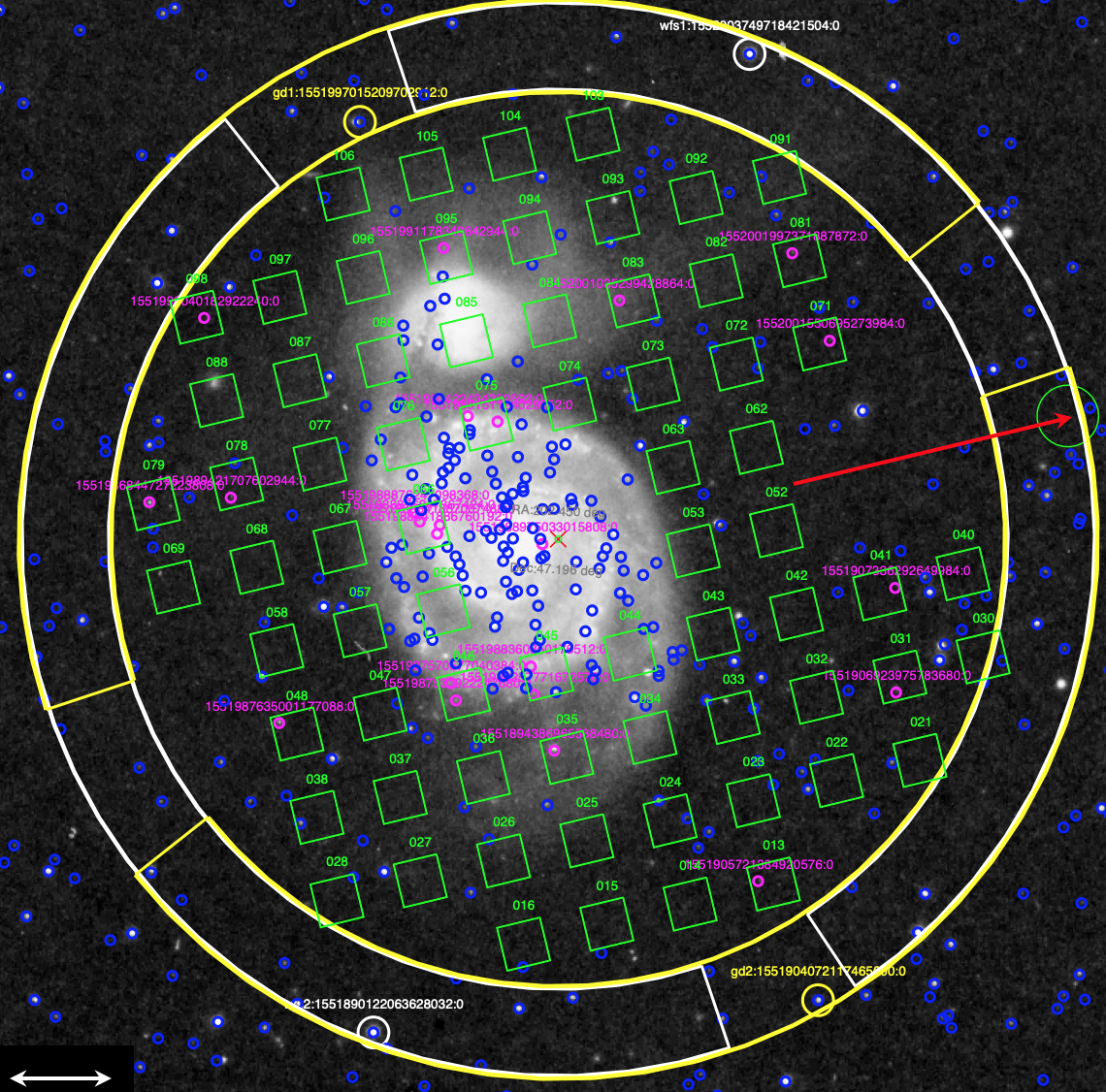}
\caption{
\label{shuffleM51} \footnotesize
Example setup of the HET using the “Shuffle” software tool on M51 to illustrate the size of the 22 arcminute diameter HET field of view and the footprint of the VIRUS IFUs. The scale indicated by the arrows in the lower left corner is 2.0 arcminutes. The setup is for a west track at telescope azimuth 310$^\circ$. The outer annulus extends from 9-11 arcminutes field radius and has the patrol regions of the two guideprobes and 2 wavefront sensors indicated, along with the identifications and coordinates of the guide and wavefront stars chosen by the software (in yellow and white, respectively). 
The available VIRUS and LRS2 IFUs and their input head mount plate seat locations are indicated as green squares (which are the correct size for VIRUS, but just indicate the centers for LRS2 in input head mount plate seat locations 056 and 066).
Blue circles indicate all objects in the reference catalog (PanSTARRS or Gaia) and magenta indicates those that fall on IFUs. A small red cross indicates the coordinate of the target or field center. The parallactic angle direction is indicated by the red arrow, which is the direction of the zenith for track-center for this particular observation, and sets the orientation of the IFU pattern on the sky. This direction also corresponds to the Y-axis of the HET tracker.
}
\end{figure}

Figure~\ref{shuffleM51} presents a Shuffle setup on M51, as an example that illustrates the sectors available to each guideprobe and operational wavefront sensor probe in the outer 2 arcminute annulus. Each probe can range through 180$^\circ$, and while their ranges overlap, they cannot physically collide \citep{vattiat14}.
Shuffle can adjust the pointing center if desired to improve guidestar selection.
Other options allow placement of a target on a specific IFU with offsets from the IFU center. The output of Shuffle is a file capturing the information to set up the HET on the target, including photometry of the guidestars which is used to reference the calculation of the transparency during the observation.
Shuffle also generates a synthetic acquisition camera image for the telescope operator that identifies the pixel positions of stars for precise blind setup, if required. 
Shuffle is generally run well prior to an observation, but executes in about 20 seconds.
Shuffle coordinates of the instruments and guideprobes are based on a coordinate system that is accurately tied to sky through observations of star fields with VIRUS.

The setup of VIRUS for HETDEX observations is based on the guide probe positions, since the accurate coordinates of the observation are derived from the data during processing \citep{geb21}. There are two usable guidestars more then 90\% of the time. One star is centered on a guide probe and becomes primary for guiding, while both guiders provide metrology streams on other parameters of the observing conditions.
Such blind setups have an accuracy of $1 \farcs 5$ rms, comparable to the separation of the fibers in the VIRUS IFUs, and more than adequate for HETDEX and most other VIRUS observations.
The primary guidestar fiducial is moved to provide the small offsets for the dithers needed to fill in the gaps between fibers in the IFUs. 

The ability to orchestrate the telescope control system has also led to improvements in efficiency. The observing conditions decision tool is a Python state machine that monitors the event stream from the metrology system to decide whether conditions are suitable for HETDEX observing. 
The observing conditions decision tool sets the exposure time for the next observation using the transparency and image quality  from the current exposure as inputs, along with the primary mirror illumination for the next observation. 
Exposure times are allowed to increase up to a factor of two, to compensate for poor conditions.   No compensation is made for better than median conditions.
When conditions are deemed acceptable and the moon is down, the observing conditions decision tool selects the best field and (if allowed) will automatically slew the telescope and perform the observation with only a setup and guiding confirmation and focus check needed from the Telescope Operator. This orchestration has improved operational efficiency for HETDEX observing with VIRUS and is accomplished within a framework that will allow other instruments to interact with the telescope as the system is developed further. 

Observing conditions metrology from the guideprobes informs target selection by the night staff or by the observing conditions decision tool. Transparency and sky background brightness are derived from photometry performed on the star images with knowledge of the guidestar brightness passed from Shuffle; image quality is measured from star profile fitting. This metrology stream is logged and displayed in real time for decision-making by the night staff. 
An interesting observing mode enabled by the common focal surface and shutter between the instruments is VIRUS operating in parallel with the primary observation with LRS2 or HPF (and HRS, when delivered). Such observations are typically not dithered, but are often of much longer exposure time than the HETDEX observations, so have significant value for many projects. This mode can cover large areas with blind spectroscopy and will have interesting utility, especially for stars and other continuum objects. 

\vfill
\subsection{Pointing, Tracking and Guiding}\label{subsec:pointingperf}

\begin{figure}
\epsscale{1.0}
\plotone{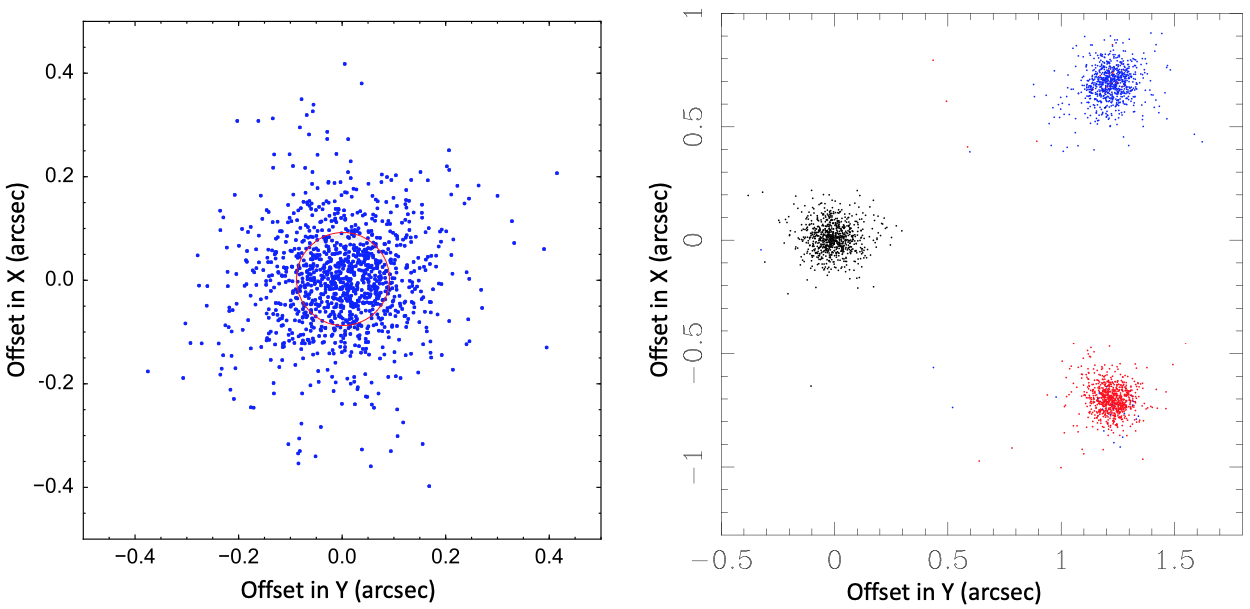}
\caption{
\label{guideoff} \footnotesize
HET WFU guiding and offsetting accuracy projected into tracker X and Y coordinates. Tracker Y is aligned with the parallactic angle.  Left panel:  example guiding residuals measured from star centroids for a 1.5 hour track, producing 0.09$\arcsec$ rms guiding, indicated by the red circle.  Right panel:  accuracy of dither pattern in tracker coordinates, under guider control, for 864 observations from 2018.  The pattern is an equilateral triangle of side $1 \farcs 46$ and the offsets are accurate to 0.06$\arcsec$ rms.
}
\end{figure}

The WFU has added detailed metrology on all degrees of freedom of HET positioning.  As an example of guiding performance, Figure~\ref{guideoff} shows on-sky residuals during a full 1.5-hour track in the north. The left panel presents the centroids of guide images demonstrating $< 0 \farcs 1$ rms guiding accuracy. Dithering to fill in the fiber pattern within the VIRUS IFUs (\S\ref{sec:virus}) is achieved through offsetting the guide fiducial on the primary guide probe\footnote{A dither mechanism was deployed as part of PFIP \citep{vattiat14,hil16c}; however, it proved difficult to set up and the tracker offsets are so precise that it was ultimately not needed. It has hence been disabled.}. This process achieves a precision in the dither pattern of $0 \farcs 06$ rms as illustrated in the right panel of Figure~\ref{guideoff}.

\begin{figure}
\epsscale{1.15}
\plotone{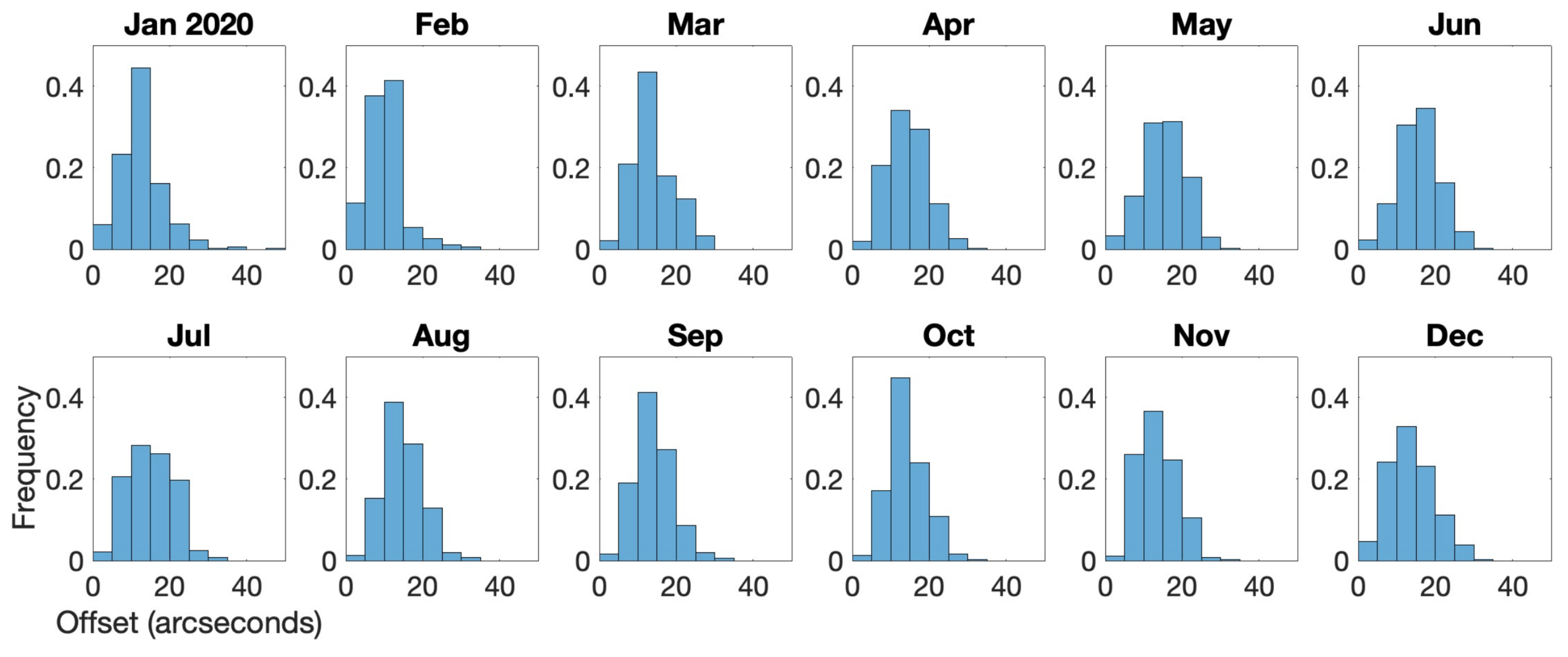}
\caption{
\label{pointstats} \footnotesize
Examples of performance metrics from the telescope control system database for 4237 target setups over 2020.  The panels show statistics by month of initial pointing corrections that indicate the magnitude of the pointing error.  All HET science observations are included, and all weather conditions, not just those for HETDEX.  
}
\end{figure}

Telescope pointing and tracking have been improved through application of a physical mount model (\S\ref{subsec:wfuintegrate}); improvements in pointing are reflected in improved tracking. A key requirement for efficient observing with VIRUS is the ability to point the telescope such that most observations start with the guidestars within the 22$\arcsec$ guide-probe fields, or sufficiently close to the edge to be seen in the guide image by the telescope operator. 
This goal is achieved most of the time, and allows setup to be achieved simply by centering the guidestar without the overhead needed to deploy the acquisition camera.  Figure~\ref{pointstats} shows initial pointing residuals by month over 2020. These data allow any degradation in pointing to be identified quickly, and indicate a stable mode for pointing accuracy of around 10-15$\arcsec$.
In fact the Telescope Operators find there are still zero-point offsets in pointing of around 10\arcsec\ that persist for long periods of time and contribute to the distributions in Figure~\ref{pointstats}. When these offsets are taken into account the pointing is 10\arcsec\ rms, which usually results in immediate acquisition of the guide stars.

\subsection{Field acquisition time}\label{subsec:fieldaq}

\begin{figure}
\epsscale{1.0}
\plotone{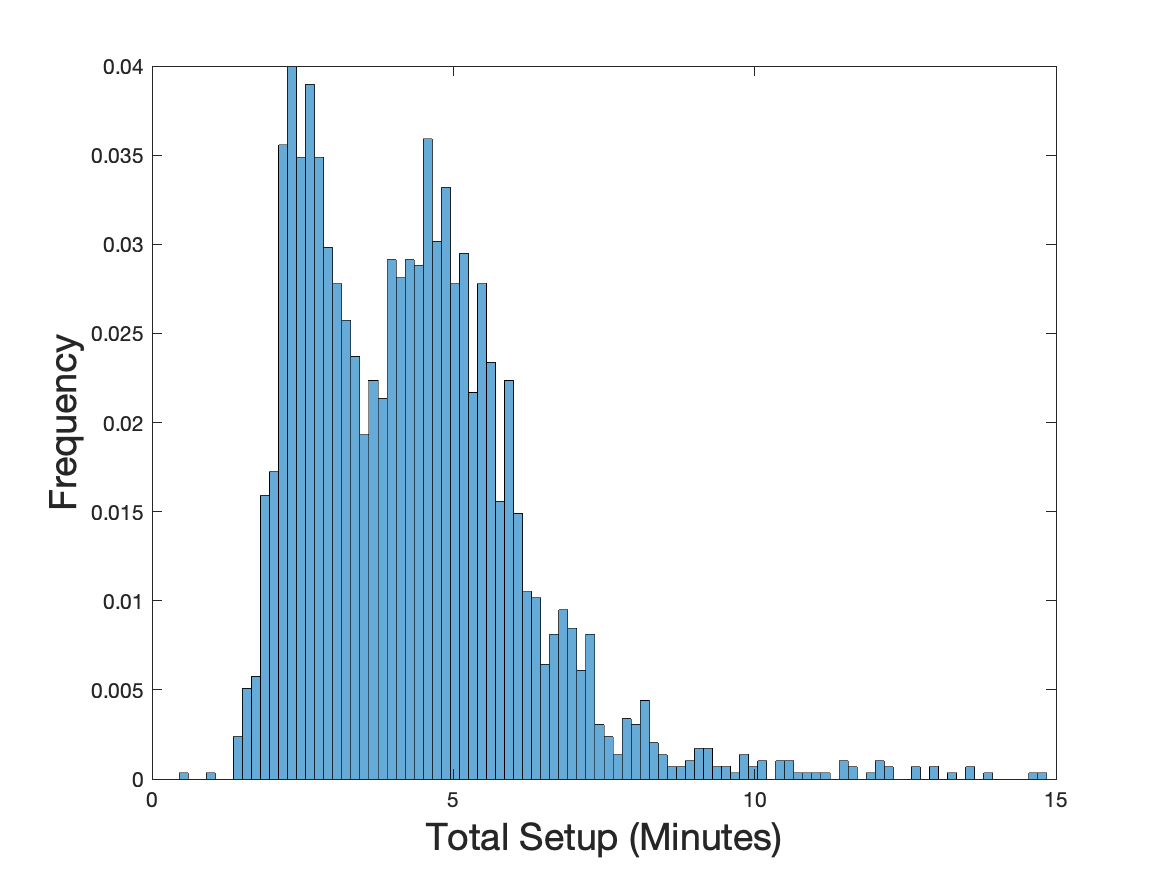}
\caption{
\label{setuptime} \footnotesize
Setup time in minutes (measured from the command to slew the tracker to being ready to open the shutter on the next observation) for all observations over the months of January to September 2020. The data on 2952 setups are not filtered by HETDEX observation, and show two peaks. The peak at around three minutes is associated with HETDEX observations, run with the observing conditions decision tool automation, while the second peak around five minutes is associated with LRS2 observations and the tail to longer times is mostly associated with more exacting setups for HPF. The few instances with setup times beyond 10 minutes are associated with instances where conditions were difficult or where errors occurred. 
}
\end{figure}

The telescope control system logging allows tagging and monitoring of the duration of all steps in the observational setup process, which has proven helpful in identifying inefficiencies and in driving down the setup time. The most important aspect of reducing setup time has been improvements in pointing. When the majority of setups start with the guidestars in the guide-probe field of view, the interactive part of the setup requires only 10-20 seconds; it is usually not necessary for the telescope operator to observe the field with the acquisition camera. Figure~\ref{setuptime} shows the distribution of setup time between observations (from command to go to next observation to being ready to open the shutter on that observation). These statistics include all observations, not just for HETDEX. The peak close to 200 seconds is dominated by HETDEX setups, while the tail is primarily due to setups for the other instruments.
Total setup time between observations (shutter close to shutter open on the next observation) includes all overheads, as charged to observing programs.
Median total setup times for accepted observations are 4 minutes for VIRUS, 6 minutes for LRS2, and 7 minutes for HPF.
These are a little longer than the setup times in Figure~\ref{setuptime} due to some additional overhead. About 20 seconds additional overhead is due to tracker shutdown and preparation book-keeping that happens after the shutter closes and readout of the previous exposure begins, and for HETDEX, and there is an additional 20 seconds due to executing an observing conditions decision tool optimization to choose the next target from the full database of survey observations. These overheads may be amenable to additional optimization.

While general observational overheads have been reduced significantly compared to the 10 minutes typical for setups on HET before the upgrade, and meet requirements, the goal of 1.5 minute setups has not been achieved for HETDEX observing, when no azimuth change is required. 
Detailed examination of the events stream from the telescope control system will lead to further incremental improvements, but setup time is now limited by the fundamental motion control of the tracker, with median move time of $\sim$90 seconds \citep{good18,rams18}, 
so further significant reductions below 4 minutes median total setup time are not expected. The consequence of this setup time for the three dithered 360 second exposures plus readout time that make up a HETDEX observation is a 7\% increase in the total HETDEX observing time over the goal, for the expected number of observations requiring Azimuth moves.

\subsection{Optical performance of WFU and Delivered Image Quality}\label{subsec:imagequal}

Operations have provided extensive data logging of the delivered images from the guide cameras (over a million images each year), allowing further diagnostic information to monitor the wave-front and optical image quality, and records of weather conditions. The wealth of data available from the metrology systems allow trends in image quality with temperature, wind speed and direction, and other quantities to be investigated. In particular the delivered image quality can be broken up into contributions from the corrector optics, the primary mirror, the dome environment and the column of atmosphere outside the dome, through analysis of data gathered from the wavefront sensors, the center of curvature tower instrumentation, and from other metrology subsystems.

Wave-front sensor tests of the optical system during commissioning confirmed that the WFC and the metrology systems deliver the required optical image quality over the full field of view \citep{lee16a}. 
Those measurements indicate a floor of $\sim$1$\arcsec$ FWHM in the best site seeing conditions, and $1 \farcs 3$ FWHM in $1 \farcs 0$ site seeing conditions, as expected from the design. The WFC image quality varies a small amount between field center and edge of field where the guide probes measure the image quality for each observation. In median seeing, this change amounts to only $0 \farcs 1$ larger images at field edge. 
Direct tests between the acquisition camera on axis and the guide probes at the edge of the field demonstrate image quality is quite constant over the 22 arcminute diameter field of view in typical conditions. 
These measurements indicate that the upgraded HET optical system is performing to specification.
The improved image quality of the WFC over the much enlarged field of view of the new HET still needs to be convolved with the performance of the primary mirror, any in-dome seeing component, and the atmospheric contributions to the delivered image quality. 

\edit1{The new hardware for the WFU and the support system for VIRUS were designed with integrated circulating glycol heat removal systems to extract excess heat from the dome environment that could impact dome seeing. The goal was to maintain surfaces at, or a few degrees below, ambient temperature. 
The VIRUS enclosures are well insulated and employ air circulation and glycol heat exchangers to control internal temperature \citep{prochaska14, spencer18}. 
%The internal temperature tracks ambient up to a maximum of 10 $^\circ$C to limit the total temperature range and support the management of the VIRUS CCD operating temperatures. If the dew point is above 10 $^\circ$C, the glycol temperature is adjusted to be one degree higher than the dew point. 
Heat removal from the tracker is via heat exchange in glycol jackets on all motors, along with insulation jackets \citep{zier12}. Heat generated by cameras and other equipment in the prime focus instrumentation payload is removed via glycol heat exchangers as part of the air circulation and filtration system \citep{vattiat12}. All external surfaces in the payload are carbon fiber laminated foam insulation panels to reduce heat conduction.
Measurements from temperature sensors deployed on the telescope and from a thermal camera have been used to evaluate the performance of the heat removal systems, by examining skin temperatures for subsystems within the dome environment. 
The VIRUS enclosure skin temperature is observed to closely track the ambient temperature while the enclosure steel frame is a few degrees colder. The tracker and payload skin temperatures track ambient or are 1-2 degrees below, due to radiation to the sky when the dome is open at night. The surfaces of the glycol supply lines are a few degrees below ambient. Examination of the thermal imaging does not reveal areas that are systematically warmer than the dome environment, so it is not expected that the hardware associated with the WFU and VIRUS will contribute significantly to any dome seeing component of the delivered image quality. 
}

\begin{figure}
\epsscale{0.8}
\plotone{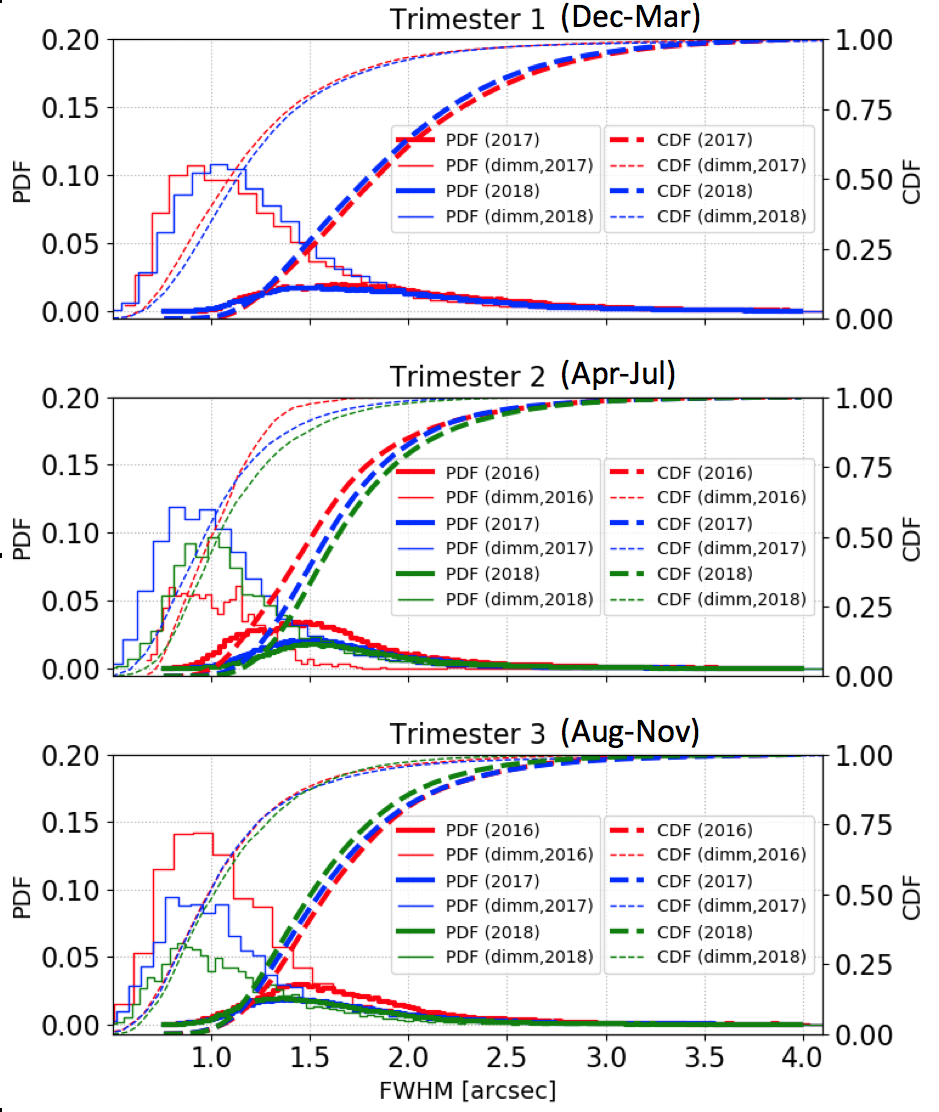}
\caption{
\label{imagequal} \footnotesize
HET delivered image quality measured from guide probes (field edge). Panels display seasonal variations in the distribution of \cite{moffat69} FWHM over the three trimesters (T1 = Dec-Mar; T2 = Apr-Jul; T3 = Aug-Nov). Both frequency and cumulative distributions are shown. The DIMM (differential image motion monitor) data representing the site seeing are also indicated in each panel.
}
\end{figure}

As with other ground-based telescopes, the delivered image point spread functions (PSFs) of HET are well fitted by \cite{moffat69} profiles\footnote{$I_r$ = $I_o$ [1+ (r$/$$\theta)$$^2$]$^{-\beta}$},
characterized in terms of the Moffat $\beta$ exponent (e.g. \citealt{truj01}). Smaller values of $\beta$ are associated with more pronounced wings. Atmospheric turbulence leads to $\beta$~=~4.765, a value approached in poorer seeing (e.g. \citealt{dey14}). 
The HET metrology database allows the PSF shape and FWHM to be characterized over millions of guider images and with the acquisition camera. This analysis reveals a tight correlation between $\beta$ and FWHM, with $\beta$~$\sim$~3.5 in $1 \farcs 0$  FWHM, and asymptoting to $\beta$~$\sim$~4.7 in the poorest conditions (FWHM~$\sim$~$4 \farcs 0$). Median FWHM values  from the guide probes at field-edge, and from free air differential image motion monitor (DIMM) over five years are summarized in Table~\ref{tab-summimage}.  
Weather conditions at McDonald Observatory vary seasonally, so each year's image data are divided into the observing trimesters covering December to March, April to July, and August to November.  
Guide probe FWHM values include all observations for each trimester and are derived from
Moffat fits to the radial profiles of the guidestars with $\beta$ a free parameter.
Figure~\ref{imagequal} displays the image quality in more detail with plots of probability distribution functions and cumulative distribution functions, by trimester.
Median image quality of the HET is improved, post-upgrade, but still delivers images larger than expected from the combination of DIMM site seeing and telescope optical performance. 
Investigations with high speed cameras and accelerometers have eliminated wind shake or jitter in tracking as primary sources of the added image quality component. 

\begin{figure}
\epsscale{1.2}
\plotone{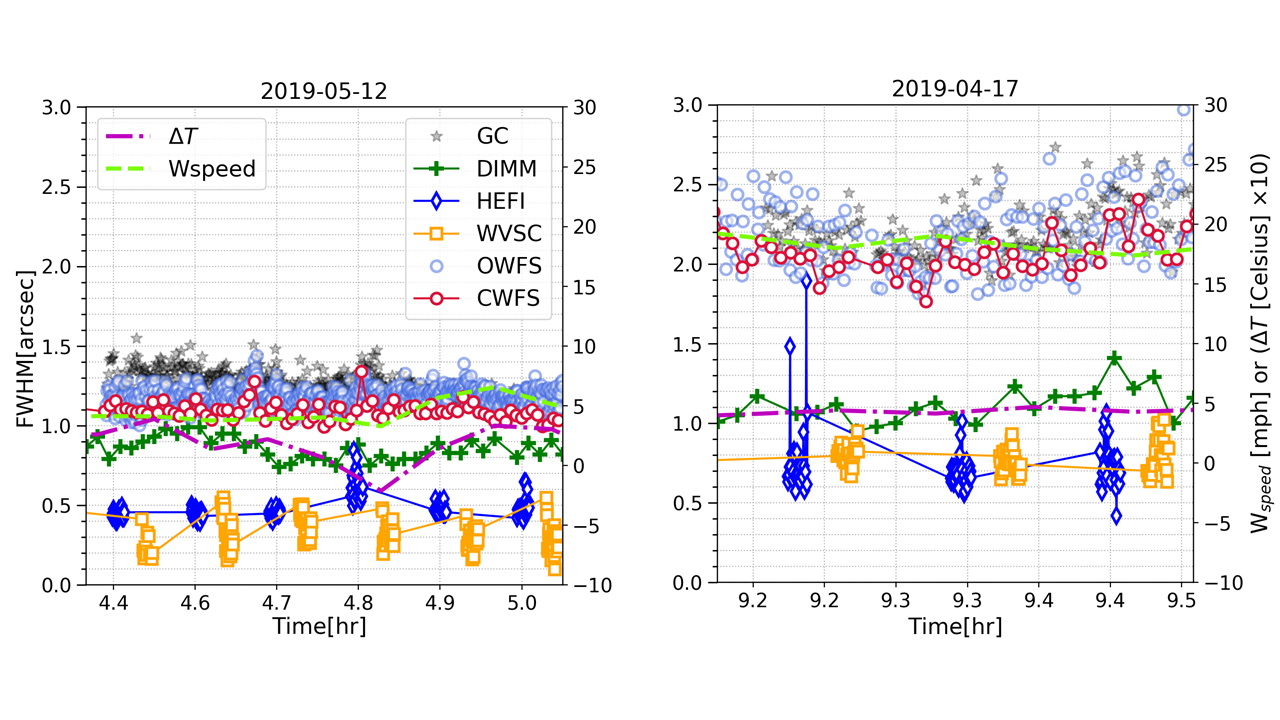}
\caption{
\label{iqanalysis} \footnotesize
Example metrology system analysis of HET delivered image quality for two tracks. Left shows a period of 65 minutes recorded on 2019-05-12 UT, during a period of good delivered image quality. Right shows 20 minutes recorded on 2019-04-17 UT, when the HET was delivering poor image quality. Time is UT. Data from the PFIP guide cameras (GC), operational wavefront sensors (OWFS), and calibration wavefront sensor (CWFS) are plotted from tracking a star on axis and 4 stars in the guide region at the periphery of the 22 arcminute diameter field of view. Simultaneous data from the Hartman Extra-focal Instrument (HEFI) and Wavescope (WVSC) located in the center of curvature tower and viewing the primary mirror are also presented.
Facility metrology on wind speed and temperature differential from the last primary mirror stack are also presented. See text for discussion.
}
\end{figure}

The ability to monitor the state of the primary mirror using instruments in the center of curvature tower, while simultaneously tracking stars and monitoring on-sky guider and wavefront sensor data,
presents an opportunity to go beyond the guideprobe observing metrology data in assessing contributions from different components of the HET image quality. 
The Hartman Extra-focal Instrument measures the primary mirror image quality, including the mirror stack plus dome seeing. The Wavescope measures these components on a small scale, without the stacking included.

Figure~\ref{iqanalysis} presents two examples of the engineering data that can be obtained from simultaneous on-sky metrology coupled with center of curvature instrumentation measurements of the primary mirror and DIMM measurements of free-air seeing. 
To accomplish this test, a star was placed on the calibration wavefront sensor at field center, while the guiders and operational wavefront sensors observed four stars. The calibration wavefront sensor provides image quality in small apertures, which is a measure of site seeing. 
When run at high speed, the calibration wavefront sensor acts as a multi-scale DIMM, including the full air column contributing to the HET image quality.
Simultaneously, the primary mirror was observed from the center of curvature tower with the Hartman Extra-focal Instrument. The in-focus image from the Extra-focal Instrument provides a measure of the contribution of the primary mirror stack plus segment image quality. 
Various components of the image quality can be separated using these different metrology systems.

Analysis of the figures of the primary mirror segments suggests that they can contribute an extra $0 \farcs 34$ in quadrature on-sky, beyond the stacking error, and this contribution can also be included.
The data in the left panel of Fig.~\ref{iqanalysis} were obtained during a period of good delivered image quality, with the DIMM showing site seeing of 0.8-$0 \farcs 9$. Wind speed was about 5 mph and the ambient temperature had been stable since primary mirror stacking. The primary mirror Hartman Extra-focal Instrument and Wavescope data indicate a stable contribution of $\sim 0 \farcs 5$. 
The calibration wavefront sensor image size indicates seeing of about 1.0-$1 \farcs 1$ through the column, including interior and exterior air, indicating only a small additional component to that seen by the DIMM, if any.
Combination of these components in quadrature accounts for the image size measured on the guide cameras of 1.2-$1 \farcs 4$. During this period HET image quality is behaving as intended, with only a small extra component indicated by the difference between DIMM and calibration wavefront sensor image size.
The right panel of the figure presents a poor period with delivered image quality above 2\arcsec, obtained with winds around 20 mph and a 5~$^\circ$C temperature difference since stacking the primary mirror. The temperature difference is reflected in the poorer primary mirror stack as indicated by the Hartman Extra-focal Instrument. The DIMM was reporting site seeing above 1\arcsec, but the calibration wavefront sensor was showing images around 2\arcsec. The difference between the HET air column and free air is a combination of the dome and the environment above the dome. However, the dome seeing component, which is sampled by the Wavescope, indicates that the developed image quality is not dominated by dome seeing. The correlation with wind speed possibly indicates a cause associated with the environment outside the dome.

Such engineering tests are being obtained on a regular basis to investigate components of the delivered image quality and to monitor the optical quality of the system to ensure there is no degradation from the current performance. These measurements are discussed in more detail in \citet{lee21}. To date, however, no trends have emerged that would indicate immediate actions to significantly improve delivered image quality.

In summary, HET is delivering a median image quality between 1.5 and $1 \farcs 8$ FWHM with a Moffat profile, with a tail to poorer images above $2 \farcs 0$, particularly in Trimester 1 (December to March). This trend with trimester has not changed since the original HET began operation, although the upgrade delivers better image quality when the site seeing is good. There is a component to the HET image quality that is attributable to a combination of the primary mirror state, dome seeing, and exterior wind speed. This component was present in the pre-upgrade HET, and was not influenced by the improved image quality of the WFC, or the careful attention to heat sources in the dome during the upgrade. The primary mirror and general dome environment were not altered within the scope of the WFU. Ongoing data collection and engineering tests will hopefully reveal more about the nature of this component and suggest some mitigations, but for the planning of HETDEX it is assumed that the statistics assembled over the past several years are indicative of image quality for the remainder of the survey. As a result, a relaxed image quality criterion for HETDEX observations has been adopted, to allow observing in up to 
$2 \farcs 5$ rather than limiting at $2 \farcs 0$ image quality, with exposure times increased by the observing conditions decision tool to compensate.

\section{Survey Results and Example Spectra}\label{sec:spectra}

As a blind spectroscopy survey of very wide area, HETDEX records spectra of any object that falls on the $\sim$35k fibers. VIRUS has also been used for targeted observations of extended objects and in parallel with the other science instruments. This section presents \edit1{an overview of data analysis, some early science results, and} a sampling of spectra to illustrate the diversity of objects accessible to the instrument.

Data are copied automatically to the Texas Advanced Computing Center (TACC) at the University of Texas at Austin, where the processing software is installed on TACC supercomputers. 
Data processing and analysis with custom software for HETDEX is described in \citet{geb21}.
In addition to this HETDEX-specific reduction, several software packages have been developed and utilized during spectrograph characterization and operations and for processing observations with VIRUS and LRS2. 
The package {\it Vaccine} was developed for the HETDEX Pilot Survey, which demonstrated the application 
of wide field integral field spectroscopy to blind surveys from instrument to software pipeline \citep{adams11,blanc11}.
The package {\it Cure} \citep{goessl06,snig12,snig14} was developed early in the project and forms the basis for the LabCure package utilized in IFU characterization (\S\ref{sec: IFU}, \S\ref{sec: alignment}) and scripts used for early spectrograph characterization \citep{indahl16}.
At HET, the {\it VIRUS health check} is a Python utility, run on every exposure for real-time feedback throughout the night, that calls {\it Cure} subroutines and acts as quality control for the raw data, raising flags should there be unexpected data properties. 
A Python software package {\it Panacea} was also developed for VIRUS and LRS2\footnote{https://github.com/grzeimann/Panacea}; the latter is run automatically on a nightly basis.
The Python package {\it Remedy} provides on-demand reductions for non-HETDEX observations with VIRUS \citep{zeim21}, including combining data for mosaics, where multiple pointings are used to fill in the gaps between VIRUS IFUs to create maps\footnote{https://github.com/grzeimann/Remedy}. 

\edit1{During an exposure, the metrology stream from the guide probes provides a measurement of the throughput of the system in the g band, used to correct each exposure for variations in telescope illumination over a track and for atmospheric transparency.
The system response and flux calibration of VIRUS observations for HETDEX are fixed relative to (tens of) stars in the VIRUS field of view \citep{geb21} and to both field stars and standard stars for {\it Remedy} data reduction \citep{zeim21}.
The shape of the system response is very stable and calibrations for HETDEX are accurate to 5\% \citep{geb21}. Comparison between flux calibrated spectra of standard stars and other continuum objects observed with both LRS2-B and VIRUS show agreement in the shape of the calibration at the 1\% level in the overlapping spectral region (3700 - 5500 \AA).  LRS2-B has a fully-filled IFU, allowing the differential atmospheric refraction (DAR) shift in image position with wavelength to be followed exactly.
This high level of agreement in system response calibration, internally (between VIRUS and LRS2-B) and externally against stars of known spectral energy distribution, demonstrates that the DAR model used in the extraction of continuum objects in VIRUS data is not a limiting factor in the flux calibration.
}

\begin{figure}[!ht]
\epsscale{0.9}
\plotone{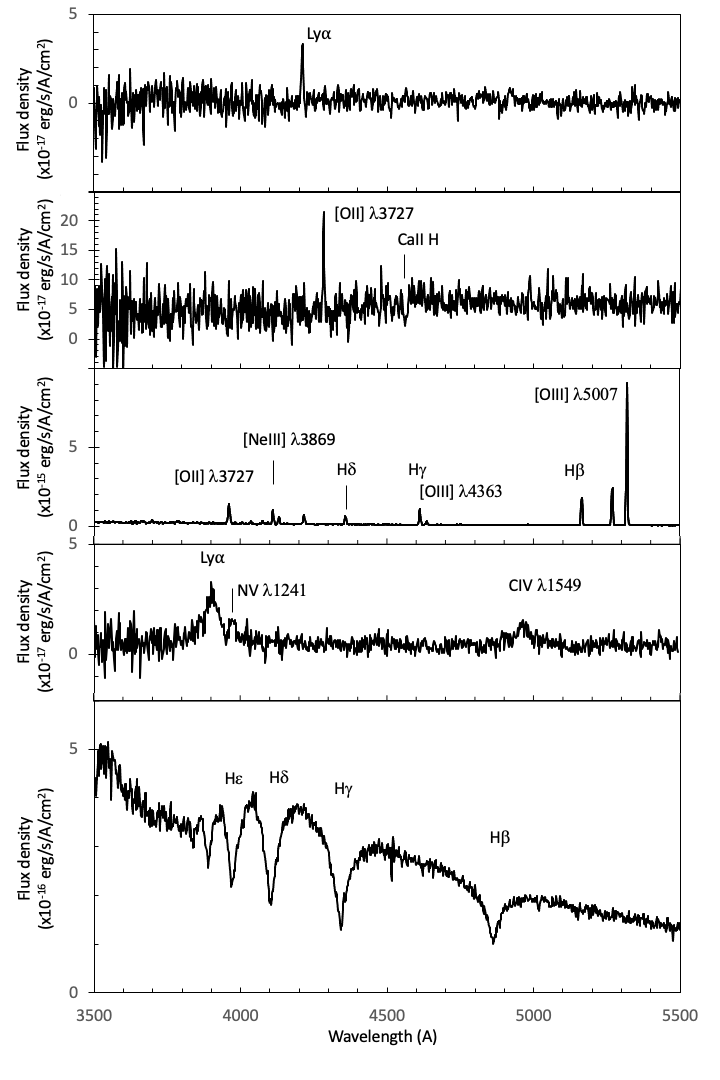}
\caption{
\label{examplespec} \footnotesize
Examples of emission line and continuum sources detected in VIRUS data for HETDEX. From top to bottom: a Lyman-$\alpha$ emitting galaxy at $z$=2.46;  an [OII] emitting galaxy at $z$=0.149. The [OII] emitter has detected continuum emission in addition to the emission line; a low mass local galaxy selected for strong [OIII] emission and large [OIII]$/$[OII] emission line ratio indicative of low metallicity 
\citep{indahl21a};
a broad line AGN at $z$=2.209 with faint continuum of g=22.5; a DA white dwarf 
\citep{hawk21}. 
}
\end{figure}

\edit1{As noted in \S\ref{subsec:target}, VIRUS can observe simultaneously with the other HET instruments in parallel mode. Any exposure of 300 seconds or longer with LRS2 or HPF as the primary instrument automatically triggers a secondary VIRUS exposure of the same duration. 
The parallel mode and the resulting continuum object spectral catalog of the HET VIRUS Parallel Survey (HETVIPS) are discussed in \citet{zeim21}.
Parallel observations are not dithered, but any object falling on a VIRUS fiber will have a spectrum recorded with the wavelength-dependent sensitivity depending on the position of the object with respect to the fiber aperture and the image quality. Differential atmospheric refraction shifts the object position about 1$\arcsec$ as a function of wavelength, so sensitivity varies from object to object.}

\edit1{Large areas are being surveyed in parallel mode, which accounts for more than half of VIRUS observations, and $\sim$50~\sqdeg~ has been observed with fiber spectra to date. The continuum object catalog, extracted at the positions of Pan-STAARS1 continuum objects, includes over 200,000 objects. Exposure times vary up to 4500 seconds, with a median of 1800 seconds, and sky brightness varies widely, driven by the primary instrument science. At g=19.5 the typical spectrum has signal-to-noise ratio of 10, averaged over the g filter. }
%HETVIPS has 3,795 shots = 268 sq degrees at ~1/4.5 fill (~60 sq deg; though not all IFUs); Greg estimates perhaps 52 sq degrees. These data obtained over about 2 years; multiply by 5 => 300 sq degrees million objects

\edit1{Parallel observations are reduced with {\it Remedy} independently from HETDEX and largely cover areas outside the HETDEX sky footprint where VIRUS is usually the primary instrument. 
%\citet{zeim21} show the throughput of the HET plus VIRUS system derived from stars in the parallel data that is in very good agreement with that in Figure~\ref{onsky}, providing an independent confirmation of the system throughput derived in \S\ref{subsec:Vthuput}.
\citet{zeim21} demonstrate automatic identification of stars, galaxies and quasars in the parallel dataset. Science projects enabled by parallel observations include the chemical history of the Galaxy, censuses of individual stellar systems such as white dwarfs, and properties of quasars and radio sources out to redshifts of 3.5.
%The database is available to all HET institutions and there will be periodic public data releases. 
It is projected that over a decade, HETVIPS will observe 300 ~\sqdeg~ of sky area spread over the entire HET observing sky area and will detect and classify a million objects in a completely blind survey. 
}

\edit1{As of 2021 September, HETDEX has completed about half the survey and has covered 45 sq. degrees area (area of IFU coverage), recording 300 million spectra and detecting more than a million emission lines and 110,000 continuum objects.} Figure~\ref{examplespec} presents several VIRUS spectra obtained during the course of HETDEX observing. LAEs and emission line galaxies dominate the detected objects, and a million of each are expected in the full survey \citep{geb21}. 
In addition, the survey will contain $\sim$300,000 stars \citep{hawk21}, 50,000 local galaxies brighter than $g$=22, and 10,000 AGN. 
All these objects will be observed without any pre-selection \edit1{and HETDEX observing is expected to be complete in 2024}.

\edit1{The main driver for HETDEX is the determination of the expansion rate of the Universe in the 1.9 $<$ z $<$ 3.5 epoch. The key science requirements that led to the technical requirements in \S\ref{sec:design} are discussed in \citet{geb21}, confirming that the survey is yielding 2.5 LAEs per IFU per observation. Position and redshift accuracy are 0\farcs35 and 100 km$/$s, respectively, well within requirements. Other important requirements that are not related to the instrument capabilities include false positive rates and separation of LAEs from low redshift [OII]$\lambda3727$ emitting galaxies, as discussed in \citet{geb21}.
}

\edit1{The blind spectroscopic nature of the HETDEX survey will yield results in many areas other than cosmology.
Early science publications based on about a third of the eventual dataset provide an indication of the content and applications of the survey.
A first analysis of the luminosity functions of LAEs and AGN in early HETDEX data is presented by \citet{zhang21}. The sample comprises 18,320 LAEs, selected from deep Subaru imaging that overlaps HETDEX,  to probe the rest-frame ultraviolet luminosity function of galaxies and AGN. The sample includes 2,126 broad-line AGN and  confirms that the bright end of the luminosity function is dominated by AGN. At fainter luminosities, the luminosity function is consistent with previous studies 
and future papers will add determinations of the LAE emission line luminosity function as well as measures of their clustering.}
%and provide overwhelming statistics on other properties of LAEs.

%\edit1{Detailed studies of the properties of LAEs and their environments will be enabled by the unprecedented sample size from HETDEX. An interesting application is the study of the intergalactic medium around LAEs probed through {H{\sc i}} Ly$\alpha$ tomography derived from spectra of background quasars and galaxies. \citet{mukae20} provide a first look by combining the HETDEX LAE distribution with {H{\sc i}} tomography in the CLAMATO survey \citep{clamato} and in EGS \citep{davis07}. Voids and density peaks within the {H{\sc i}} gas distribution were correlated with the positions of the HETDEX LAEs to show that strong {H{\sc i}} absorption exists around LAEs. The result shows that LAE host galaxies reside in {H{\sc i}}-gas overdensities with density profiles that can be traced out to 10 Mpc scales. \citet{mukae20} also identify a candidate giant {H{\sc ii}} bubble with low {H{\sc i}} density at z = 2.16 in EGS, coincident with several LAEs and quasars,  indicating a large region ionized by the quasars. Such studies hold the promise of providing detailed information on the intergalactic environments in which LAEs reside at redshifts of z $\sim$ 2-3.}

\edit1{HETDEX will contain examples of rare emission line galaxies that are extremely difficult to identify in imaging surveys, such as the population of nearby (z$<$0.1) metal-poor [OIII] emitting galaxies discovered by \cite{indahl21a}. Galaxies were selected to have high [OIII] $\lambda 5007$ / [OII] $\lambda 3727$ ratio, implying highly ionized nebular emission often indicative of low metallicity systems (Fig.~\ref{examplespec}). Follow-up with LRS2 spectroscopy confirmed their low metallicities and reveal these objects as a population of high star formation rate, low mass galaxies that would not have been selected for follow-up without the blind spectroscopy of HETDEX. }
%\cite{indahl21a} estimate that the full HETDEX survey will contain about 250 such galaxies.}

\edit1{\citet{hawk21} provide a first look at the stars detected in the HETDEX dataset, with a sample of 100,000 stars selected by cross-matching with {\it Gaia} point sources. The spectra cover {\it Gaia} magnitudes 10 $<$ G $<$ 21. This study demonstrates that accurate classifications are possible with the VIRUS spectral resolution as well as radial velocities accurate to 28 km$/$s for stars considerably fainter than the {\it Gaia} radial velocity limit (G$<$14). An interesting result is that there is sufficient information content in the relatively low resolution VIRUS spectra to uncover 416 new metal-poor candidate stars. These were isolated via machine learning methods, which demonstrate that VIRUS spectra can constrain effective temperature, surface gravity, and metallicity. Follow up is underway with higher spectral resolution to verify the accuracy of the values derived from the HETDEX spectra. Additionally, samples of white dwarfs (Fig.~\ref{examplespec}) and chemically peculiar stars with enhanced abundance ratios can also be picked out in the data.
}
%Lyman continuum \citep{davis21}
%3DHST \citep{weiss21}

\section{Summary}\label{sec:summary}

Following a decade of development, the upgraded HET saw first light in 2015, after the two-year installation of the new tracker, new wide field corrector, new prime focus instrument package with new metrology instrumentation, and the new telescope control system. 
HET now has the largest field of view of any 10 m class telescope, operating in the optical and near infrared. 
This upgrade was motivated by HETDEX science requirements and reimagined the HET as a wide field survey instrument, in combination with the replicated VIRUS integral field spectrograph, which places about 35,000 fibers on sky and observes 56 \sqam~ per dithered observation of three exposures, within a field diameter of 18 arcminutes.  
The system achieves a high degree of observing automation and has been in full queue-scheduled science operations since December 2016. HETDEX observations started in January 2017. 
Performance of the telescope plus instrumentation system is much enhanced over the original HET in field area, observational multiplex, and operational efficiency.

VIRUS has the largest grasp (A$\Omega$, collecting area of telescope $\times$ area of sky observed $\simeq$ 4.7$\times$10$^6$ m$^2$\sqas) of any spectrograph and represents a milestone in the development of highly-multiplexed instruments based on large-scale replication (defined as requiring more then 100 copies of a base instrument; \citealt{hil14}). It has a similar number of detector pixels to the largest imagers.
Other spectrographs employing significant levels of replication include the VLT MUSE integral field spectrograph with 24 spectrograph channels \citep{bacon10}, the LAMOST multi-object spectrograph with 32 spectrograph channels \citep{lamost}, and the DESI multi-object spectrograph with 30 spectrograph channels \citep{desi}.
VIRUS provides a first, and currently unique, opportunity to understand the performance of spectrographs replicated on a 100-fold scale. Such instruments have applications on future extremely large telescopes as discussed in \citet{hil14}. The experience with VIRUS is that for large-scale replication it is possible to predict the performance and, with care in specifying component requirements, achieve the expectations on-sky.

The combination of the new 10 m wide-field HET with the grasp of VIRUS creates a unique facility, that is designed as a system, able to survey vast areas of sky with un-targeted  spectroscopy for the first time.
This facility opens up sensitive wide-field blind spectroscopy as a new method to view the universe.

\clearpage

\acknowledgments
HETDEX (including the WFU of the HET) is led by the University of Texas at Austin McDonald Observatory and Department of Astronomy with participation from the Ludwig-Maximilians- Universität München, Max-Planck-Institut für Extraterrestriche Physik (MPE), Leibniz-Institut für Astrophysik Potsdam (AIP), Texas A\&M University, Pennsylvania State University, Institut für Astrophysik Göttingen, The University of Oxford, Max-Planck-Institut für Astrophysik (MPA), The University of Tokyo, and Missouri University of Science and Technology. In addition to Institutional support, HETDEX is funded by the National Science Foundation (grant AST-0926815), the State of Texas, the US Air Force (AFRL FA9451-04-2- 0355), and generous support from private individuals and foundations.

We thank the following reviewers for their valuable input at various stages in the project: 
\begin{itemize}
\item Science Requirements Review 26 June 2007, Roland Bacon, Gary Bernstein, Gerry Gilmore, Rocky Kolb, Steve Rawlings
\item Preliminary Design Review 10 April 2008, Bruce Bigelow, Gary Chanan, Richard Kurz, Adrian Russell, Ray Sharples
\item Tracker Factory Acceptance Test Plan Review 8 March 2011, Povilas Palunas, Jeffrey Kingsley, Dave Chaney
\item PFIP Integration and Alignment Review 26 July 2011, Larry Ramsey, Bruce Bigelow, Steve Smee, Mike Smith
\item Wide Field Upgrade Readiness Review, 16 July 2013, Daniel Fabricant, Fred Hearty
\item Wide Field Corrector Pre-shipment Review 22 April 2015, Daniel Fabricant, Fred Hearty 
\item VIRUS Detector System Review, 1 February 2016, Roger Smith, Ian McLean, Phillip MacQueen
\item External Review, Astronomy Department and McDonald Observatory, 19-22 March 2017, Matthew Bershady, David Charbonneau, Martha Haynes, Piero Madau
\end{itemize}

We thank the staffs of McDonald Observatory, the Hobby-Eberly Telescope, and the Center for Electromechanics, University of Texas at Austin, the College of Optical Sciences and the Imaging Technology Lab, University of Arizona, the Leibniz-Institut für Astrophysik Potsdam (AIP), the Department of Physics and Astronomy, TAMU, the Max-Planck-Institut für Extraterrestriche Physik (MPE), The University of Oxford Department of Physics, Universit\"ats-Sternwarte M\"unchen, and the Institut für Astrophysik Göttingen, for their contributions to the HET Wide Field Upgrade and VIRUS. 

We particularly thank the following individuals for their contributions over the course of the project:
Joshua Adams, Richard Allen, Heiko Anwad-Heerwart, Edmundo Balderrama, Timothy Beets, Joseph Beno, Lana Beranek, Emily Bevins, Guillermo Blanc, John Booth, Brent Buetow, Jim Burge, Maria Bustamante, John Caldwell, Dustin Davis, Doug Edmundston, Linda Elliot, Neal Evans, Daniel Farrow,  Eric Frater, Alexander Gehrt, Claus Goessl, Frank Grupp, Kaartik Gupta, Lei Hao, Christian Haubitz-Reinke, Richard Hayes, Dionne M. Haynes, Roger Haynes, James Heisler, Sarah Hinze, John Jackson, Bryan Keener Smith, Jeff Kingsley, Robert Leach, Mike Lesser, Chenxu Liu, Suvrath Mahadevan, Amanda Martin, Emily Martin, Emily McLinden, Jason Mock, Nicholas Mollison, Omar Molina, Brian Murphy, Jeremy Murphy, Chang Jin Oh, David Ouellette, Justen Pautzke, Eric Peng, Dave Perry, Andrew Peterson, Emil Popow, Marc Rafal, Jean-Philippe Rheault, Christer Sandin, Richard Savage, Logan Schoolcraft, David Sheikh, Greg Smith, Michael Smith, Katie Smither, Ian Soukup, Brian South, Mike Tacon, Eusebio Terrazas, Rodrigo Viveros, Douglas Wardell, Gordon Wesley, Gregory Wedeking, Amy Westfall, Michael Worthington, Joseph Zierer.

The Low Resolution Spectrograph 2 (LRS2) was developed and funded by the University of Texas at Austin McDonald Observatory and Department of Astronomy and by Pennsylvania State University. We thank the Leibniz-Institut f\"ur Astrophysik Potsdam (AIP) and the Institut f\"ur Astrophysik Göttingen (IAG) for their contributions to the construction of the integral field units.

We acknowledge the Texas Advanced Computing Center (TACC) at The University of Texas at Austin for providing high performance computing, visualization, and storage resources that have contributed to the results reported within this paper. 
This work makes use of the Pan-STARRS1 Surveys (PS1) and the PS1 public science archive, which have been made possible through contributions by the Institute for Astronomy, the University of Hawaii, the Pan-STARRS Project Office, the Max-Planck Society and its participating institutes. 
This work makes use of data from the European Space Agency (ESA) mission {\it Gaia} (https://www.cosmos.esa.int/gaia), processed by the {\it Gaia} Data Processing and Analysis Consortium (DPAC, https://www.cosmos.esa.int/web/gaia/dpac/consortium). Funding for the DPAC has been provided by national institutions, in particular the institutions participating in the {\it Gaia} Multilateral Agreement.
This work makes use of the Sloan Digital Sky Survey IV, with funding provided by the Alfred P. Sloan Foundation, the U.S. Department of Energy Office of Science, and the Participating Institutions. SDSS-IV acknowledges support and resources from the Center for High-Performance Computing at the University of Utah. The SDSS web site is www.sdss.org.

We thank Frank Bash and Mary Ann Rankin for supporting the HET upgrade and VIRUS at a formative stage, and the members of the HET Board, who over the years supported the project from concept to completion.

We especially acknowledge the role of David L. Lambert who, as McDonald Observatory Director, provided crucial leadership and support for the HETDEX project for a decade.\\
\\
\appendix
\label{appendix}

\section{Glossary and Acronyms used in this paper}
\label{sec:acronyms}
{\tiny
\begin{center}
%\begin{tabular}{lllll}
%\begin{tabular}{lll|ll}
\begin{tabular}{llp{1cm}ll}
AGN	&	Active Galactic Nucleus	& &	MPA	&	Max-Planck-Institut f\"ur Astrophysik	\\
AIP	&	Leibniz-Institut f\"ur Astrophysik Potsdam (AIP)	& &	MPE	&	Max-Planck-Institut f\"ur Extraterrestrische Physik	\\
ARC	&	Astronomical Research Cameras Inc	& &	MUSE	&	Multi-Unit Spectroscopic Explorer	\\
AVT	&	Allied Vision Technology	& &	{\it Panacea}	&	Reduction package for VIRUS and LRS2 data	\\
CCD	&	Charge Coupled Device	& &	OWFS	&	Operational Wavefront Sensor	\\
CWFS	&	Calibration Wavefront Sensor	& &	PCI	&	Peripheral Component Interconnect (VIRUS interface card)	\\
{\it Cure}	&	Reduction package for VIRUS data	& &	PCIe	&	PCI-Express	\\
DAR	&	Differential Atmospheric Refraction	& &	PFIP	&	Prime Focus Instrument Package	\\
DIMM	&	Differential Image Motion Monitor	& &	PSF	&	Point Spread Function	\\
FRD	&	Focal Ratio Degradation	& &	PSU	&	Pennsylvania State University	\\
FWHM	&	Full Width at Half Maximum	& &	{\it Remedy}	&	Reduction package for VIRUS data	\\
GC	&	Guide Camera	& &	Rho axis	&	Axis of rotation about the optical axis of PFIP	\\
HEFI	&	Hartmann Extra Focal Instrument	& &	SALT	&	South African Large Telescope	\\
HET	&	Hobby-Eberly Telescope	& &	STA	&	Semiconductor Technology Associates, Inc	\\
HETDEX	&	Hobby-Eberly Telescope Dark Energy Experiment	& &	{\it Shuffle}	&	Target setup utility	\\
HETVIPS	&	Hobby-Eberly Telescope VIRUS Parallel Survey	& &	TACC	&	Texas Advanced Computing Center	\\
HPF	&	Habitable-zone Planet Finder	& &	TAMU	&	Texas A\&M University	\\
HRS	&	High Resolution Spectrograph	& &	{\it Vaccine}	&	Reduction package for VIRUS-P data	\\
IFU	&	Integral Field Unit	& &	VIRUS	&	Visible Integral-field Replicable Unit Spectrograph	\\
ITL	&	University of Arizona Imaging Technology Laboratory	& &	VIRUS-P	&	VIRUS Prototype	\\
LAE	&	Lyman-$\alpha$ Emitter	& &	{\it VIRUS Health Check}	&	Python utility to check the integrity of VIRUS exposures	\\
LRS2	&	Second generation Low Resolution Spectrograph	& &	VLT	&	Very Large Telescope	\\
LRS2-B	&	LRS2 – Blue unit	& &	WFC	&	Wide Field Corrector	\\
LRS2-R	&	LRS2 – Red unit	& &	WFU	&	Wide Field Upgrade	\\
M2-5	&	WFC Mirrors	& &	WVSC	&	Wavescope	\\
MDO	&	McDonald Observatory	& &		&		\\
\end{tabular}
\end{center}
}

%\begin{figure}[!ht]
%\epsscale{0.9}
%\plotone{acronyms.png}
%\end{figure}

\clearpage

%Chris' bibtex version
%%%%%%%%%%%%%%%%%% BIBLIOGRAPHY %%%%%%%%%%%%%%%%%%%%%%%%%%%%%%%%%%%%%%%%%%
%\clearpage
%\bibliographystyle{apj} % it doesn't work with this definition!
\bibliography{hetdex.bib}

%%%%%%%%%%%%%%%%%%%%%%%%%%%%%%%%%%%%%%%%%%%%%%%%%%%%%%%%%%%%%%%%%%%%%%%%%%

%bibtex is whining about lack of publisher in the SPIE papers; I looked at a couple of your ApJ papers with SPIE citations, and there is no publisher given.  Additionally, in looking on line this seems to be a common headache; let's ignore it

%Need to resolve the difference between Book and Proceedings for SPIE\\

%\end{thebibliography}



\clearpage

% upgraded HET properties
\begin{center}
\begin{deluxetable}{@{\extracolsep{0.6in}}lcc}
\tabletypesize{\normalsize}
\tablewidth{0pt}
\tablecaption{HET Properties \label{tab-het}}
\tablecolumns{3}
\tablehead{
\colhead{Property}					&
\colhead{Original HET}  &
\colhead{Upgraded HET}
}
\startdata
Aperture on axis & 9.2 m & 10.0 m \\
Focal length @ Focal-ratio & 42,718 @ f$/$4.64 & 36,500 mm @ f$/$3.65 \\
Plate scale (arcsec.$/$mm) & 4.83 & 5.65 \\
Field of View (arcmin. diameter)  & 4.0 (science and guiding)  & 18.0 science, 22.0 guiding \\
\hline
Accessible declination range (note 1) & \multicolumn{2}{c}{-10.3$^\circ$ to +71.6$^\circ$} \\
Maximum declination range (note 2) & \multicolumn{2}{c}{-12$^\circ$ to +74$^\circ$} \\
Minimum track time (@ declination, note 3) & \multicolumn{2}{c}{50 minutes (@-10.0$^\circ$)} \\
Maximum track time (@ declination) & \multicolumn{2}{c}{2.8 hours (@+67.2$^\circ$)} \\
Maximum time on target (note 4) & \multicolumn{2}{c}{5.07 hours (@+63.5$^\circ$)} \\
\enddata
Note 1: for tracks with the pupil passing through the primary mirror center at track center; \\
Note 2: for tracks of 30 minutes that do not have a fully illuminated pupil at track center; \\
Note 3: minimum track time where the pupil center passes through the primary mirror center; \\
Note 4: in one night on the same target with a combination of east and west tracks that do not overlap in time.
\end{deluxetable}
\end{center}

\begin{center}
\begin{deluxetable}{@{\extracolsep{0.6in}}ll}
\tabletypesize{\normalsize}
\tablewidth{0pt}
\tablecaption{HET Wide Field Upgrade Chronology\label{tab-chron}}
\tablecolumns{2}
\tablehead{
\colhead{Event or Milestone}					&
\colhead{Date}
}
\startdata
HET taken off line & Sep-2013 \\
New Tracker install completed & Feb-2014 \\
Wide Field Corrector installation & May-2015 \\
PFIP installation & Jul-2015 \\
First Light & 29-Jul-2015 \\
New Low Resolution Spectrograph (LRS2) installed & Nov-2015 \\
First 16 VIRUS spectrograph units installed & May-2016 \\
Early science operations commence & Jul-2016 \\
Full queue science operations with LRS2 (2 weeks per month) & Dec-2016 \\
HETDEX Survey Starts & Jan-2017 \\
Queue science operations transition to 3 weeks per month & Jul-2017 \\
Habitable-zone Planet Finder (HPF) delivered & Oct-2017 \\
HPF early science operations commence & May-2018 \\
VIRUS deployment reaches 40 units (half of the total, 18k fibers) & Jun-2018 \\
Full science operations commence without bright-time engineering periods & Jun-2018 \\
%VIRUS deployment reaches 74 units required for HETDEX (95\% of total, 33k fibers) & May-2021 \\
VIRUS deployment reaches full 78 units on sky (35k fibers) & Aug-2021 \\
% reached 74 on sky on 28 May, 2021
% reached 77 on sky on 9 July 2021, UT
\enddata
\end{deluxetable}
\end{center}

\begin{center}
\begin{deluxetable}{@{\extracolsep{0.3in}}lcl}
\tabletypesize{\normalsize}
\tablewidth{0pt}
\tablecaption{VIRUS Spectrograph Channel Properties \label{tab-virus}}
\tablecolumns{3}
\tablehead{
\colhead{Property}					&
\colhead{Value}  &
\colhead{Note}
}
\startdata
Fiber core diameter & 266 $\mu$m = 1.5 arcsec & fed at f$/$3.65 by WFC \\
Number of fibers $/$ channel & 224 & 2 channels per IFU and per VIRUS unit \\
Collimator focal ratio & f$/$3.33 & oversized to allow for FRD \\
Beam size & 125 mm & including the margin for FRD \\
Grating fringe frequency & 930 lines$/$mm & light incident at 12.15$^\circ$ to grating normal\\
%  & & \hspace{0.05in} to grating normal \\
Schmidt Camera focal ratio & f$/$1.25 & f/1.33 on axis \\
CCD format & 2064x2064 & with 15 $\mu$m pixels \\
Wavelength coverage & 3500 - 5500\AA & Resolving power $\simeq$ 800 at 4500\AA \\
\enddata
%Note 1: Volume phase holographic gratings \\
\end{deluxetable}
\end{center}

\begin{center}
\begin{deluxetable}{lllll}
\tabletypesize{\scriptsize}
\tablewidth{0pt}
\tablecaption{HET Wide Field Upgrade Performance\label{tab-wfupgrade}}
\tablecolumns{5}
\tablehead{
\colhead{Performance Area} 			&
\colhead{Requirement}				&
\colhead{Original HET}				&
\colhead{Upgraded HET}				&
\colhead{Comment}                           
}
\startdata
Pupil diameter						&  10 m class						&  9.2 m					&  10.0 m								&  At center of track \\
								&								&						&									& \\
Field of view (diameter)						&  22 arcmin				&  4 arcmin				&  22 arcmin							&  70 times larger area at same \\
								&								&						&									&  \hspace{0.05in} level of field vignetting \\
																&								&						&									& \\
Median on-axis image quality 			&  1.25 arcsec						&  1.7 arcsec				&  1.3 arcsec							&  50\% encircled energy diameter  \\
 											&						&									& & \hspace{0.05in} in 1.0 arcsec site seeing \\
Open-loop pointing				&  25 arcsec rms  						&  30 arcsec rms				&  10 arcsec rms 			&  Achieved over full tracker range\\
								&  \hspace{0.05in} (goal 9 arcsec rms)		&						&  \hspace{0.05in} 75\% $<$ 12 arcsec absolute									&   \\
																&								&						&									& \\
Setup time\tablenotemark{a} 			&  $<$ 5 minutes 90\%				&  10-20 mins				&  4-7 minutes 							&  Upgrade can set up blind on invisible \\
								&								&						&   \hspace{0.05in} ($<$ 5 min 50\% of time)	&  \hspace{0.05in} targets in same setup time \\
																&								&						&									& \\
HETDEX setup time					&  5 min with Az move;				&  \nodata					&  Median 4 min independent of			&  With observing conditions decision tool \\
								&  \hspace{0.05in} (goal 1.5 min, no move)		&						&   Az move; $<$5 min 90\% of time			& \hspace{0.05in} scheduling automation   \\
																&								&						&									& \\
Setup accuracy (rms)				&  0.25 arcsec 						&  0.5 arcsec				&  0.2 arcsec							&  Ability to center target on fiber, \\
								&  \hspace{0.05in} (goal 0.1 arcsec)		&						&									&  \hspace{0.05in} slit, or IFU\tablenotemark{b} \\
																&								&						&									& \\
Metrology System					& Full sensing of all					&  Only guiding				& All degrees							&  Wavefront sensing feedback completes \\
								&  \hspace{0.05in} degrees of freedom	&						&									&  \hspace{0.05in} all degrees of freedom, particularly focus \\
																&								&						&									& \\
Guiding residuals					&  0.25 arcsec rms					&  0.5 arcsec				&  0.10 arcsec rms						&  Including field rotation and plate scale \\
								&								&						&									& \\
focus tracking						& 15 $\mu$m rms					&  200 $\mu$m	&   17 $\mu$m rms								&  Under wavefront sensor control \\
								&								&						&									& \\
tip/tilt residuals						& 10 arcsec rms					&  30 arcsec				& 0.6 arcsec rms						&  Tip/tilt tracked by metrology system \\
								&								&						&									& \\
\enddata

\vspace*{0.1in}
\tablenotetext{a}{defined as the end of one exposure to start of the next one}
\tablenotetext{b}{these have been measured on the old LRS and new LRS2 spectrographs for the original and upgraded HET, respectively}
\end{deluxetable}
\end{center}

%\clearpage
\begin{center}
\begin{deluxetable}{@{\extracolsep{0.6in}}cccccc}
\tabletypesize{\normalsize}
\tablewidth{0pt}
\tablecaption{HET Wide Field Upgrade Image Quality\label{tab-summimage}}
\tablecolumns{6}
\tablehead{
\colhead{Trimester}					&
\colhead{2016}						&
\colhead{2017}						&
\colhead{2018}						&
\colhead{2019}                      &
\colhead{2020}
}
\startdata
\multicolumn{4}{c}{Median Delivered Image Quality (FWHM)}	\\
T-1	&  \nodata			&  1.82$\arcsec$	&  1.78$\arcsec$  &  1.87$\arcsec$ &  1.85$\arcsec$ \\
T-2	&  1.50$\arcsec$	&  1.57$\arcsec$	&  1.62$\arcsec$  &  1.64$\arcsec$ &  1.66$\arcsec$ \\
T-3	&  1.57$\arcsec$	&  1.53$\arcsec$	&  1.49$\arcsec$  &  1.55$\arcsec$ &  1.57$\arcsec$ \\
\multicolumn{4}{c}{Median DIMM\tablenotemark{a} (FWHM)}	\\
T-1	&  \nodata			&  1.13$\arcsec$	&  1.17$\arcsec$  &  1.15$\arcsec$ &  1.23$\arcsec$ \\
T-2	&  1.01$\arcsec$	&  1.00$\arcsec$	&  1.06$\arcsec$  &  1.03$\arcsec$ &  1.08$\arcsec$ \\
T-3	&  1.03$\arcsec$	&  1.01$\arcsec$	&  1.03$\arcsec$  &  0.94$\arcsec$ &  1.02$\arcsec$ \\
\enddata
\vspace*{0.3in}
\tablenotetext{a}{DIMM = differential image motion monitor}
\end{deluxetable}
\end{center}

%\clearpage
%\begin{center}
%\begin{deluxetable}{lcccl}
%\tabletypesize{\scriptsize}
%\tablewidth{0pt}
%\tablecaption{HET Wide Field Upgrade Image Quality\label{tab-iqcomponent}}
%\tablecolumns{4}
%\tablehead{
%\colhead{Factor}					&
%\colhead{Track 1}					&
%\colhead{Track 2}					&
%\colhead{Track 3}					&
%\colhead{Image Quality (IQ) Component Sampled by Data}				                         
%}
%\startdata
%Guider		&    1.0	&   1.18	&   1.58	&  delivered IQ from the guider images \\
%CWFS		&   0.78	&   1.09	&   1.68	&  site seeing measured in sub-aperture of calibration wavefront sensor (DIMM surrogate) \\
%HEFI			&  0.498	&   0.486	&   0.531	&  CCAS extrafocal imager measures overall PM IQ plus dome seeing \\
%Wavescope	&   0.32	&   0.32	&   0.568	&  sub-aperture PM IQ (without stacking component) \\
%implied stack	&   0.38	&   0.37	&\nodata	&  quadrature difference between HEFI and wavescope measures \\
%predicted IQ	&   0.99	&   1.24	&   1.79	&  quadrature sum of CWFS, HEFI, \& PM figure (compared to Guider) \\
%\enddata
%\end{deluxetable}
%\end{center}

\end{document}